\documentclass[12pt,a4paper]{article}       
\usepackage[latin1]{inputenc}   
\usepackage{amsmath} 
\usepackage{amsfonts}      
\usepackage{amssymb}  
\usepackage{graphicx} 
\usepackage{epstopdf} 
\usepackage{color}
\usepackage{here}
\usepackage{subcaption}
\usepackage{mathpazo}
\usepackage{multirow}
\usepackage{array}
\usepackage{hyperref}

\usepackage[left=2cm,right=2cm,top=2cm,bottom=2cm]{geometry} 

\newcommand{\alphab}{\alpha_\text{b}}
\newcommand{\alphad}{\alpha_\text{d}}
\newcommand{\alphabold}{\boldsymbol{\alpha}}

\newcommand{\Beq}{B_\text{eq}}

\newcommand{\Ccal}{\mathcal{C}}

\newcommand{\Dcal}{\mathcal{D}}

\newcommand{\ebold}{\mathbf{e}}

\newcommand{\Fcal}{\mathcal{F}}

\newcommand{\Feq}{F_\text{eq}}

\newcommand{\Ftilde}{\widetilde{F}}

\newcommand{\Ffrakeq}{\Fcal_\text{eq}}

\newcommand{\Gammabold}{\boldsymbol{\Gamma}}

\newcommand{\Hfrakeq}{\Hcaleq}
\newcommand{\Heq}{H_\text{eq}}

\newcommand{\Htilde}{\widetilde{H}}

\newcommand{\Hcaleq}{\mathcal{H}_\text{eq}}

\newcommand{\Ical}{\mathcal{I}}

\newcommand{\Ibold}{\mathbf{I}}

\newcommand{\Jb}{J_\text{b}}
\newcommand{\Jd}{J_\text{d}}
\newcommand{\Jbar}{\bar{J}}
\newcommand{\Jtilde}{\tilde{J}}
\newcommand{\Jbold}{\mathbf{J}}

\newcommand{\Jboldstar}{\mathbf{J}_{\ast}}

\newcommand{\Jtildebold}{\mathbf{\tilde{J}}}

\newcommand{\kpar}{k_{\parallel}}

\newcommand{\Mcal}{\mathcal{M}}

\newcommand{\Ncal}{\mathcal{N}}
\newcommand{\Ncaleq}{\mathcal{N}_\text{eq}}

\newcommand{\nb}{n_\text{b}}
\newcommand{\nbold}{\mathbf{n}}

\newcommand{\nueff}{\nu_\text{eff}}
\newcommand{\nustar}{\nu_{\ast}}

\newcommand{\nubar}{\bar{\nu}}

\newcommand{\Ocal}{\mathcal{O}}
\newcommand{\omegab}{\omega_\text{b}}
\newcommand{\Omegab}{\Omega_\text{b}}
\newcommand{\Omegad}{\Omega_\text{d}}

\newcommand{\Omegabold}{\boldsymbol{\Omega}}

\newcommand{\phieq}{\phi_\text{eq}}

\newcommand{\Phieq}{\Phi_\text{eq}}

\newcommand{\Scal}{\mathcal{S}}
\newcommand{\Scalres}{\mathcal{S}_\text{res}}
\newcommand{\Scalcoll}{\mathcal{S}_\text{coll}}
\newcommand{\Sdot}{\dot{S}}

\newcommand{\Sresdot}{\dot{S}_\text{res}}
\newcommand{\Scolldot}{\dot{S}_\text{coll}}

\newcommand{\Stdot}{\dot{S}_t}

\newcommand{\Tcal}{\mathcal{T}}
\newcommand{\thetab}{\theta_\text{b}}

\newcommand{\thetabold}{\boldsymbol{\theta}}

\newcommand{\Teq}{T_\text{eq}}

\newcommand{\tbold}{\mathbf{t}}

\newcommand{\Ucal}{\mathcal{U}}
\newcommand{\Ueq}{U_\text{eq}}
\newcommand{\Ueqdag}{U_\text{eq}^{\dag}}
\newcommand{\Utilde}{\widetilde{U}}

\newcommand{\Ufrakeq}{\Ucal_\text{eq}}
\newcommand{\Ufrakeqdag}{\Ucal_\text{eq}^{\dag}}

\newcommand{\vpar}{v_{\parallel}}

\newcommand{\Vpareq}{V_{\parallel\text{ eq}}}

\newcommand{\Wpareq}{W_{\parallel\text{ eq}}}

\begin{document}
\title{\textbf{A reduced model for neoclassical transport in non-axisymmetric magnetic configurations}}
\author{Y. Ban$^{1}$, X. Garbet$^{1,2}$, R. Varennes$^{2}$, Z. Qu$^{1}$, K. Lim$^{1}$, Y. Sarazin$^{2}$, A. Ibrahim$^{1}$,\\ Y.C. Tay$^{1}$, B.B. Ang$^{1}$}
\date{%
    $^1${\small School of Physical and Mathematical Sciences, 
		Nanyang Technological University, 637371 Singapore}\\ %
    $^2${\small CEA, IRFM, Saint-Paul-lez-Durance, F-13108, France}\\ 
    \vspace{5mm}%
    {\small \today}
}
\maketitle
\abstract 
A unified framework is proposed to compute a neoclassical equilibrium for tokamaks with non-axisymmetric magnetic perturbations, based on a Hamiltonian description combined with a variational approach that relies on a principle of minimum entropy production. 
This approach allows for a transparent identification of the respective contributions from different classes of trapped particles in magnetic well minima. 
Analytic expressions are derived for specific collisionality regimes, and a fast and efficient numerical procedure is devised to compute the entropy production rate. 
Finding the extremum of the entropy production leads to tractable expressions of transport coefficients that can be computed quickly and efficiently. 

\section{Introduction}
\label{Introduction}

Neoclassical transport in three-dimensional magnetic configurations is of considerable interest in fusion research. 
Increasingly, attention is being brought to the use of externally applied helical magnetic perturbations in tokamaks \cite{Evans_2004, Park_2009, Shaing_2015, Park_2021}, the impact of magnetic ripple and error fields \cite{Kovrizhnykh_1999, Kovrizhnykh_2003, Saibene_2007, Nave_2010, Varennes_2023}, and optimisation of stellarators \cite{Nuehrenberg_1988, Mynick_2006, Boozer_2008, Helander_2014}, to list some of the most prominent examples. 
This work focuses on tokamaks with externally applied magnetic ripple or non-resonant helical perturbations. Neoclassical transport in such complex geometries is usually explored using drift-kinetic codes \cite{Hirshman_1986, Satake_2010, Landreman_2014, Belli_2015}; while these are highly comprehensive, they do not always provide straightforward physical insight into the competing mechanisms involved.

This work proposes a unified framework for computing neoclassical equilibria, based on combining a Hamiltonian description with a variational formulation using a principle of minimum entropy production \cite{Rosenbluth_1972,Samain_1977, Requin_1981, Samain_1997, Garbet_2009, Garbet_2010b}. A key advantage of this method is the enablement of transparent identification of respective contributions of different classes of trapped particles in magnetic well minima. It thereby provides a clear interpretation of the mechanisms responsible for the largest transport coefficients.

In tokamaks, we consider the $1/R$ toroidal perturbation combining with the helical perturbation. Starting with one perturbation, the first source of entropy production arises from particle trapping, which corresponds to neoclassical transport in configurations possessing a symmetry, here referred to as the ``primary'' perturbation. 
When adding the effect of the other perturbation, denoted the ``secondary'' perturbation, further transport contributions are introduced through uncompensated magnetic drifts. 
In the case of a rippled tokamak, or a tokamak with a single helical perturbation, we recover the four basic contributions to neoclassical fluxes \cite{Mynick_2006}, these being the axisymmetric, helically symmetric, superbanana, and banana drift transport contributions. 
In the absence of precession, the latter two are known to lead to the detrimental ``$1/\nu$'' regime, where transport is inversely proportional to collision frequency $\nu$ due to the breakdown of magnetic drift compensation. 
Fortunately, magnetic and $E\times B$ drifts, primarily the latter, improve this scaling successively to $\sqrt{\nu}$ and ultimately $\nu$ at sufficiently low collisionality. The $\nu$ limit is discussed but not implemented in our model, as it corresponds to a range of parameters that are not applicable to tokamaks. Neither will we cover the Pfirsch-Schl\"uter regime in this paper, as the regime only prevails at very high collisionality and is not related to particle trapping. All other expected collisionality scalings are recovered together with a clear physical interpretation. 
The present approach also provides a consistent description of particle, momentum, and heat transport.

This methodology leads to a reduced model which enables a fast evaluation of neoclassical transport. 
The objective is not highly precise predictions, but rather the identification of transport trends in arbitrary three-dimensional configurations. 
Comparisons with simulations performed using the full gyrokinetic code GYSELA \cite{Grandgirard_2016} and the drift-kinetic code NEO3D \cite{Belli_2015} have already shown fair agreement in the case of rippled tokamaks \cite{Varennes_2023}. 
The present work opens the way to similar benchmarks for helical perturbations.

The rest of this paper is organised as follows. 
A general Hamiltonian framework in the presence of two static perturbations is presented in Section \ref{sec:Hamiltonian framework}, where a variational principle based on minimum entropy production is computed from a mean-field Fokker-Plank equation.
Entropy production rates associated with primary and secondary perturbations are derived in Sections \ref{sec:Primary_entropy_production_rate} and \ref{sec:Secondary_entropy_production_rate}, respectively. 
This model is then applied to tokamaks, where explicit forms of the entropy production and a method for the computation of neoclassical fluxes are given in Section \ref{sec:Computation of neoclassical fluxes}.
Lastly, conclusions are presented in Section \ref{Conclusion}.


\section{Hamiltonian framework}
\label{sec:Hamiltonian framework}
\subsection{Hamilton equations in angle/action coordinates}
\label{Generalisation to arbitrary helicities}

We address a general magnetized plasma physics problem in angle/action variables, written as $(\alphabold,\Jbold)$, where the Hamiltonian is of the form 
\begin{equation}
	\nonumber
    H (\alphabold , \Jbold , t )  = \Heq (\Jbold) + 
			\Htilde (\alphabold , \Jbold ) +
			\Htilde' (\alphabold , \Jbold )			
	\label{eq:total_Hamiltonian_double_perturbation}
\end{equation}
The equilibrium Hamiltonian $\Heq$ depends only on actions which satisfy the following equations of motion
\begin{eqnarray} 
   \frac{d\alpha_i}{dt} & = & \frac{\partial \Heq}{\partial J_i}
	  = \Omega_i
      \label{eq:hamilton_eq_angle}
      \\
		\frac{dJ_i}{dt} & = & 
		- \frac{\partial \Heq}{\partial \alpha_i} = 0
	\label{hamilton_eq_action}
\end{eqnarray}  
where $\Omega_i(\Jbold)$ are the Hamiltonian angular frequencies. 
The perturbation Hamiltonians are 
\begin{equation}
	\nonumber
    \Htilde ( \alphabold , \Jbold )  = 
		   - h(\Jbold) \cos 
			 \left( \nbold \cdot \alphabold \right) 
	\label{eq:primary_perturbation}
\end{equation} 
and
\begin{equation}
	\nonumber
    \Htilde' (\alphabold , \Jbold )  = 
		   - h'(\Jbold) \cos 
			 \left( \nbold' \cdot \alphabold \right) 
	\label{eq:second_perturbation}
\end{equation} 
with $\nbold = (n_1=0, n_2, n_3)$  and $\nbold' = (n_1'=0, n_2', n_3')$ for all cases of interest.
The amplitude perturbations $h$ and $h'$ are assumed to depend slowly on the action variables $\Jbold$, 
and the meaning of ``slow'' will be further explained in later sections. 
This general problem can be applied to tokamaks by choosing an appropriate equilibrium Hamiltonian $\Heq$ and parameters $(h, \nbold)$ and $(h', \nbold')$ characterising the perturbations.
This is done in Section \ref{sec:Computation of neoclassical fluxes}.

\subsection{Fokker-Planck equation}
\label{Fokker-Planck equation}
The distribution function $F(\alphabold , \Jbold , t)$ of a given species is solution of the Fokker-Planck equation
\begin{equation}
	\frac{\partial F }{\partial t} - 
		  \left\{ H , F \right\} = C[F]
 \label{eq:Fokker_Planck_equation_xp}
\end{equation} 
where $C[F]$ is the collision operator. 
The Poisson bracket $\left\{ . , . \right\}$  in $(\alphabold , \Jbold)$ variables is
\begin{equation}
	\nonumber
    \left\{ H , F \right\} = 
		\frac{\partial H}{\partial \alphabold } \cdot
		\frac{\partial F}{\partial \Jbold} -
		\frac{\partial H}{\partial \Jbold} \cdot
		\frac{\partial F}{\partial \alphabold } 
	\label{eq:Poisson bracket_xp}
\end{equation}

The unperturbed distribution function $\Feq$ is then a steady-state solution of Eq.(\ref{eq:Fokker_Planck_equation_xp}) with $H = \Heq(\Jbold)$. 
In the absence of collisions, the solutions are arbitrary functions of the actions $\Jbold$; these are the ``canonical'' distribution functions. 

Anticipating that solutions of the problem will be close to Maxwellian, we choose a generic form of the unperturbed distribution function such that
\begin{equation}
  \nonumber
   	\Feq(\Jbold,t) = F_0 \exp\left\{ - \frac{ \Heq(\Jbold) - \Ueq(\Jbold,t) }{T_\ast} \right\}
	\label{eq:distribution_potential}
\end{equation}
where $F_0$ is a constant, and $\Ueq$ is a ``thermodynamic potential'' defined up to a constant that can be absorbed into $F_0$. 
The ratio $\Ueq / T_\ast$ measures the distance to thermodynamic equilibrium, which we will assume to be $\Ueq / T_\ast \ll 1 $, i.e. close to equilibrium.

\subsection{Mean field theory}
\label{Mean field theory}
The distribution function and Hamiltonian are split into an angle-averaged part and a ``perturbed'' part
\begin{eqnarray} 
	  F( \alphabold , \Jbold , t ) & = & 
	     \Feq( \Jbold , t ) + 
		   \Ftilde( \alphabold , \Jbold, t ) 
	\label{eq:split_F_perturbed_unperturbed}	\\
		 H( \alphabold , \Jbold ) & = & 
	   \Heq( \Jbold) + 
		 \Htilde( \alphabold , \Jbold )
 \label{eq:split_H_perturbed_unperturbed}		
\end{eqnarray}
The collision operator $\Ccal[F]$ is linearised in $F$ near a Maxwellian distribution. The equations over $\Feq$ and $\Ftilde$ are then readily obtained
\begin{eqnarray} 
	  & & \frac{\partial \Feq }{\partial t } 
		- \langle 
				\{ \Htilde , \Ftilde \} 
			\rangle_{\alphabold}	
		= \Ccal[\Feq]
	\label{eq:kinetic_equation_mean} \\
	  & & - \{ H , \Ftilde \}  
	   + \langle \{ \Htilde , \Ftilde \} \rangle_{\alphabold}	 
		 - \Ccal[\Ftilde] = 
		\{ \Htilde , \Feq \}
  \label{eq:kinetic_equation_perturbed}		
\end{eqnarray}
where the brackets $\langle . \rangle_{\alphabold}$ denote an average over the angles $\alphabold$. 
The term $\langle \{ \Htilde , \Ftilde \} \rangle_{\alphabold} $ governs the evolution of the equilibrium distribution function $\Feq$, as seen from Eq.(\ref{eq:kinetic_equation_mean}). This evolution is mostly diffusive, with a diffusion coefficient of the order of $D \simeq \Htilde ^2 \tau_c $, where $\Htilde $ is a representative amplitude of the Hamiltonian perturbation, and $\tau_c$ is a an interaction time determined by resonant processes (for a fusion device, transit motion along field lines) or collisions. The transport coefficient $D$ thus corresponds to diffusion in the action space, and is therefore homogeneous to the square of an action divided by time. Consequently, the equilibrium distribution transport $\Feq$ evolves over a diffusive time scale $\tau_D \simeq J_F^2 / D$, where $J_F$ is the typical gradient scale of the equilibrium distribution function $\Feq$ in the action space. The perturbed distribution function $\Ftilde $ evolves over a much shorter time scale $\tau_c \ll \tau_D$. Focusing on Eq.(\ref{eq:kinetic_equation_perturbed}), comparing $\{ H , \Ftilde \} $ to $\langle \{ \Htilde , \Ftilde \} \rangle_{\alphabold} $, the ratio is
\begin{equation}
    R = \frac{\{ H , \Ftilde \}}
       {\langle \{ \Htilde , \Ftilde \} \rangle_{\alphabold}}
       \simeq \frac{\Ftilde}{\Feq} \frac{\tau_D}{\tau_c}
 \label{eq:ratio_term_Vlasov}
\end{equation}
Anticipating that $R\gg 1$, the balance between the first and third terms of Eq.(\ref{eq:kinetic_equation_perturbed}) to its r.h.s. provides an estimate of the perturbed distribution function
\begin{equation}
 \nonumber
    \Ftilde \simeq \Htilde \tau_c  \frac{\Feq}{ J_F}
      \simeq \Feq \sqrt{\frac{\tau_c}{ \tau_D}}
 \label{eq:perturbed_distribuion_function}
\end{equation}
where the scalings derived from the transport equation Eq.(\ref{eq:kinetic_equation_mean}) have been used. Plugging this estimate in Eq.(\ref{eq:ratio_term_Vlasov}) provides the estimate $R\simeq \sqrt{\tau_D/\tau_c} \gg 1$. Thus, it appears that the transport term can be ignored in the equation governing the perturbed distribution function Eq.(\ref{eq:kinetic_equation_perturbed}).
The set of equations Eqs.(\ref{eq:kinetic_equation_mean}, \ref{eq:kinetic_equation_perturbed}) lines up with the rationale of a mean field theory, where fast varying fluctuations are driven in a system that is not at full thermodynamic equilibrium. In turn, these fluctuations produce a slow relaxation of the distribution function towards its equilibrium value. 
\subsection{Entropy production and variational principle}
\label{subsec:Entropy_production_and_variational_principle}
We will attempt to compute transport coefficients by using a variational principle based on a minimum of entropy production, due to its theoretical utility and instructiveness. 
The entropy for each species is defined as
\begin{equation}
 \nonumber
    S = - \int d\gamma \: F \ln F
 \label{eq:entropy_definition}
\end{equation}
where $d\gamma = d^3\alphabold\ d^3\Jbold$ is the volume element in phase space. 
For a given total number of particles, the entropy production rate reads
\begin{equation}
 \nonumber
    \Sdot = - \int d\gamma  \: \frac{\partial F}{\partial t} \ln F
 \label{eq:entropy_production_rate}
\end{equation}
In the absence of collisions, the entropy production rate is exactly zero. Let us write the full distribution in the form
\begin{equation}
    F = \exp\left(- \frac{ H - U }{T_\ast}\right)
 \label{Distribution_thermodynamic_potential}
\end{equation}
where $U$ is a generalisation of the thermodynamic potential $\Ueq$, and can be written as the sum of the mean and perturbed contributions, $U = \Ueq + \Utilde$, and $T_\ast$ is an arbitrary reference temperature. The entropy production rate can then be rewritten as
\begin{equation}
   \Sdot = - \frac{1}{T_\ast} \int d\gamma  \:
		        \frac{\partial F}{\partial t} \: U
 \label{eq:entropy_production_rate_U}
\end{equation}
when energy is conserved for each species, such that temperatures are approximately equal. Let us consider the implications for a mean field theory. 
A mean entropy production rate is
\begin{equation}
    \Stdot = -   
		 \frac{1}{T_\ast}  \int d\gamma \: 
		 \frac{\partial \Feq}{\partial t}\:  \Ueq = -   
		 \frac{1}{T_\ast^2}  \int d\gamma \: \Feq \:
		 \frac{\partial \Ueq}{\partial t} \Ueq
 \label{eq:entropy_production_rate_equilibrium}
\end{equation}
Given the structure of Eq.(\ref{eq:entropy_production_rate_equilibrium}), a reformulation in terms of the thermodynamic potential alone is useful. The flux $\Gammabold$ can be redefined as a function of the perturbed potential $\Utilde$ as follows
\begin{equation}
   \nabla_\Jbold \cdot \Gammabold = 	- \left\langle 
				\left\{ \Htilde , \Ftilde \right\} 
			\right\rangle  = 	- \frac{\Feq}{T_\ast}  \left\langle 
				 \left\{ \Htilde , \Utilde \right\} 
			\right\rangle =  \frac{\Feq}{T_\ast}  \nabla_\Jbold \cdot \Gammabold_U
 \label{eq:value_flux_modified}
\end{equation} 
with 
\begin{equation}
   \Gammabold_U = \int \frac{d^3\alphabold}{\left(2\pi\right)^3} 
			\left( - \frac{\partial \Htilde }
			       {\partial \alphabold }
			\right) \Utilde  
 \label{eq:definition_flux_gamma_U}
\end{equation} 
This identity is approximate as it relies on the relationship $\left\{ \Htilde , \Ftilde \right\} \simeq \Feq / T_\ast \left\{ \Htilde , \Utilde \right\}$, which holds if gradients of the perturbed potential $\Utilde $ in the action space are steeper than derivatives of $\Feq$. This condition can also be read as $\left|\nabla_\Jbold \Utilde \right| \gg \left|\nabla_\Jbold \Ueq \right|$. As the collision operator is linearised near an unperturbed Maxwellian distribution, the linear collision operator that applies on $\Ueq$ is defined as
\begin{equation}
 \nonumber
    \Ccal[\Ueq] =  \frac{T_\ast}{\Feq} C[\Feq]
 \label{eq:collision_operator_vs_U}
\end{equation} 
The transport equation Eq.(\ref{eq:kinetic_equation_mean}) can then be written as an equation that rules the potential $\Ueq$ 
\begin{equation}
    \frac{\partial \Ueq }{\partial t } 
		+  \nabla_\Jbold \cdot \Gammabold_U 
		= \Ccal[\Ueq]
 \label{eq:kinetic_equation_mean_U}
\end{equation}
The entropy production rate $\Stdot$, which describes the entropy dynamics, thus contains two contributions that come from transport and collisions respectively
\begin{equation}
  \Stdot =  \Sresdot + \Scolldot 
 \label{eq:entropy_production_rate_split_eq}
\end{equation}
where
\begin{eqnarray} 
	\Sresdot & = & -  \frac{1}{T_\ast^2}  
	 \int d\gamma \: \Feq \: 
		   \Gammabold_U \cdot 
	     \frac{\partial \Ueq}{\partial \Jbold}
	\label{eq:resonant_entropy_production}	\\
	\Scolldot & = & -  \frac{1}{T_\ast^2} 
		\int d\gamma \: \Feq \: \Ueq \Ccal[\Ueq]
 \label{eq:collisional_entropy_production}		
\end{eqnarray}
assuming a thermodynamically insulated plasma, i.e. no entropy flux through the outer boundary.

A powerful way to derive an equation of the thermodynamic potential $\Ueq$, and subsequently the transport equations, is to devise a variational principle based on the entropy production rate. 

To this aim, a functional $\Scal$ is built with time-evolving, resonant, and collisional contributions

\begin{equation}
  \Scal_t + \Scalres + \Scalcoll = \Scal
 \label{eq:entropy_production_generalised}
\end{equation}

Further details can be found in Appendix \ref{appendix:functional_entropy_production}. The result is that the dominant contribution, which is responsible for the neoclassical fluxes, is the resonant part that reads:

\begin{equation}
  \Scalres\left(\partial_\Jbold\Ueqdag \right) =
	 -  \frac{1}{T_\ast^2}  
	  \int d\gamma \: \Feq 
		\frac{\partial \Ueqdag}{\partial \Jbold} 
		  \cdot \Gammabold_U[\Ueqdag]  
  \label{eq:entropy_production_generalised_resonant_bis}
\end{equation} 

The evolution equation Eq.(\ref{eq:kinetic_equation_mean}) on $\Feq$ is equivalent to computing an extremum of the functional $\Scalres[\Ueq , \partial_\Jbold\Ueqdag]$ with respect to $\Ueqdag $, calculated at $\Ueqdag = \Ueq$.
As the unperturbed distribution function $\Feq$ and the Hamiltonian $H = \Heq+\Htilde$ are prescribed and $\partial_{\Jbold} \Ueqdag$ is a parameter, the unknown is the flux $\Gammabold_U[\Ueqdag]$ which depends on the perturbed thermodynamic potential $\Utilde = T_\ast \Ftilde + \Htilde$ as shown in Eq.(\ref{eq:definition_flux_gamma_U}).
Sections \ref{sec:Primary_entropy_production_rate} and \ref{sec:Secondary_entropy_production_rate} respectively aim to derive this flux for a single perturbation and two perturbations concurrently.

\section{Primary entropy production rate}
\label{sec:Primary_entropy_production_rate}
\subsection{Hamiltonian dynamics on a resonant surface}
\label{subsec:Hamiltonian dynamics on a resonant surface}
We first consider a single perturbation of the Hamiltonian, which we will refer to as the ``primary'' perturbation, of the form
\begin{equation}
	\nonumber
    \Htilde (\alphabold , \Jbold )  = 
		   - h(\Jbold) \cos \left( \nbold \cdot \alphabold \right)
	\label{eq:perturbed Hamiltonian}
\end{equation}
The actions evolve according to
\begin{equation}
    \frac{ d \Jbold }{ dt }  =  
		 - \frac{\partial \Htilde}{\partial \alphabold} =
		 - h \nbold \sin \xi 
	\label{eq:action_island}
\end{equation}
where the phase $\xi = \nbold \cdot \alphabold $ is governed by the equation 
\begin{equation}
	\nonumber
    \frac{ d \xi }{ dt }  =  
		    \nbold \cdot \frac{\partial \Heq}{\partial \Jbold} = \nbold \cdot \Omegabold(\Jbold) = \Omega(\Jbold) 
	\label{eq:phase_island}
\end{equation}
assuming that the evolution of $\Htilde$ is slow with $\Jbold$, which will be defined later.

In order to make the problem more tractable, it is useful to change variables to obtain a Hamiltonian system that is analogous to a simple pendulum. 
Defining a resonant surface in the space of actions $\Jbold$ as satisfying the resonance condition $\Omega(\Jbold)=0$,
we constrain the problem to cases where $h(\Jbold)$ depends on one action only, say $J_2$, such that the resonance condition $ \Omega(J_2 , J_3) = 0 $ defines a curve $J_3 = G(J_2)$ in the 2-dimensional$(J_2,J_3)$ space, where $G$ is the function that satisfies 
\begin{equation}
    \Omega(J_2 , G(J_2)) = 0 
	\label{eq:def_G}
\end{equation}	

This change of variable is detailed in Appendix \ref{Primary perturbation and island in the phase space}.
This lets us obtain a Hamiltonian that is homogeneous to the square of a frequency 
\begin{equation}
    H_\Omega = \frac{1}{2} \Omega^2 - \omegab^2 \cos(\xi)
    \label{eq:H_Omega}
\end{equation}
where we have introduced the bounce frequency 
\begin{equation}
    \omegab = \left(C h\right)^{1/2}
    \label{eq:bounce_frequency}
\end{equation}
with the Hamiltonian ``curvature'' $C$ defined as
\begin{equation}
    C(\Jbold_R)  =  \left. n_i n_j 
					  \frac{\partial^2 \Heq}
		             {\partial J_i J_j} 
				  \right|_{\Jbold = \Jbold_R} 
	\label{eq:curvature}
\end{equation} 
wherein $\Jbold_R$ is a point on the resonant surface.
Thus, the total Hamiltonian is $H = H_{\rm eq} +\frac{1}{C} H_\Omega$.
Then, the equations of motion for the coordinates $(\xi , \Omega)$  read
\begin{eqnarray} 
    \frac{d\xi}{dt} & = & 
	     \frac{\partial H_\Omega }{\partial \Omega} = \Omega \\
	\label{eq:angle_equation_motion_island_pendulum}
		\frac{d \Omega}{dt} & = & 
		  - \frac{\partial H_\Omega }{\partial \xi} =
	   	- \omegab^2 \sin \xi  
	\label{eq:frequency_equation_motion_island_pendulum}
\end{eqnarray}
The advantage of this formulation is that the quantities of interest $(\Omega,\omegab)$ are frequencies, and the bounce frequency $\omegab$ naturally emerges as a convenient frequency unit. Its meaning is elucidated in the following. 
\subsection{Island in the phase space}
\label{Island in the phase space}
\begin{figure}[!ht]
\centering
	\centerline{\includegraphics[width=8cm]
	{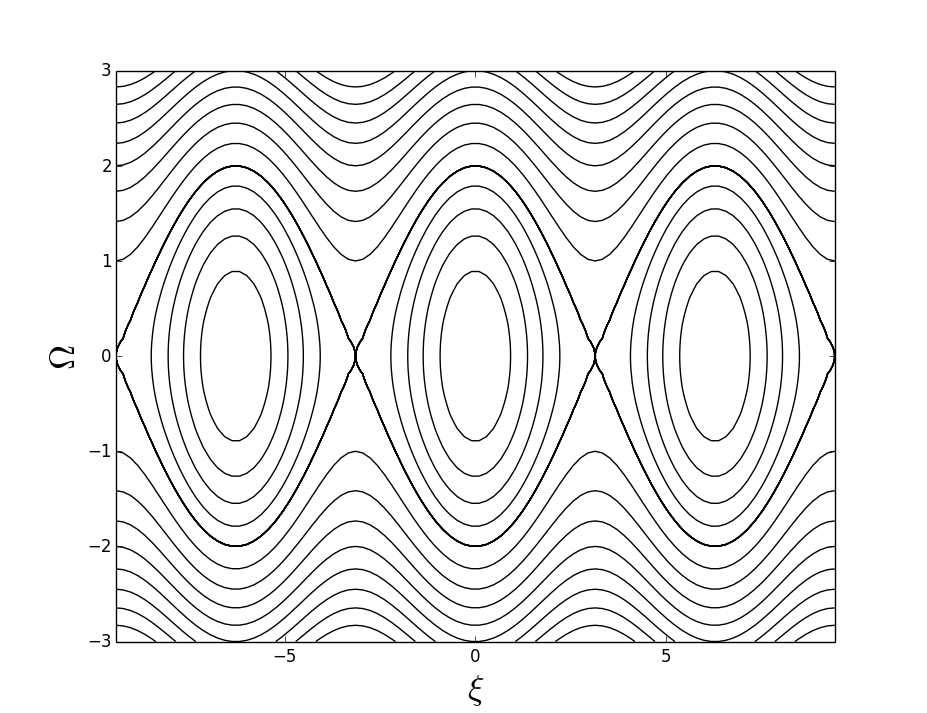}}
	\caption{\small{Contour lines of the Hamiltonian $ H_\Omega = 
	\frac{1}{2} \Omega^2 - \omegab^2 \cos(\xi)$.}}
	 \label{Fig:Island_contours}
\end{figure}

The structure of the Hamiltonian $H_\Omega$ is similar to that of a simple pendulum. 
Correspondingly, lines of constant $H_\Omega$ draw the usual shape of an island (or cat-eye) in the phase space, as shown in Fig.\ref{Fig:Island_contours}. 
Particles are trapped when $ -\omegab^2 \le  H_\Omega \le \omegab^2$, and passing when $ \omegab^2 \le  H_\Omega \le +\infty$. 
This means that trapped particles travel between two angle positions $ - \xi_b$ and $\xi_b$ (modulo $2\pi)$ such that $H_\Omega = - \omegab^2 \cos \xi_b$, whereas passing particles span the entire $\xi$ domain. 
The trapped and passing domains are separated by a separatrix, defined in the space $(\xi , \Omega)$ by the condition $H_\Omega = \omegab^2$, which defines the separatrix equation $ \Omega_\text{sep} = \pm 2 \omegab \cos \left(\xi / 2 \right)$. 
Thus, the island half-width is given by $\delta_\Omega = 2 \omegab$, in frequency units.
We are now in position to define ``slow'' variations of the perturbation amplitude $h(\Jbold)$: when the length scale of $h$ along the $\nbold$ direction is much larger than the island width $\delta_\Omega$.
%
\subsection{Kinetic equilibrium near an island}
\label{Kinetic equilibrium near an island}
\begin{figure}[!ht]
\centering
	\centerline{\includegraphics[width=14cm]
	{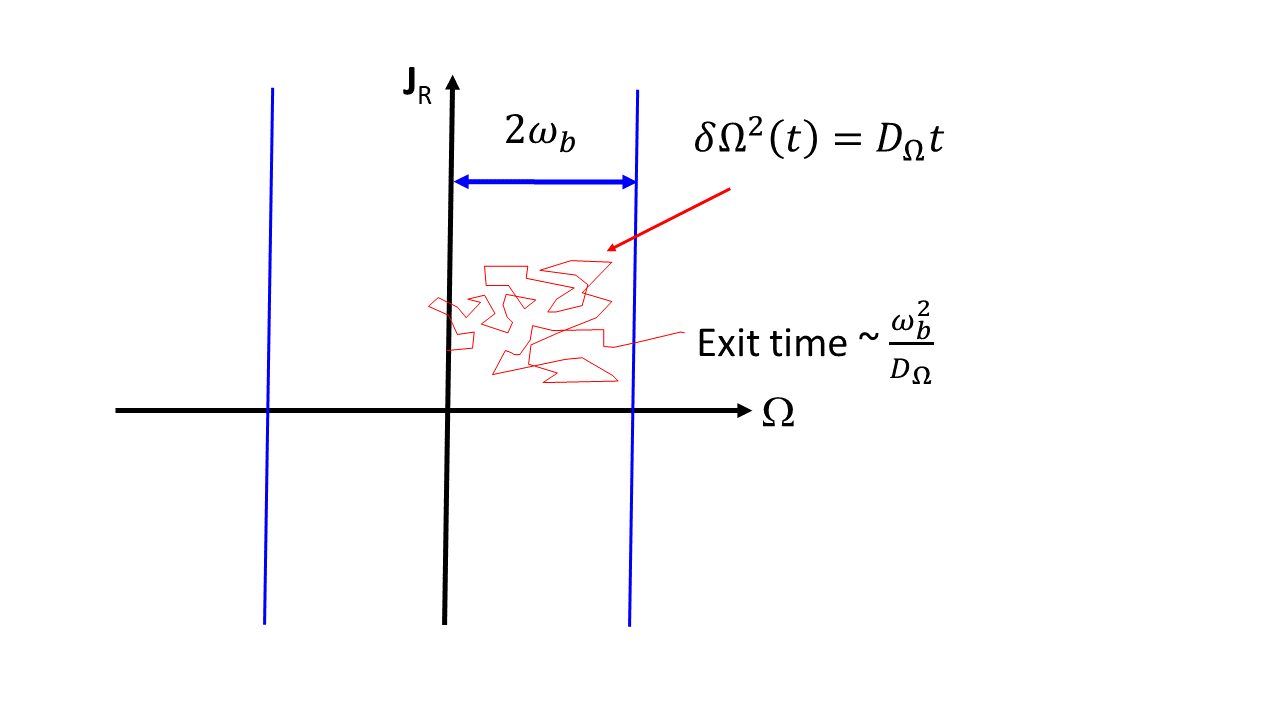}}
	\caption{\small{Schematic random motion of a particle due to collisions. A particle that starts from the resonant surface $\Omega=0$ diffuses as $\delta\Omega^2(t) \sim D_\Omega t $. The island half-width is $ 2 \omegab $, so that it escapes the island after a typical detrapping time $ \tau_{d} \sim \omegab^2 / D_\Omega $, due to collisions.}}
	 \label{Fig:Particle_diffusion_island}
\end{figure}

Following the pioneering work of Zakharov and Karpman \cite{Zakharov_Karpman_1963}, the first step to characterise the thermodynamic equilibrium near an island in the phase space is to solve Eq.(\ref{eq:kinetic_equation_perturbed})
\begin{equation}
   \left\{ H_\Omega , 
	          \Feq+\Ftilde 
		 \right\}
		 + \Ccal[\Ftilde] = 0
 \label{eq:phase_island_equation_Utilde}
\end{equation}
\begin{figure}[!ht]
\centering
	\centerline{\includegraphics[width=12cm]
	{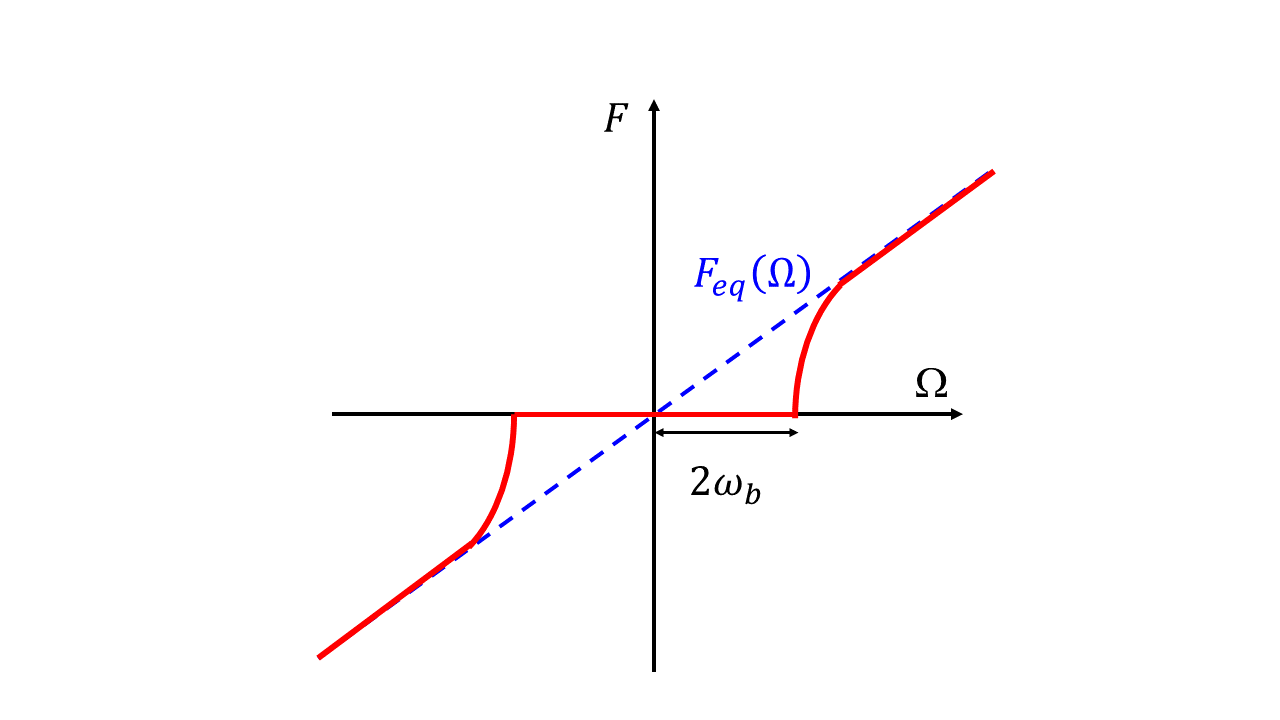}}
	\caption{\small{Typical shape of a distribution function near an island in the action space (in red).}}
	 \label{Fig:Distribution_function_near_island}
\end{figure}
The resonant singularity of the distribution function near $\Omega=0$, as illustrated in Fig.\ref{Fig:Distribution_function_near_island}, is expected to be regularised by collisions.
Hence, it makes sense to use a simple diffusion expression for the collision operator
\begin{equation}
	\nonumber
    \Ccal[\Ftilde] = 
		  \frac{\partial  }
		       {\partial \Omega} 
			\left( D_\Omega 
			      \frac{\partial \Ftilde }
		        {\partial \Omega } 
					 \right)
	\label{eq:collision_operator_near_island}
\end{equation} 
where 
\begin{equation}
    D_\Omega = \frac{1}{2}
      \frac{\left\langle \Delta \Omega^2 \right\rangle}{\Delta t} 
    \label{eq:D_Omega}
\end{equation}
is a diffusion coefficient in $\Omega$ that has units of cubic frequency, and the bracket $\langle.\rangle$ is defined according to the convention for Fokker-Planck equations \cite{Helander_2014}. 
This coefficient is evaluated at the resonant surface $\Omega=0$, as its variations usually play no role. 
Then, the dimensionless Fokker-Planck equation to solve reads:
\begin{equation}
   \frac{1}{\omegab^3}\left\{ H_\Omega , 
	          \Feq+\Ftilde 
		 \right\}
		 + \nu_\Omega \frac{\partial^2 \Ftilde  }
		       {\partial \Omega^2}
         = 0
 \label{eq:phase_island_equation_Utilde_adim}
\end{equation}
where the dimensionless parameter $\nu_\Omega$, referred to as ``collisionality'', is defined as
\begin{equation}
   \nu_\Omega = \frac{D_\Omega}{\omegab^3}
 \label{eq:nu_Omega}
\end{equation}
As illustrated in Fig.\ref{Fig:Particle_diffusion_island}, since collisions are responsible for random walk in the phase space, the displacement $ \delta \Omega = \Omega(t) - \Omega(0) $ of a particle that leaves the resonant surface $\Omega=0$ at the initial time $t=0$ evolves according to $\delta \Omega^2(t) \sim D_\Omega t $. 
As the island half-width is $ 2 \omegab $, the particle will escape the island after a typical detrapping time $ \tau_{d} \sim \omegab^2 / D_\Omega $. 
Hence, the physical meaning of $ \nu_\Omega = \tau_b / \tau_{d} $ can be understood as the ratio of a bounce time $\tau_b = \omegab^{-1}$ to the detrapping time $\tau_{d}$. 
If the particle gets detrapped by collisions before traversing the island, then $\nu_\Omega \gg 1$; conversely, in the case $\nu_\Omega \ll 1$, the particle will bounce many times before it escapes the island. 
In that case, trapped particles will traverse the entire constant energy surface in the island before undergoing a collision, causing mixing throughout the island region, such that no gradient can persist within it. 
As a result, the distribution function in the island is flattened, as shown in Fig.\ref{Fig:Distribution_function_near_island}.
\subsection{Entropy production rate for a single perturbation}
\label{Entropy production rate for a single perturbation}
Solving the Fokker-Planck equation Eq.(\ref{eq:phase_island_equation_Utilde_adim}) yields the perturbed distribution function $\Ftilde$ and enables us to compute the functional associated with the resonant entropy production rate Eq.(\ref{eq:entropy_production_generalised_resonant_bis}).
The final form, as derived in \cite{Garbet_2009}, is

\begin{equation}
  \Scalres \left[\partial_\Jbold\Ueqdag \right] =  
	\frac{1}{T_\ast^2}  
	 \int d\gamma \Feq(\Jbold) 
	   \delta( \nbold\cdot\Omegabold ) h^2
	    \Lambda_{\rm I}(\Jbold)
			\left( \nbold \cdot 
	          \frac{\partial \Ueqdag}{\partial \Jbold} 
			\right)^2
 \label{eq:value_entropy_production_collisional_summary}
\end{equation}
where $\Lambda_{\rm I}(\Jbold)$ is a positive function
\begin{equation}
 \nonumber
  \Lambda_{\rm I}(\Jbold) = \nu_\Omega
			\int_{-\infty} ^{+\infty} \frac{d\Omega}{\omegab} 
	\int_{-\pi}^{\pi} \frac{ d \xi }{ 2 \pi } 
	\left(\frac{ \partial \Ftilde(\xi,\Omega / \omegab) }
		         { \partial \Omega } \right)^{2}
\label{eq:value_Lambda_weakly_collisional_alternate_summary}
\end{equation}
In the limit of either small or large $\nu_\Omega$, the calculation of the entropy production rate is analytical, with  
\begin{equation}
\nonumber
	\Lambda_{\rm I}(\Jbold) = 
	\left\{
	  \begin{array}{cl}
      \dfrac{\pi}{2} &  \text{~~if~~} \nu_\Omega \gg 1   
      \\[8pt]
      2 \Ical \nu_\Omega  &  \text{~~if~~} \nu_\Omega \ll 1  \text{~~;~~} \Ical \simeq 1.38
    \end{array}
	\right.
  \label{Lambda_vs_eta}
\end{equation}
However, the expression for $\nu_\Omega \gg 1$ breaks down for very large values of $\nu_\Omega$ where the Pfirsch-Schl\"uter regime would prevail, and $\Lambda_I$ ultimately decreases as $1/\nu_\Omega$.
In standard neoclassical theory, for tokamaks, $\nu_\Omega \ll 1 $ denotes the banana regime, and $\nu_\Omega \gg 1 $ denotes the plateau regime. This result is consistent with the classical analysis of Zakharov and Karpman \cite{Zakharov_Karpman_1963}. 

In the following section, we assess the effect of a secondary perturbation on top of the new equilibrium obtained with the primary perturbation.

\section{Secondary entropy production rate}
\label{sec:Secondary_entropy_production_rate}

In this section, the methodology to derive the neoclassical contribution of a secondary perturbation on top of an equilibrium obtained with the primary perturbation is described in detail.
We will not consider chaos arising from the islands associated with each perturbation overlapping \cite{Lichtenberg_Lieberman_83} in the present work.

\subsection{Switching to equilibrium associated with primary perturbation}
\label{Switching to equilibrium associated with primary perturbation}

When computing the effect of a secondary perturbation
\begin{equation}
	\nonumber
    \Htilde' (\alphabold , \Jbold )  = 
		   - h'(\Jbold) \cos \left( \nbold' \cdot \alphabold \right)
	\label{eq:secondary_perturbation_Hamiltonian}
\end{equation}
we will first assume that the equilibrium being perturbed is the same one obtained from previously considering the primary perturbation.
Then, it is useful to change from variables $( \alpha_2 , \alpha_3 ,  J_2 , J_3 )$ to a new set of conjugate variables $( \alphab , \alphad ,  \Jb , \Jd  )$ that are more suitable for computing the secondary perturbation, and express the Hamiltonian and distribution function in this set of variables.

\subsubsection{Equilibrium Hamiltonian}

We introduce the trapping parameter $K$, defined as
\begin{equation}
  \nonumber
    K = \frac{1}{2\omegab^2}  \left( H_\Omega + \omegab^2 \right)
	\label{eq:definition_K_parameter}
\end{equation}
and spanning $0 \le K \le 1$. The limit $K \rightarrow 0$ corresponds to deeply trapped particles, and $K \rightarrow 1$ to barely trapped particles. Let us note that 
\begin{equation}
    K = \frac{\Omega^2}{4\omegab^2}  + \sin^2\left( \frac{\theta_2}{2} \right)
	\label{eq:expression_K}
\end{equation}
The invariance of Poincar\'e integrals under change of conjugate variables implies that
\begin{equation}
    \Jb = \oint \frac{d \theta_2}{2\pi}  I_2\left( H_\Omega,\thetab \right) = 
         \frac{\sqrt{2}}{C} \oint \frac{d \theta_2}{2\pi} 
         \sqrt{ H_\Omega + \omegab^2 \cos\theta_2}
	\label{eq:action_Jb}
\end{equation}
An expression of the action $\Jb$, valid for all values of $K$, is 
\begin{equation}
   \Jb = \lambda_m Q(K)
  \label{eq:expression_Jbar2}
\end{equation}
where
\begin{equation}
  \nonumber
   \lambda_m = \frac{2\omegab}{C} = 2 \sqrt{\frac{h}{C}}
\label{eq:expression_lambdam}
\end{equation}
measures the action amplitude, and
\begin{equation}
  \nonumber
   Q(K) = \oint \frac{d \theta_2}{2\pi} 
	           \sqrt{ K - \sin^2\left( \frac{\theta_2}{2} \right)}
\label{eq:expression_QvsK}
\end{equation}
The integral covers the interval $- \thetab \le \theta_2 \le + \thetab$ forwards and backwards, a total of twice, where $\thetab$ is the bounce angle, defined by $K = \sin^2\left( \thetab/2 \right)$. 
The associated pulsation reads
\begin{equation}
   \Omegab = \frac{2 \omegab} { R(K)} 	             
   \label{eq:expression_Omegabar2}     
\end{equation}
where
\begin{equation}
  \nonumber
    R(K) = \oint \frac{d \theta_2}{2\pi} 
	              \frac{1}
								     { \sqrt{ K - \sin^2\left( \frac{\theta_2}{2} \right)} 
										 }             
   \label{eq:expression_RvsK} 
\end{equation}
Using $\partial_K Q = R/2 $ and normalisations, it can be verified that $\partial_H \Jb = 1/\Omegab$. Moreover, the functions $Q$ and $R$ can be related to the elliptic functions, namely
\begin{equation}
   R(K) = \frac{4}{\pi} \mathbb{K}(K)
 \label{eq:expression_Rvselliptic}
\end{equation}
\begin{equation}
  \nonumber
   Q(K) = \frac{4}{\pi} \left\{ \mathbb{E}(K) - 
         (1 - K) \mathbb{K}(K)   \right\}
\label{eq:expression_Qvselliptic}
\end{equation}
where
\begin{equation}
  \nonumber
   \mathbb{K}\left( K \right)= \int_{0}^{\frac{\pi}{2}}
	  \frac{d u}{\sqrt{1 - K \sin^2u}}
   \label{complete_expression_elliptical_function_K_bis}
\end{equation}
\begin{equation}
  \nonumber
   \mathbb{E}\left( K \right)= \int_{0}^{\frac{\pi}{2}}
	  du \: \sqrt{1 - K \sin^2u}
   \label{complete_expression_elliptical_function_E_bis}
\end{equation}
are, respectively, the complete elliptic functions of the first and second kind.
Useful limits to keep in mind are $\mathbb{E}(0) = \mathbb{K}(0) = \pi/2$, $\mathbb{E}(1) = 1$, and $\mathbb{K}(K)\simeq \frac{1}{2}\ln \left(\frac{16}{1 - K}\right)$ when $K \rightarrow 1$. 
Finally, we need to find the relationship between the initial variables $\Omega,\theta_2$ and the new ones $K,\alphab$.
As the dynamics are the same as for a simple pendulum, the variables can be related by Jacobi elliptic functions, such that 
\begin{eqnarray} 
  \sin\left( \frac{\theta_2 }{2}\right) & = & 
    \sqrt{K} \text{sn} \left[ \frac{2}{\pi} \mathbb{K}(K) \alphab, K \right]
    \label{eq:theta_vs_alphab_Jacobi}
    \\
	\Omega & = &  2 \sqrt{K} \omegab 
	\text{cn} \left[ \frac{2}{\pi} \mathbb{K}(K) \alphab, K \right]
    \label{eq:Omega_vs_alphab_Jacobi}
\end{eqnarray}
which are clearly consistent with Eq.(\ref{eq:expression_K}), since $\text{cn}^2 + \text{sn}^2 = 1$. These relations are well defined for the interval $\left[-\pi/2,\pi/2\right]$ in $\alphab$, for which $\theta_2$ spans $\left[-\thetab,\thetab\right]$. The reverse relationship that yields the bounce angle $\alphab$ and the initial angular variable $\theta_2$ is also known 
\begin{equation}
    \alphab=  \frac{\pi}{2} \frac{1}{\mathbb{K}(K)}
      \mathbb{F}
       \left\{ \arcsin
          \left[ 
            \frac{1}{\sqrt{K}} 
            \sin\left( \frac{\theta_2}{2} \right) 
          \right] , K  
       \right\}
	\label{eq:alpha_b_vs_theta_2}
\end{equation}
where $\mathbb{F}$ is the incomplete elliptic function of the first kind
\begin{equation}
  \nonumber
    \mathbb{F}\left( \phi , K \right)= \int_{0}^{\phi}
	   \frac{d u}{\sqrt{1 - K \sin^2u}}
   \label{incomplete_expression_elliptical_function_K_bis}
\end{equation}
Complete and incomplete elliptic functions of the first kind are linked by $\mathbb{K}(K) = \mathbb{F}\left( \frac{\pi}{2}  , K \right)$.
For deeply trapped particles (small bounce angle $\thetab \ll 1$), a Taylor expansion readily proves that $K = \thetab^2/4 + \Ocal(\thetab^4)$, and 
\begin{eqnarray} 
 \nonumber
  \theta_2 & = & \thetab \sin \alphab            \\
 \nonumber
	\Omega & = & \omegab \thetab \cos \alphab
 \label{eq:theta_vs_alphabar}
\end{eqnarray}
where $\omegab = \sqrt{ C h }$. The Hamiltonian $H = H_\Omega $ reads
\begin{equation}
	\nonumber
    H_\Omega = \frac{1}{2}\omegab^2 \thetab^2 
		         = C \Jb \Omegab
	\label{eq:Hamiltonian_omega}
\end{equation}
It is tempting to use the deeply trapped approximation for all particles, since it can be used to derive tractable asymptotic expressions, but doing so will quickly lead to an inconsistency. Obviously, the maximum value of the bounce angle is $\thetab=\pi$; however, the relationship $K = \thetab^2/4$ implies that the maximum value of the trapping parameter $K$ should be $\pi^2/4$, which is clearly larger than $1$. This contradiction arises because the deeply trapped approximation is used beyond its range of validity. A quick fix would consist of introducing a cut-off in the bounce angle $\thetab$ at $\theta_m = 2 $, such that the trapping parameter $K$ spans the interval $[0,1]$ accordingly. Additionally, note that in the deeply trapped limit, 
\begin{equation}
    \Jb = \lambda_m K =\lambda_m \frac{\thetab^2}{4}
	\label{eq:Jbar2_deeply_trapped}
\end{equation}
The corresponding resonant frequency is the bounce frequency $\Omegab $.

We now rename the angles $(\theta_3 , I_3)$ to $(\alphad , \Jd)$ with ``d'' for ``drift'', where
\begin{equation}
  \nonumber
     \Jd = n_3 \left( J_2 - J_{\ast 2} \right) 
         - n_2 \left( J_3 - J_{\ast 3} \right)
	\label{eq:Jd_3rd_drift_action}
\end{equation}
wherein $J_{\ast 2}$ and $J_{\ast 3}$ are the second and third components respectively of a point $\Jboldstar$ on the resonant surface.
The associated resonant angular frequency $\Omegad$ can be split in contributions from the dependences of $\Heq (\Jbold_R)$ and $\Jb$ on $\Jd$.
To compute the latter, we first define a function $\mathcal{J}$ that depends only on $\Jb$ and $\Jd$, satisfying the implicit equation on $(\Jb,\Jd)$
\begin{equation}
     \Jb - \mathcal{J} \left[ H_\Omega(\Jb,\Jd) , \Jd \right]  = 0
	\label{eq:implicit_equation_actions}
\end{equation}
Then, one can use the relationships $\Omegab = \left. \partial_{\Jb} H_\Omega \right|_{\Jd}$ and $\Omegad(\Jd) = \left. \partial_{\Jd} H_\Omega\right|_{\Jb}$ so that the partial derivatives of the relationship Eq.(\ref{eq:implicit_equation_actions}) provide two independent relationships
\begin{eqnarray} 
 \nonumber
   1 & = &  \left. \frac{\partial \mathcal{J}}{\partial H_\Omega}  \right|_{\Jd} \Omegab      \\
 \nonumber
	 0 & = & \left. \frac{\partial \mathcal{J}}{\partial H_\Omega}  \right|_{\Jd} 
	          \left. \frac{\partial H_\Omega}{\partial \Jd} \right|_{\Jb} +
	        \left. \frac{\partial \mathcal{J}}{\partial \Jd}  \right|_{H_\Omega}
 \label{eq:independent_relationships}
\end{eqnarray}
from which we obtain the total frequency $\Omegad(\Jd)$
\begin{equation}
   \Omegad(\Jd) = \left. \frac{\partial \Heq}{\partial \Jd} \right|_{\Jbold=\Jbold_R} 
        - \Omegab \left. \frac{\partial \mathcal{J}}{\partial \Jd}  \right|_{H_\Omega}
 \label{eq:Omega_d}
\end{equation}
Despite the explicit dependence on the bounce frequency $\Omegab$ in the second term of the expression, the precession frequency $\Omegad$ remains much smaller than $\Omegab$ in most practical cases (one notable exception being that of energetic particles in a tokamak). 

The new set of variables $( \alphab, \alphad ,  \Jb , \Jd )$ is conjugate to the original set $( \alpha_2 , \alpha_3 ,  J_2 , J_3 )$, and provides a convenient set of coordinates in which to render the secondary perturbation. The system is integrable, and the new ``unperturbed'' Hamiltonian, written as $\Hfrakeq(\Jb , \Jd)$, is defined as 
\begin{equation}
    \Hfrakeq(\Jb , \Jd) = \Heq(\Jd) + \frac{1}{C} H_\Omega(\Jb , \Jd)
	\label{eq:new_hamiltonian}
\end{equation}
Note that the Hamiltonian perturbation amplitude $h$ has been taken to be constant. 
In some cases, $h$ does depend on the actions, but the above calculation will still be valid as long as this dependence is slow compared the island width, such that it can be considered to be dependent on the ``slow'' resonant variable $\Jd = I_3$. 

\subsubsection{Perturbed Hamiltonian}

Following the same procedure as in Section \ref{subsec:Hamiltonian dynamics on a resonant surface}, we start by calculating the frequencies associated with the primary and secondary perturbations on the primary resonant surface.
From derivations found in Appendix \ref{appendix:Expressing_the_secondary_frequency_at_the_primary_resonant_surface}, the link between these frequencies are constrained by the relation:
\begin{equation}
   \Omega'(\Jbold) = \Omega'(\Jbold_R) + \sigma \Omega(\Jbold)
	\label{eq:link_omega_omega'}
\end{equation}
with
\begin{eqnarray} 
    \Omega  &=& C I_2              \\
    \nonumber
    \Omega'  &=& \Omega'(\Jbold_R) + C^{\ast} I_2 \approx C^{\ast} I_2
 \label{eq:Omega_and_Omega_prime}
\end{eqnarray}
for which the rightmost equality is obtained by noting that $\Omega'(\Jbold_R)$ is small compared to $C^{\ast} I_2$ in most cases of interest. (For more detail, see section \ref{Drift frequency}.) We then introduce a ``hybrid'' curvature
\begin{equation}
    C^{\ast} =
            \left. n_i n'_j 
					  \frac{\partial^2 \Heq}
		             {\partial J_i \partial J_j} 
				  \right|_{\Jbold = \Jbold_R} 
    \label{eq:curvature_hybrid}
\end{equation} 
and a ``curvature ratio''
\begin{equation}
    \sigma = \frac{C^{\ast}}{C} \approx \frac{\Omega'}{\Omega}
	\label{eq:sigma_vs_C_Cstar}
\end{equation} 
In our cases of interest, both $C$ and $C^{\ast}$ are constant. 

The perturbation $\Htilde' ( \alphabold , \Jbold )$ can be rewritten as a function of the new angles and actions that is periodic in $( \alphab, \alphad )$. 
Its expansion in Fourier series then reads
\begin{equation}
    \Htilde'(\alphab, \alphad, \Jb, \Jd) =  e^{i n_d \alphad}
		   \sum_{\nb = - \infty}^{+\infty} H'_{\nb} (\Jb, \Jd)
			 e^{i \nb \alphab } + C.C.
	\label{eq:fourier_series_vs_alphabar}
\end{equation}
where 
\begin{equation}
    n_d = n'_2 n_3 - n_2 n'_3
    \label{eq:n_d}
\end{equation}
The Hamiltonian Fourier harmonics formally read
\begin{equation}
	\nonumber
    H'_{\nb} = - \frac{h'}{2} \int_{-\pi}^{\pi} \frac{d\alphab}{2\pi}
		   \exp \left[ i \sigma \theta_2(\Jbold,\alphab) - i \nb \alphab\right]	   
	\label{eq:Hamiltonian_Fourier_harmonic}
\end{equation}
Using the odd parity of the expression of $\theta_2 $ vs $\alphab$, these components can also be written as
\begin{equation}
    H'_{\nb} = - \frac{h'}{2} \int_{-\pi}^{\pi} \frac{d\alphab}{2\pi}
		   \cos\left[ \sigma \theta_2(\Jbold,\alphab)  - \nb \alphab \right]		   
	\label{eq:Hamiltonian_Fourier_bounce_average}
\end{equation}
The relationship between the angles $\theta_2$ and $\alphab$ is given by Eq.(\ref{eq:theta_vs_alphab_Jacobi}).
These are related to time via the following relationship 
\begin{equation}
  \nonumber
		dt = \frac{d\alphab}{\Omegab} = 
		  \frac{d\theta_2}{\Omega} = 
		  \frac{d\theta_2}
		       {2\omegab \sqrt{K - \sin^2\left( \frac{\theta_2}{2}
		        \right) } }
  \label{eq:relation_time_alphab_theta2}
\end{equation}
Then, an alternative expression can be readily found 
\begin{equation}
     H'_{\nb} = - \frac{h'}{2} L_{\sigma,\nb}(K) 
  \label{eq:expression_hamiltonian_secondary_bounceaverage}
\end{equation}
where
\begin{equation}
    L_{\sigma,\nb}(K) =  \frac{1}{R(K)} \frac{2}{\pi }
        \int_0^{\frac{\thetab}{2} } du \: 
          \cos\left(2 \sigma u - \nb \alphab(K,2u) \right)
	           \frac{1}{ \sqrt{ K - \sin^2 u } }
  \label{eq:expression_LvsK}
\end{equation}
We note again here that $K = \sin^2(\thetab/2)$, and as such, the bounce angle $\thetab$ can be written as a function of $K$. Obviously, $L_{0,0}(K) = 1$. 
The bounce-averaged component of $\tilde{H}'$ is given by the component $H'_{\nb=0}$. As the relationship between $\alphab$ and $\theta_2$ is given by the incomplete elliptic function of the first kind Eq.(\ref{eq:alpha_b_vs_theta_2}), $L_{\sigma,\nb}(K)$ can be computed numerically with standard integration methods per quadrature employing Eq.(\ref{eq:expression_LvsK}) or through identification with Eq.(\ref{eq:Hamiltonian_Fourier_bounce_average}), such that
\begin{equation}
    L_{\sigma,\nb}(K) =  \int_{-\pi}^{\pi} \frac{d\alphab}{2\pi}
		   \cos\left[ \sigma \theta_2(\Jbold,\alphab)  - \nb \alphab \right]		  
  \label{eq:expression_LvsK_bis}
\end{equation}

In the limit where only deeply trapped particles in the primary perturbation are considered, $\omegab = \Omegab$ and $\theta_2 = \thetab \sin\alphab$. Then, we obtain 
\begin{equation}
	\nonumber
     H'_{\nb}(\Jb, \Jd) = - \frac{h'}{2} J_{\nb}(\sigma \thetab)
	\label{eq:fourier_component_trapped}
\end{equation}
where $\thetab$ must be understood as a function of $(\Jb,\Jd)$, and $\sigma$ is a function of $\Jd$. For large values of $\sigma$, i.e. $\sigma \gg 1$, we can use the following asymptotic form of the Bessel functions
\begin{equation}
    J_{\nb}(\sigma \thetab) \simeq 
		   \sqrt{\frac{ 2 }{ \pi \sigma \thetab } }
		   \cos \left( \sigma \thetab 
			             - \nb \frac{\pi}{2} 
			             - \frac{\pi}{4} 
						 \right)
	\label{eq:Bessel_function_asymptotics}
\end{equation}

\subsection{Discriminating trapping in primary and secondary perturbations}
\label{Discriminating trapping in primary and secondary perturbations}
When adding a secondary perturbation on top of a primary, it becomes important to discriminate particles trapped in the newly added perturbation from those still trapped in the primary. 
In addition, collisional detrapping processes are unequal between the perturbations, as a particle in one type of trapping might transition to another type or to passing, depending on its energy, pitch-angle, and the ratio $h'/h$. 
Also, we will show that the effective amplitude of the trapping associated with the secondary perturbation is modified compared with the primary perturbation. The derivation follows closely previous calculations done for a rippled tokamak \cite{Connor_1973,Yushmanov_1990}.

The trapping condition in the secondary perturbation can be understood as follows. 
The phase evolves in the primary perturbation following $d \xi/ dt = \Omega $, while that in the secondary perturbation follows $d \xi' / dt  = \Omega' $, implying that $d\xi' = \sigma d\xi$.
Whenever $\sigma$ evolves slowly, the solution can be simply written as
\begin{equation}
  \nonumber
     \xi' = \sigma \xi + \xi'_0
	\label{eq:xiprime_vs_xi}
\end{equation} 
where $\xi'_0 = \xi'(\xi = 0)$, wherein freedom in the choice of $\xi = 0$ and thus of $\xi'_0$ allows us to choose $\xi'_0$ to coincide with a local minimum of the primary perturbed Hamiltonian.

\begin{figure}[!ht]
\centering
	\centerline{\includegraphics[width=10cm]
	{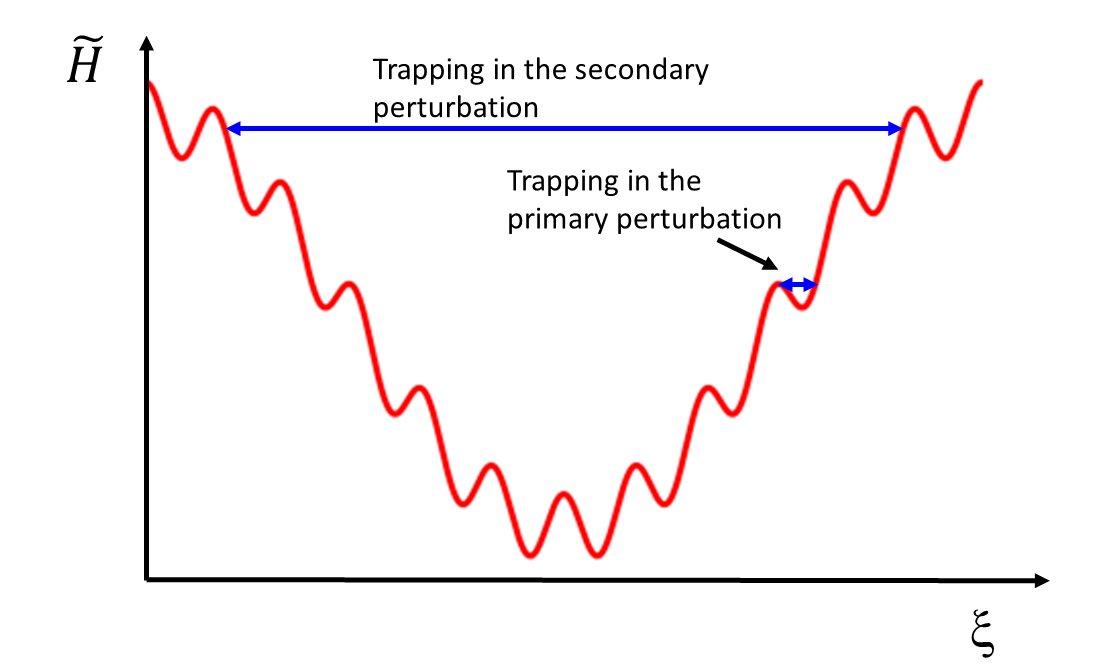}}
	\caption{\small{Illustration of energy condition for particle trapping in primary and secondary perturbations.}}
	 \label{Fig:Trapping_primary_secondary}
\end{figure}

Let us suppose that $\sigma \ll 1$, which occurs when the wavenumber of the primary perturbation $n_h = n_2 + n_3q$ is much larger than the wavenumber of the secondary perturbation $n'_h = n'_2 + n'_3 q$
such that the primary perturbation oscillates faster than the secondary.
The Hamiltonian perturbation can be expressed with the secondary phase $\xi'$ as 
\begin{equation}
    \Htilde (\Jbold , \xi' )  = 
			- h  \cos \left( \frac{1}{\sigma} ( \xi' - \xi'_0 ) \right) 		
			- h' \cos \left( \xi' \right) 			
	\label{eq:total_Hamiltonian_double_perturbation_perturbed}
\end{equation}
Trapping is subject to the existence of local extrema of $\Htilde$ in $\xi'$, as illustrated in Fig. \ref{Fig:Trapping_primary_secondary}.
At an extremum, $d \Htilde /d \xi'=0$, which means
\begin{equation}
          \sin \left[ \frac{1}{\sigma} ( \xi' - \xi'_0 ) \right]	=
          - \alpha^{\ast} \sin \left( \xi' \right) 
	\label{eq:condition_trapping_secondary}
\end{equation} 
where the parameter $ \alpha^{\ast} = \sigma h' / h$ is a measure of the relative amplitude of perturbations.
It appears that trapping in the secondary perturbation requires  $\left| Y \right| \le 1 $, where $Y$ is defined as
\begin{equation}
	\nonumber
   Y(\xi') = - \alpha^{\ast} \sin \left( \xi' \right) 				
	\label{eq:local_trapping_condition}
\end{equation}
Focusing on a local minimum $\xi' = \xi'_0$, this condition defines a trapping domain in $\xi'_0$, i.e. $\left| Y_0 \right| \le 1$, where $Y_0 = Y(\xi'_0) $. 
Let us stress that the parameter $\alpha^{\ast}$ depends on $\sigma$, but also on the ratio of amplitudes $h'/h$, so even if $\sigma \ll 1$, it can very well be that $\alpha^{\ast} > 1$, such that the condition $\left| Y_0 \right|\le 1$ defines an interval in $\xi'_0$ smaller than $2\pi$.

Another important point the modification of the depth of the Hamiltonian perturbation.
For the trapping associated with the secondary perturbation, it can be calculated as follows.
Let us consider, again, the case $\sigma \ll 1$. 
For vanishing secondary perturbation $\alpha^{\ast} =0 $, i.e. $Y = 0$, the minima and maxima correspond respectively to $ \xi'_{-} = \xi'_0$ and $\xi'_{+} = \xi'_0 + \pi \sigma$. The depth is just $\Delta H = 2 h$. 
Let us now assume that $Y$ is small but finite, and focus on the Hamiltonian well near the new local minimum located at $\xi' = \xi'_{-}$, in the vicinity of $\xi'_0$ assumed positive. 
Eq.(\ref{eq:condition_trapping_secondary}) implies that 2 successive minimum and maximum of $\Htilde (\Jbold , \xi' )$ are reached at $\xi'_{-}$ and $\xi'_{+}$ respectively such that
\begin{equation}
    \frac{1}{\sigma} ( \xi'_{-}	- \xi'_0 ) =  \arcsin \left[ Y(\xi'_{-}) \right]		
	\label{eq:xiprime_minus}
\end{equation}
\begin{equation}
    \frac{1}{\sigma} ( \xi'_{+}	- \xi'_0 ) = \pi - \arcsin \left[ Y(\xi'_{+}) \right]	
	\label{eq:xiprime_plus}
\end{equation}
Because $\sigma \ll 1$, a calculation can be constrained to first order in $\sigma$. 
At such order, the function $Y(\xi'_{\pm})$ can be calculated at the former local minimum $\xi'_0$. 
The function $\arcsin$ being multivalued, we focus on the value that is the closest to $0$. that we call $\beta_0 = \arcsin Y_0$.
The equations that govern $\xi'_{\pm}$ then read
\begin{equation}
	\nonumber
    \frac{1}{\sigma} ( \xi'_{-}	- \xi'_0 ) =  \beta_0		
	\label{eq:xiprime_minus_approximate}
\end{equation}
\begin{equation}
	\nonumber
    \frac{1}{\sigma} ( \xi'_{+}	- \xi'_0 ) = \pi - \beta_0		
	\label{eq:xiprime_plus_approximate}
\end{equation}
Consequently, the distance between $\xi'_{+}$ and $\xi'_{-}$ satisfies the following relationship
\begin{equation}
	\nonumber
     \frac{\xi'_{+} - \xi'_{-}}{2\sigma}  = \frac{\pi}{2} - \beta_0 = \arccos Y_0
	\label{eq:distance_maxima}
\end{equation}
with $Y_0 = - \alpha^{\ast} \sin \left( \xi'_0 \right)$.  Using Eq.(\ref{eq:total_Hamiltonian_double_perturbation_perturbed}), the depth of the Hamiltonian well, defined as $\Delta H  =  \Htilde (\Jbold , \xi'_{+}  )  - \Htilde (\Jbold , \xi'_{-}  ) $, reads
\begin{equation}
	\nonumber
   \Delta H  =  
		   - h \left[  \cos \left( \frac{1}{\sigma} ( \xi'_{+} - \xi'_0 ) \right) 
          -  \cos \left( \frac{1}{\sigma} ( \xi'_{-} - \xi'_0 ) \right) \right]
		  - h' \left[ \cos \left(\xi'_{+}\right) -  \cos \left(\xi'_{-}\right)  \right]
	\label{eq:Hamiltonian_well_intermediate}
\end{equation}
The first term in the r.h.s. is equal to $2h \cos\beta_0 = 2 h \sqrt{1 - Y_0^2 }$. 
The second term is estimated via a Taylor expansion 
\begin{equation}
	\nonumber
       - h' \left[ \cos \left(\xi'_{+}\right) -  \cos \left(\xi'_{-}\right) \right] 
         \simeq  h' \sin(\xi'_0) (\xi'_{+} - \xi'_{-})
          \simeq - 2 h Y_0 \arccos Y_0
	\label{eq:Taylor_expansion_2nd_term}
\end{equation}
Wrapping-up, the depth of the potential well is thus
\begin{equation}
	\nonumber
   \Delta H 
		  \simeq 2 h 
	     \left[ \sqrt{1 - Y_0^2} 
			 - Y_0 \arccos Y_0
			 \right] 	 			
	\label{eq:Hamiltonian_well}
\end{equation}
This expression holds when $\xi'_0<0$. For shallow wells, it is more accurate to replace $ h$ by $\Delta H /2$. 
To simplify the notations, we proceed with the notation $h$ such that
\begin{equation}
    h =
    \left\{
    \begin{array}{cc}
         &  h \text{ if } Y_0 \ge 1 \text{ (unchanged)} \\
         &  \Delta H/2 \text{ if } Y_0 < 1
    \end{array}
    \right.
\end{equation}

If $\sigma \gg 1$, the respective roles of primary and secondary perturbations swap, i.e. it is better to write the Hamiltonian perturbation as
\begin{equation}
	\nonumber
    \Htilde (\Jbold , \xi )  = 
			- h  \cos \left( \xi \right) 		
			- h' \cos \left( \sigma ( \xi - \xi_0 ) \right) 			
	\label{eq:total_Hamiltonian_double_perturbation_perturbed_bis}
\end{equation}
In this case, the secondary perturbation oscillates faster than the primary, and the trapping condition becomes $ \left|\frac{1}{\alpha^{\ast} }\sin \left( \xi_0 \right)  \right|\le 1$. 

\subsection{Characteristic frequencies}
\label{Effective collision frequency}

Before approaching the derivation of the secondary entropy production rate, it is useful to introduce new frequencies of interest that emerge when addressing Hamiltonian dynamics with two perturbations.

\subsubsection{Detrapping collision frequency}
Entropy production comes mainly from particles trapped in the primary perturbation that have their motion affected by the secondary perturbation and undergo scattering in the action space.
An important parameter to consider is the effective detrapping frequency in the primary perturbation and how it is modified by the secondary perturbation. 
A change of sign of the displacement of the action due to $\Htilde'$ occurs when the phase $\xi'$ changes by an amount of $~\pi$. 
It is this change of sign that matters for the random walk responsible for diffusion.   

Let us recall that the two phases are related via the relationship $\xi' - \xi'_0 = \sigma \xi $. The phase $\xi'_0$ evolves slowly in time, so that its exact value does not matter much in this discussion. For simplicity, we set $\xi'_0 = 0$. 

When $\sigma \ll 1 $, the effect of the secondary perturbation on the force experienced by the particle in the primary perturbation is weak, since decorrelation occurs when $\xi' \simeq \pi$ which corresponds to $\xi \simeq \pi/\sigma \gg \pi$. This thus comes well after the change of sign in the first perturbation $\xi \simeq \pi$. Therefore, the effective detrapping time in the primary perturbation is unchanged, since decorrelation in the secondary perturbation occurs whilst the primary phase has already undergone a shift of several times $\pi$.  The relation $H_\Omega = \frac{1}{2} \Omega^2 - \omegab^2 \cos\xi$ implies that this corresponds to a variation of $\delta\Omega \simeq \omegab$. 
The effective detrapping frequency, defined as the inverse of the time needed by a particle to cross the trapped domain associated with the primary perturbation, is therefore $ \nueff \simeq D_\Omega / \omegab^2$. 

In the opposite case, i.e. $ \sigma \gg 1 $, the secondary perturbation plays an important role, since decorrelation occurs when $\xi' \simeq \pi$, and thus $\xi \simeq \pi / \sigma \ll \pi$. A phase shift of $ \xi \simeq \pi / \sigma$ implies a shift of $\delta\Omega \simeq \omegab/\sigma $ for deeply trapped particles. The effective detrapping frequency is thus enhanced by a factor $\sigma^2$, such that $ \nueff  \simeq \sigma^2 D_\Omega / \omegab^2 $. 
The general expression of the effective detrapping frequency is then defined as 
\begin{equation}
   \nueff \simeq \nu_d \max(1 , \sigma^2)
 \label{eq:effective_collision_frequency}
\end{equation}
where 
\begin{equation}
    \nu_d = \frac{1}{2}\frac{D_\Omega}{\omegab^2}
    \label{eq:nud}
\end{equation}
is a reference detrapping frequency.

\subsubsection{Drift frequency}
\label{Drift frequency}
%
Finally, there is the drift frequency of the secondary phase on the primary resonant surface. 
This frequency comes from the fact that, in the relation $\xi' - \xi'_0 = \sigma (\xi - \xi_0)$, $\xi_0$ and $\xi'_0$ can move slowly in time, related to the dynamics along the resonant surface, while the term $\sigma \xi$ is instead related to the motion across the resonant surface.

The time derivative of the relation $\xi' = n_d \theta_3 + \sigma \theta_2$ reads $\Omega' = n_d \Omegad + \sigma \Omega$, consistently with Eq.(\ref{eq:link_omega_omega'}). 
It allows the identification of $n_d \Omegad$ with $\Omega'(\Jbold_R)$, and $n_d \theta_3$ with a reference phase $\xi'_0$ such that $d_t\xi'_0 = \Omega'(\Jbold_R) = n_d \Omegad$. The approximation $\Omega'(\Jbold_R) \ll  \sigma \Omega$ used in Eq.(\ref{eq:Omega_and_Omega_prime}) thus holds as long as $|n_d \Omegad| \ll |\sigma \Omega|$. The number $n_d$ is an integer that fully defined by $(\nbold , \nbold')$.  In the following a "drift frequency" is defined as
\begin{equation}
\nonumber
    \omega_d = n_d \Omegad 
	\label{eq:drift_frequency_definition}
\end{equation} 
which can be recast, as detailed in Appendix \ref{appendix:Covariant_expression_of_the_entropy_production_rate}, as
 \begin{equation}
    \omega_d = (\nbold' - \sigma \nbold)  \cdot \Omegabold
 \label{eq:drift_frequency}
\end{equation}

Note that, formally, $\omega_d$ depends on the actions $\Jbold$.
However, as we will see in the application to tokamaks in Section \ref{sec:Computation of neoclassical fluxes}, $\omega_d$ depends weakly on $\Jbold$ in limits of interest. 


\subsection{Derivation of the secondary entropy production rate}
\label{Entropy production rate for a secondary perturbation}
As in the previous Section, the general approach is to solve the Fokker-Planck equation and to plug the solution into the entropy production rate functional.

\subsubsection{Solving the Fokker-Planck equation}

The effect of the primary perturbation is to produce a new equilibrium such that the distribution function, noted $\Ffrakeq$, is a function of the invariants of motion on the resonant surface, i.e. $\Jbold_R = (\Jb, \Jd)$ at given $J_1$, and so are the Hamiltonian $\Hfrakeq$ and thermodynamic potential $\Ufrakeq$. 
At low collisionality, the effect of the secondary perturbation is essentially to modify the motion of trapped particles in the primary Hamiltonian perturbation.
However, if the drift frequency is smaller than the bounce frequency, an adiabatic theory can be used --- dissipation will then come exclusively from collisions. 
When the collision frequency gets higher, the effect of trapping vanishes, and the secondary perturbation mainly affects passing particles --- the response is essentially of the linear Landau type. 
The Fokker-Planck equation to solve is
\begin{equation}
    - \left\{ \Htilde' , \Ffrakeq \right\}  
		- \left\{ \Hfrakeq , \Ftilde' \right\} 
		- \left\{ \Htilde' , \Ftilde' \right\}
		= \Ccal[\Ftilde']
	\label{eq:Fokker_Planck_secondary}
\end{equation}
written in the system of variables $(\alphab, \alphad, J_1, \Jb, \Jd)$. 
If the amplitude of the secondary perturbation is small enough, the Fokker-Planck equation becomes
\begin{equation}
    - \left\{ \Htilde' , \Ffrakeq \right\}  
		- \left\{ \Hfrakeq , \Ftilde' \right\}
		= \Ccal[\Ftilde']
	\label{eq:Fokker_Planck_linearised}
\end{equation}
where the collision operator is linearised in $\Ftilde$. 

\subsubsection{Entropy production rate}

The resonant entropy production rate functional is given by Eqs.(\ref{eq:entropy_production_generalised_resonant_bis}, \ref{eq:definition_flux_gamma_U})
\begin{equation}
  \nonumber
	  \Scalres \left(\partial_\Jbold\Ufrakeqdag \right)  = 
	 -  \frac{2}{T_\ast}  
	  \int d\gamma \;
		\Ueqdag
		  \left\{ \Htilde' , \Ftilde' \right\} 
	   = \frac{2}{T_\ast^2}  
	  \int d\gamma \; \Ffrakeq \;
		  \left\{ \Htilde' ,  \Ufrakeqdag\right\} \Utilde'
  \label{eq:entropy_production_resonant_secondary_0}
\end{equation}
Because $\Utilde'$ responds linearly with $\Ufrakeq$, the functional can be as well expressed as a bi-linear function of $\Ufrakeqdag$ just by removing the pre-factor $2$, i.e.
\begin{equation}
  \nonumber
	  \Scalres \left(\Ufrakeq , \partial_\Jbold\Ufrakeqdag \right)
	   = \frac{1}{T_\ast^2}  
	  \int d\gamma \; \Ffrakeq \;
		  \left\{ \Htilde' ,  \Ufrakeqdag\right\} \Utilde^{'\dag}
  \label{eq:entropy_production_resonant_secondary_reshaped}
\end{equation}
where it is understood that $\Utilde^{'\dag}$ is calculated for given $\Ufrakeqdag$.

The entropy production rate for a given set of wavenumbers $\nbold$ and $\nbold'$ is fully characterized by four dimensionless parameters:
\begin{itemize}
    \item The ratio of the amplitudes of the primary and secondary perturbations $h^\prime / h$;
    \item The ratio of Hamiltonian curvatures $\sigma$ (see Eq.(\ref{eq:sigma_vs_C_Cstar}));
    \item The normalised detrapping collision frequency (Eqs.(\ref{eq:drift_frequency}, \ref{eq:nud}))
    \begin{equation}
        \hat{\nu} = \frac{\nu_{\rm d}}{\omega_d}
        \label{eq:nuhat}
    \end{equation}
    \item The normalised bounce frequency (Eqs.(\ref{eq:bounce_frequency},\ref{eq:drift_frequency}))
    \begin{equation}
        \hat{\omega}_b = \frac{\omegab}{\omega_d}
        \label{eq:hat_omega_b}
    \end{equation}
\end{itemize}
Let us note that the normalised collisionality Eq.(\ref{eq:nu_Omega}) introduced to characterise the neoclassical transport due to the primary perturbation can be related to the frequencies defined above, i.e. $\nu_\Omega = D_\Omega/\omegab^3 = 2 \nu_d / \omegab = 2 \hat{\nu}/\hat{\omega}_b$.

The entropy production rate due to the secondary perturbation can be calculated in two collisionality regimes.

\subsection[]{Highly collisional case $\nu_\Omega \max(1 , \sigma^2) \gg 1 $}
\label{subsec:Highly_collisional_case_secondary}

Let us recall that the effective detrapping collision frequency in the primary perturbation modified by the effect of the secondary perturbation is given by Eq.(\ref{eq:effective_collision_frequency}), i.e. $ \nueff \simeq \left[D_\Omega / \omegab^2 \right] \: \max(1 , \sigma^2) $. 
If the regime is collisional $\nueff \gg \omegab $, or equivalently $ \nu_\Omega \max(1 , \sigma^2) >> 1 $, the Fokker-Planck equation Eq.(\ref{eq:Fokker_Planck_secondary}) is readily solved via a method already employed for a single perturbation. 
Indeed it is known \cite{Helander_Sigmar_2002} that in this regime, the full collision operator can be replaced by a simple Krook operator to yield the Fourier components of the distribution function
\begin{equation}
		F'_{\nbold'} = - H'_{\nbold'} 
			   \frac{ \nbold' \cdot \frac{\partial \Ffrakeq} {\partial \Jbold} }
				      { \nbold' \cdot \Omegabold - i 0^{+} }	
	   \label{eq:Landau_regime_secondary}
\end{equation}
However, one important difference is that the distribution function $\Ffrakeq$ is the one for a modified equilibrium in the presence of the primary perturbation.  


The entropy production rate is thus of the Landau type in the plateau regime, similar to the single perturbation case in the same collisional regime, and one uses the set of variables $(\alphabold,\Jbold)$ to find

\begin{equation}
  \Scalres  \left[\partial_\Jbold\Ufrakeqdag \right] =  
	   \frac{\pi}{2} \frac{1}{T_\ast^2} 
	   \int d\gamma \; \Feq(\Jbold) \;
	     \delta( \nbold'\cdot \Omegabold ) h'^2
			  \left( \nbold' \cdot 
	          \frac{\partial \Ufrakeqdag}{\partial \Jbold} 
			  \right)^2
 \label{eq:value_entropy_production_collisional_secondary collisional}
\end{equation}

The distribution function $\Ffrakeq$ has been replaced by $\Feq$ since the integration over the resonant response function $\delta(\nbold \cdot \Omegabold)$ enforces $\Feq$ to be a function of $\Jbold_R$ only, i.e. a function of the actions that lay on the resonant surface. 
Note also that the entropy production rate goes like the square of the gradient of the thermodynamic potential, as required. 

As detailed in Appendix \ref{appendix:Covariant_expression_of_the_entropy_production_rate}, this entropy production rate functional can be re-expressed in a more convenient covariant form

\begin{equation}
    \Scalres \left[\partial_\Jbold\Ueqdag \right] =  
	    \frac{1}{T_\ast^2} \frac{\pi}{2}
	    \int d\gamma \; \delta( \nbold' \cdot \Omegabold ) \; 
	    \Feq(\Jbold)  \; h'^2 \;      
			   \left( (\nbold' - \sigma \nbold)  \cdot 
	          \frac{\partial \Ueqdag}{\partial \Jbold} 
			   \right)^2 
   \label{eq:value_entropy_production_collisional_secondary_collisional}
\end{equation}
The entropy production rate does not depend on the collision frequency in this case, corresponding to a plateau regime for the secondary perturbation. However, this regime is physically meaningful only when a particle trapped by the primary perturbation can become a passing particle as a result of collisions. This occurs when $h>h'$.

This subtle point can be illustrated by considering a rippled tokamak, where a particle locally trapped in a ripple minimum becomes a banana orbit following a detrapping collision. In this case, there is no plateau regime, since passing particles are not involved. Conversely, a banana orbit can become a passing particle through collisions. Consequently, a banana-drift plateau regime emerges at sufficiently high collision frequency.

\subsection[]{Weakly collisional case $\nu_\Omega \max(1 , \sigma^2) \ll 1 $}
\label{subsec:Weakly_collisional_case_secondary}

In the weakly collisional case, where $\nueff \ll \omegab $, or equivalently $ \nu_\Omega \max(1 , \sigma^2) < 1 $, the derivation is more involved and details are readily found in Appendix \ref{appendix:weakly_collisional_regime_exact}.

As demonstrated in Appendix \ref{appendix:Solving_Fokker_Planck_equation_weakly_coll_exact}, the normalized Fourier components $\hat{F}_{\nb}$ of the distribution function is solution the following Fokker-Planck equation
\begin{equation}
    i \hat{\nu} \left( G(K)   
		    \frac{ \partial^2 \hat{F}_{\nb}(K) }{\partial K^2}
		      + \frac{ \partial \hat{F}_{\nb}(K) }{ \partial K} \right) + 
		      \left( 1 + \frac{2 \nb  \hat{\omega}_b}{R(K)}  \right)  \hat{F}_{\nb}(K)
		= L_{\sigma,\nb}(K) 
	\label{eq:full_FP_equation_normalised_main}
\end{equation}
where $L_{\sigma,\nb}$ is from Eq. \ref{eq:expression_LvsK_bis}, and $G(K) = 2 Q(K)/R(K)$.

The solution of this equation can then be plugged into the expression of the entropy production rate functional, as detailed in Appendix \ref{appendix:secondary_entropy_rate_exact}, to yield, in its covariant form:
\begin{equation} 
	  \Scalres  \left[\partial_\Jbold\Ueqdag \right]  =
	   \frac{1}{T_\ast^2}  
	  \int \: d\gamma \:  \delta(\nbold\cdot\Omegabold) \: 
	    \Feq \:  h'^2 \: \hat{\omega}_b 
	    \Im\left[W
	         ( 
		           \sigma , 
		           \hat{\nu} 
		       )
		     \right]
		\left( 
	  	\left(\nbold' - \sigma \nbold \right)  \cdot 
			  \frac{\partial \Ueqdag}{\partial \Jbold} 
		\right)^2
  \label{eq:entropy_production_resonant_secondary_final_covariant}
\end{equation}
Where the dimensionless ``weight function'' $W(\sigma,\hat{\nu})$ is defined as
\begin{equation}
	  W(\sigma,\hat{\nu}) = \sum_{\nb=-\infty}^{+\infty} 
	      W_{\nb}(\sigma,\hat{\nu})
	 \label{eq:def_Phi_main}
\end{equation}
with
\begin{equation}
	  W_{\nb}(\sigma,\hat{\nu}) = \frac{1}{2} 
	    \int_0^{1} dK R(K)
	         L_{\sigma,\nb}(K) \hat{F}_{\nb}(K)
  \label{eq:def_Phi_nb_main}
\end{equation}
The source term $L_{\sigma,\nb}(K)$ is given by Eq.(\ref{eq:expression_LvsK_bis}).
Similar formulations are available in the literature, though not in the form of an entropy production rate \cite{Shaing_2010, Sun_2019}.

These expressions hold as long as islands in the phase space are well separated. 
If island widths become large compared to the distance between resonant surfaces, the motion of the particle starts being stochastic, and losses must be computed differently, e.g. using a quasi-linear theory. 
In practice, this situation only occurs for energetic particles \cite{Goldston_1981,Grua_1990,Varennes_2023}.

In principle, the function $W(\sigma,\hat{\nu})$ must be computed numerically, which requires solving the Fokker-Planck equation for each Fourier component $\nb$. 
Fortunately, fast convergence is usually granted with $\nb$, as shown in the next subsection \ref{sec:Numerical_computation_of_the_weight_function}, so that only a few components are needed.  

\subsubsection[]{Analytical expression in the adiabatic limit $\omegab \gg \omega_d$}
\label{Entropy production rate for a secondary perturbation adiabatic}

When the bounce frequency $\omegab$ is much greater than the drift frequency $\omega_d$, the bounce average component $\nb=0$ of the distribution function dominates, which to simplify the formulation. The weight function $W(\sigma,\hat{\nu}) = W_0(\sigma,\hat{\nu})$ 
then depends on the $\nb=0$ component of the distribution function
which is solution of the normalised bounce-averaged Fokker-Planck equation 
\begin{equation}
    i \hat{\nu} \left( G(K)   
		    \frac{\partial^2 \hat{F}_{\nb=0}(K)}{\partial K^2}
		      + \frac{\partial \hat{F}_{\nb=0}(K)}{\partial K} \right) + \hat{F}_{\nb=0}(K)
		= L_{\sigma}(K) 
	\label{eq:bounce_average_FP_equation_normalised}
\end{equation}
and on the source term that reduces to
\begin{equation}
  \nonumber
    L_{\sigma}(K) = 
    \int_{-\pi}^{\pi} \frac{d\alphab}{2\pi}
		   \cos\left[ \sigma \theta_2(\Jbold,\alphab) \right] =	
     \frac{1}{R(K)} \frac{2}{\pi } 
        \int_0^{\frac{\thetab}{2} } du \: \cos\left(2 \sigma u\right)
	           \frac{1}{ \sqrt{ K - \sin^2 u } }
\label{eq:expression_LvsK_main}
\end{equation}


As detailed in Appendix \ref{appendix:secondary_entropy_rate_analytical}, the weight function carries two asymptotic limits when $\sigma \ll 1$ and $\sigma \gg 1$.
Both limits can be connected in the following fitting expression
\begin{equation}
    W(\sigma,\hat{\nu}) \simeq   \frac{i \Lambda_{\rm II}(\sigma)}{\hat{\nu} \max{( 1 , \sigma^2 )}} 
	\label{eq:value_Phi_sigma_all_sigma_main}
\end{equation}
where 
\begin{equation}
    \Lambda_{\rm II}(\sigma) = \left(\frac{1}{\Lambda_0} + \frac{1}{\Lambda_{\infty}(\sigma)} \right)^{-1} 
\end{equation}
such that 
\begin{equation}
  \nonumber
   \Lambda_{\rm II}(\sigma) = 
	   \left\{
	     \begin{array}{lc}
          \Lambda_0& = \dfrac{ 16 }{ 9 \pi } \text{  if } \sigma \ll 1  
          \\[8pt]
		   \Lambda_{\infty}(\sigma)& = \dfrac{1}{\pi \sigma} \text{  if } \sigma \gg 1
     \end{array}
    \right.
  \label{eq:definition_Lambda_covariant_main}
\end{equation}
Details on the derivation of the $\Lambda_{\rm II}$ function are present in the Appendix \ref{appendix:Analytical_form_of_the_weight_function_in_the_bounce_average_limit_at_high_collisionality}.

A fully analytical expression of the entropy production rate functional can thus be written as
\begin{equation}
	  \Scalres\left[\partial_\Jbold \Ueqdag \right]  = 
	   \frac{1}{T_\ast^2}  
	    \int \: d\gamma \:  \delta( \nbold\cdot\Omegabold ) 
		    \Feq h'^2  \Lambda_{\rm II}(\sigma) 
			     \frac{\omegab}{\nueff} 
		   \left( 
		     	\left(\nbold' - \nbold \sigma \right) \cdot 
							   	\frac{\partial \Ueqdag}{\partial \Jbold} 
		  \right)^2
  \label{eq:entropy_production_resonant_secondary_covariant_main}
\end{equation}

In the next Section, the weight function Eq.\ref{eq:def_Phi_main} is numerically computed and compared against well-known results in the neoclassical theory.
In addition, the range of validity of the analytical estimation Eq.(\ref{eq:entropy_production_resonant_secondary_covariant_main}) is assessed.

\subsection[]{Numerical computation of the weight function $W$}
\label{sec:Numerical_computation_of_the_weight_function}

The Fokker-Planck equation Eq.(\ref{eq:full_FP_equation_normalised_main}) is a second-order differential equation over the complex Fourier components $ \hat{F}_{\nb}(K)$ of the distribution function. 
The bounce average Fokker-Planck equation, which is the same as the component $\nb=0$ of Eq.(\ref{eq:full_FP_equation_normalised_main}), has been widely discussed in the literature \cite{Galeev_1969, Mynick_1983, Beidler_1987, Kovrizhnykh_Shasharina_1990, Kovrizhnykh_1991, Shaing_2010, Satake_2011, Sun_2019}. 
Let us thus focus first on the bounce average component $\hat{F}_{0}(K)$ of the distribution function. 
Following the procedure proposed by Beidler \cite{Beidler_1987}, the bounce average distribution function can be split in real and imaginary parts, i.e. $\hat{F}{0}(K) = X(K) + i Y(K)$. 
In consequence, Eq.(\ref{eq:full_FP_equation_normalised_main}) leads to two second-order differential equations on $X$ and $Y$ that are coupled. 
Hence, 4 different boundary conditions are required to solve the system. 

Two boundary conditions are prescribed by regularity conditions at the origin: indeed, the functions $R(K)$, $Q(K)$ and $G(K)$ behave near $K=0$ as
\begin{equation}
	\begin{array} {ccccc}
	        R(K) \rightarrow 2 & \; &
			Q(K)  \rightarrow K  & \; &
            G(K)  \rightarrow K 
		\end{array} 
	  \label{eq:behavior_near_origin}
\end{equation}
Eq.(\ref{eq:full_FP_equation_normalised_main}) thus exhibits a regular singular point at $K=0$, and so does its real and imaginary parts. 
One solution of the homogeneous equation (r.h.s.=0) is singular at $K=0$ (logarithmic singularity), while the other one is regular. 
Only the regular solution should be selected, which provides the requested constraint. 
This is easily implemented using an auxiliary variable $x=\sqrt{K}$ and imposing a Von Neumann boundary condition at $x=0$.  

The situation at the other end of the trapped particle domain $K=1$ is more intricate. 
A common prescription is of the Dirichlet type $\hat{F}_{0}(K=1)=0$, and therefore $X(1)= Y(1) =0$. 
A collisional boundary layer develops at low collision frequency near $K=1$, of width $\Delta K_{\rm coll} \simeq \sqrt{\nueff/\omega_d}$, where $\nueff = \nu_d \: \text{Max}(1,\sigma^2)$. 
This boundary layer is responsible for a $\sqrt{\nu}$ scaling of the neoclassical transport coefficients with collision frequency $\nu$ \cite{Galeev_1969, Mynick_1983, Beidler_1987, Kovrizhnykh_Shasharina_1990, Kovrizhnykh_1991, Shaing_2010, Sun_2019}. 

Moreover, several authors \cite{Mynick_1983, Beidler_1987, Kovrizhnykh_Shasharina_1990, Shaing_2010} mention that particles that are trapped in the primary perturbation can transition continuously to particles trapped in the secondary perturbation (and vice-versa), in the absence of any collision. 
The contribution from these ``transit'' particles prevails in neoclassical fluxes at very low collision frequencies. 
This feature is responsible for a scaling of transport coefficients that is linear with $\nu$. 
Following \cite{Beidler_1987}, it appears that while $Y(K)$ still satisfies a Dirichlet boundary condition, the real part $X(K)$ overspills in the $K>1$ area. The boundary condition at $K=1$ is of the form   
\begin{equation}
  \nonumber
    \left. \frac{1}{X} \frac{dX}{dK} \right|_{K=1} = 
      - 2 \frac{h}{h'}
  \label{eq:BC_transit_Kgt1}
\end{equation}
This condition imposes an exponential decay of the distribution function in the region $K>1$. 
It makes sense only if the condition $h \gg h'$ is fulfilled. The exponential decay length is thus $\Delta K_{\rm trans} \sim h'/h$, to be compared with the collisional boundary layer near $K=1$, of the order of $\Delta K_{\rm coll} = \sqrt{\nueff/\omega_d}$. 
If $\Delta K_{\rm coll} >\Delta K_{\rm trans}$, the collisional boundary layer matters most, resulting in a $\sqrt{\nu}$ regime. 
In the opposite situation $\Delta K_{\rm coll} < \Delta K_{\rm trans}$, transit particles play a decisive role and transport is in the $\nu$ regime. 

In practice the $\nu$ regime is rarely reached in rippled tokamaks or with external magnetic perturbations. We will thus limit ourselves to the $\sqrt{\nu}$ regime, keeping in mind that only collision frequencies that verify $\Delta K_{\rm coll} \gg \Delta K_{\rm trans}$ should be considered.  
Under this restriction, the distribution function can be requested to vanish at the border of the trapped domain $K=1$, so that $\hat{F}_{\nb}(K=1)=0$. We will admit that this logic applies to all Fourier components $\nb \neq 0$, though the literature is not entirely clear on this point.

In this simplified framework, the Fokker-Planck equation can thus be treated as a single second-order partial differential equation with complex coefficients, with a solution that is regular at $K=0$ and satisfies a Dirichlet boundary condition $\hat{F}_{0}(1)=0$ at the other end of the trapped domain. 

At lower collisionality $\hat{\nu} \ll 1$ the physics of the boundary layer must be accounted for, thus leading to the $\sqrt{\nu}$ regime. As explained in the previous section, the $\nu$ regime cannot be recovered because of simplified boundary conditions on the distribution function at $K=1$. A cut-off must thus be introduced at $\hat{\nu}(\sigma) = (h'/h)^2$ whenever $h'<h$.

\subsubsection{General case - impact of bounce resonances}
\label{General case - impact of bounce resonances}
As seen in previous sections, transport depends sensitively on the parameter $\sigma$, the ratio of helical wavenumbers. 
In order to assess its impact, it is first set to a small value $0.1$. 
The other relevant parameters are the bounce average frequency normalised to the drift frequency $\hat{\omega}_b = \omegab/\omega_d$, which we choose to be a value of 10, and the collisionality $\hat{\nu}$, which we similarly choose to set as 10. 

\begin{figure}[H]
\centering
	\centerline{\includegraphics[width=16cm]
	{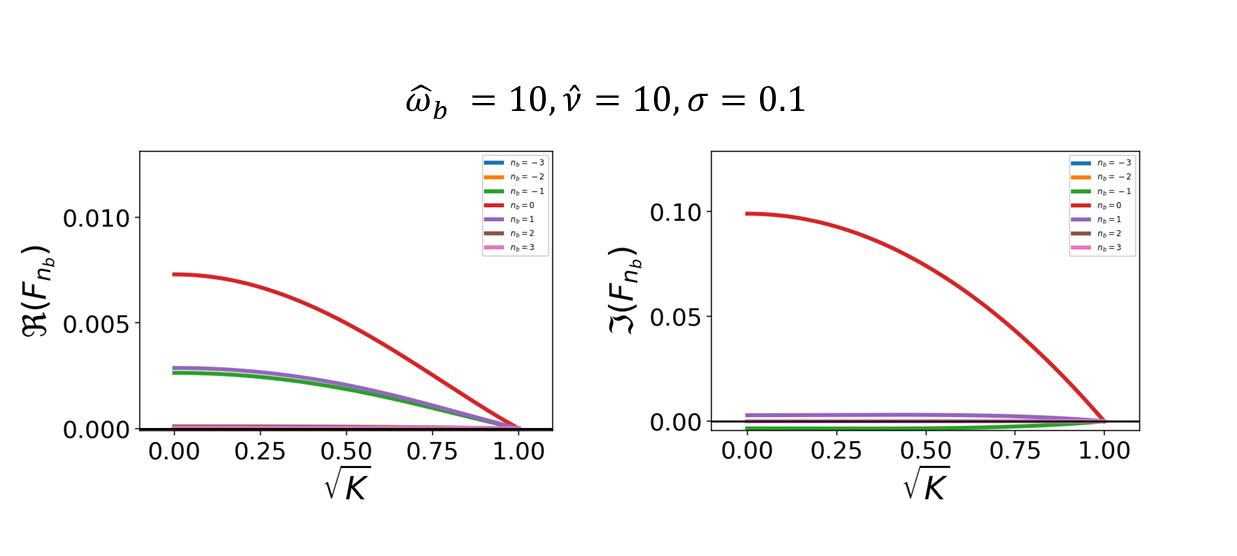}}
	\caption{\small{ Solutions of the Fokker-Planck equation Eq.(\ref{eq:full_FP_equation_normalised_main}) versus the trapping parameter $K$, for several bounce resonances, with parameters $\sigma=0.1$, $\hat{\nu} =  10$, and $\hat{\omega}_b = 10$. The first sidebands $\nb=\pm 1$ are not small compared to the bounce average component $\nb=0$. }}
	 \label{Fig:Solution_FP_full_low_omegab}
\end{figure}

\begin{figure}[H]
\centering
	\centerline{\includegraphics[width=10cm]
	{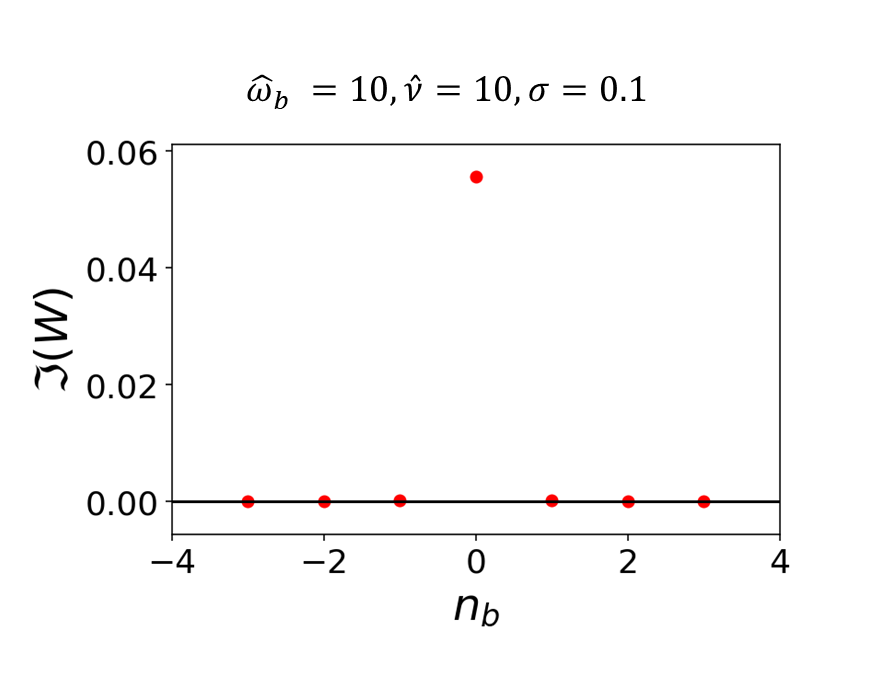}}
	\caption{\small{ Imaginary part of the weight function $W_{\nb}(\sigma,\hat{\nu})$ as given by Eq.(\ref{eq:def_Phi_nb_main}), versus the bounce resonance index $\nb$, with parameters $\sigma=0.1$, $\hat{\nu} =  10$, and $\hat{\omega}_b = 10$. Contributions from sidebands $\nb \neq 0$ are negligible compared to that from the bounce average $\nb = 0$. }}
	 \label{Fig:Weight_function_Phi_FP_full_low_omegab}
\end{figure}

Fig.\ref{Fig:Solution_FP_full_low_omegab} shows the harmonics $ \hat{F}_{\nb}(K)$ of the distribution function due to the secondary Hamiltonian perturbation acting on particles that are trapped in the primary perturbation. 
Even for this relatively low value of the normalised bounce frequency $\hat{\omega}_b = 10$, the sidebands are small compared to the bounce-averaged distribution function $\nb=0$. 
Let us stress that no boundary layer develops near $K=1$ for this high value of the collisionality. 
Nevertheless, this property persists for lower values of the parameter $\hat{\nu}$. 
Thus, the weight function $W(\sigma,\hat{\nu})$, which is the sum of all harmonics Eqs.(\ref{eq:def_Phi_main}, \ref{eq:def_Phi_nb_main}) is governed mainly by the bounce average component of the distribution function. 
This feature can be seen in Fig.\ref{Fig:Weight_function_Phi_FP_full_low_omegab}. Only the imaginary part of the weight function is shown since it alone contributes to the entropy production rate. 
As expected, contributions from sidebands $\nb \neq 0$ are negligible compared to those from the bounce average $\nb = 0$. 

\begin{figure}[H]
\centering
	\centerline{\includegraphics[width=16cm]
	{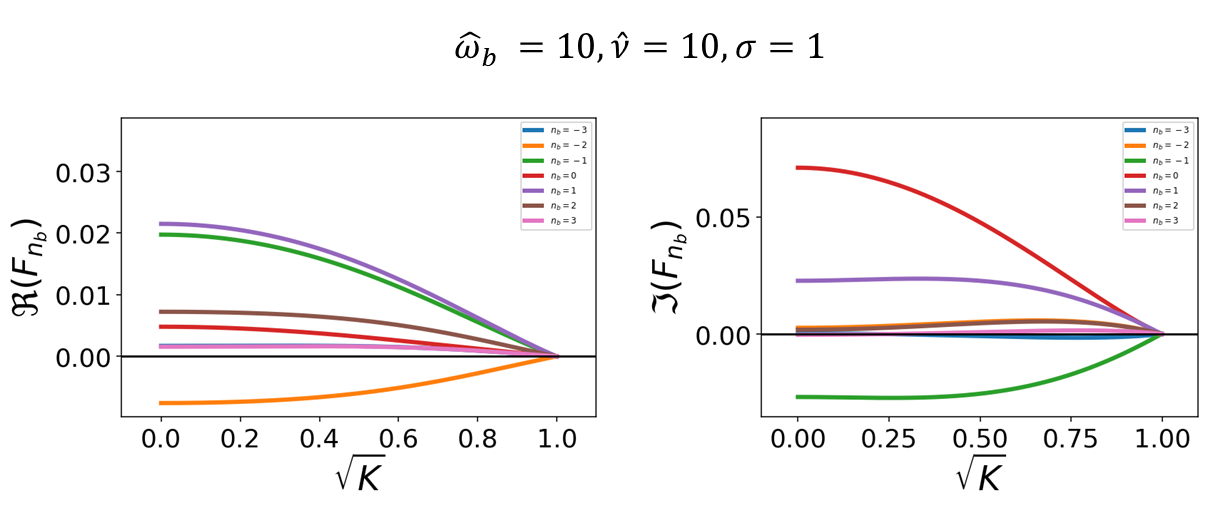}}
	\caption{\small{ Solutions of the Fokker-Planck equation Eq.(\ref{eq:full_FP_equation_normalised_main}) versus trapping parameter $K$, for several bounce resonances, with $\sigma=1$, $\hat{\nu} =  10$, and $\hat{\omega}_b = 10$. The first sidebands $\nb=\pm 1$ of the imaginary component are significant compared to the bounce average component $\nb=0$. }}
	 \label{Fig:Solution_FP_full_sigma_one}
\end{figure}

\begin{figure}[H]
\centering
	\centerline{\includegraphics[width=10cm]
	{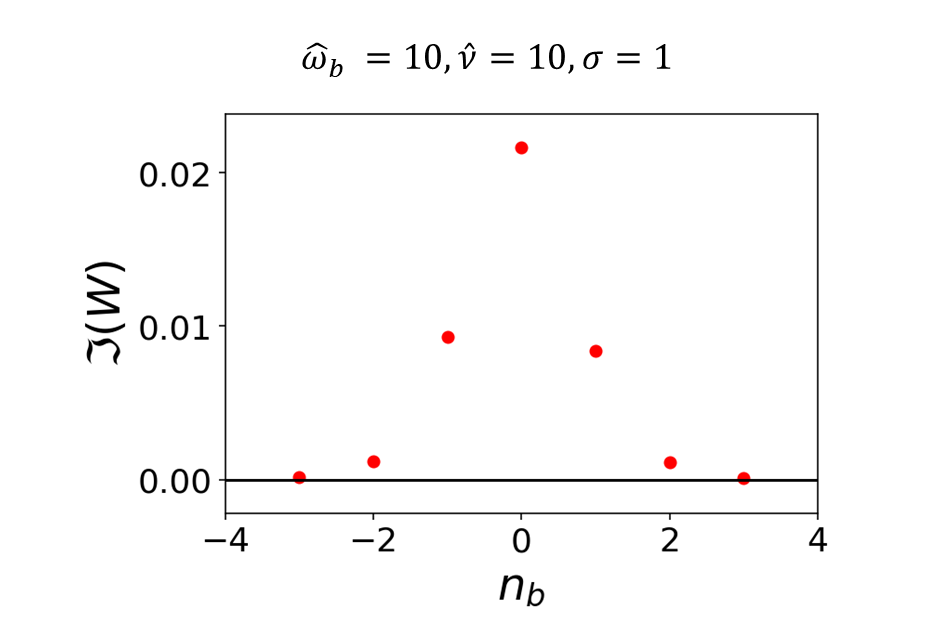}}
	\caption{\small{Imaginary part of the weight function $W_{\nb}(\sigma,\hat{\nu})$ as given by Eq.(\ref{eq:def_Phi_nb_main}), versus the bounce resonance index $\nb$, with $\sigma=1$, $\hat{\nu} =  10$, and $\hat{\omega}_b = 10$. Sidebands are not negligible compared to the bounce average contribution $\nb=0$. }}
	 \label{Fig:Weight_function_Phi_FP_full_sigma_one}
\end{figure}

It has been mentioned by Sun and collaborators  \cite{Sun_2019} that sidebands may contribute significantly to neoclassical transport, and specifically to neoclassical toroidal viscosity (NTV). 
We are able to recover this result by looking at 2 key parameters, $\sigma$ and $\hat{\omega}_b$. 
In Fig.\ref{Fig:Solution_FP_full_sigma_one}, we set $\sigma = 1$ while keeping $\hat{\omega}_b$ and $\hat{\nu}$ unchanged, and we can see that the sidebands $\nb \neq 0$ of the distribution function are now significant, particularly the imaginary component, which is what matters for entropy production and therefore transport. 
We also see that the overall shape of the harmonics $ \hat{F}_{\nb}(K)$ remains smooth in $K$. As expected, the weight function $W(\sigma,\hat{\nu})$ spreads over several sidebands $W_{\nb}(\sigma,\hat{\nu})$, as shown in Fig.\ref{Fig:Weight_function_Phi_FP_full_sigma_one}. 
However, increasing the value of the second key parameter, the bounce frequency $\hat{\omega}_b$, up to, for instance, a value of $50$, which is a more realistic value for rippled tokamaks, re-establishes the dominance of the bounce average component.

\begin{figure}[H]
\centering
	\centerline{\includegraphics[width=16cm]
	{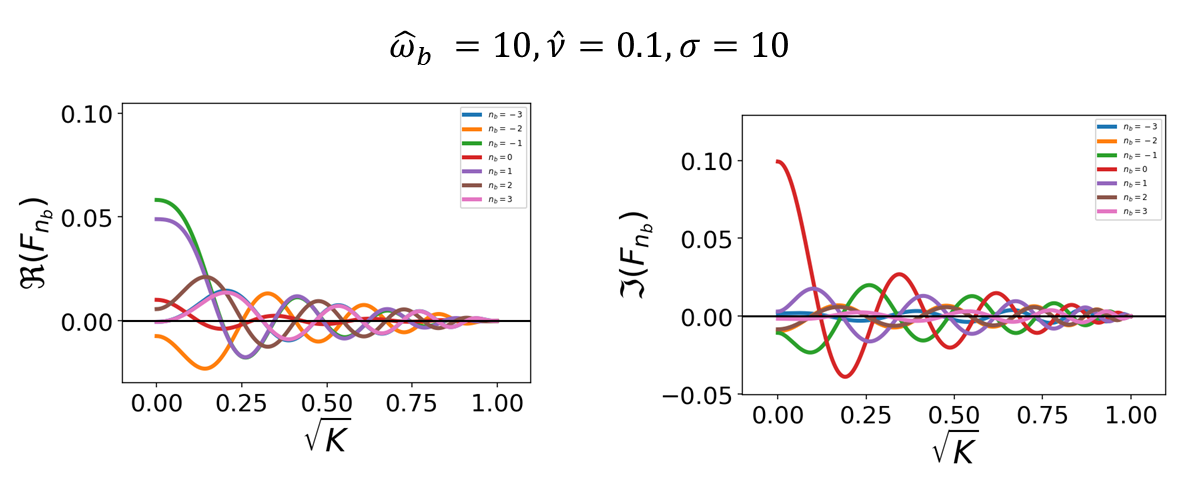}}
	\caption{\small{ Solutions of the Fokker-Planck equation Eq.(\ref{eq:full_FP_equation_normalised_main}) versus the trapping parameter $K$, for the first $\pm 3$ sideband bounce resonances, with $\sigma=10$, $\hat{\nu} =  0.1$, and $\hat{\omega}_b = 10$. The bounce average component $\nb=0$ is the largest near $K=0$, though sidebands are comparable in amplitude for larger $K$. Moreover, a large number of harmonics (here $21$) are necessary to achieve convergence. }}
	 \label{Fig:Solution_FP_full_low_omegab_highsigma}
\end{figure}

The low value of $\sigma=0.1$ that we started with typically corresponds to primary ripple or helically trapped particles. In order to analyse the outcome of banana particles under the effect of a helical perturbation, i.e. with the helical perturbation as secondary, we invert the value of $\sigma$, which thus becomes $\sigma=10$. 
The effective detrapping frequency is then multiplied by $\sigma^2$, so, to remain in the same collisional regime, we set $\hat{\nu} =  0.1$. The bounce frequency is kept constant for now at $\hat{\omega}_b = 10$. Then, Fig.\ref{Fig:Solution_FP_full_low_omegab_highsigma} shows that the bounce average component $\nb=0$ of the distribution function still dominates over the other components $\nb \neq 0$, though not by a large factor. For clarity, only the $\pm 3$ sideband bounce resonances are shown, although all harmonics up to $-10< \nb < 10$ were computed. Many of these harmonics are similar in amplitude, which implies that the range of sidebands that must be included to achieve convergence is large. 

Lastly, we see that these solutions have an oscillatory character with respect to the trapping parameter $K$, which is a direct consequence of the large value of $\sigma$. 

\begin{figure}[H]
\centering
	\centerline{\includegraphics[width=10cm]
	{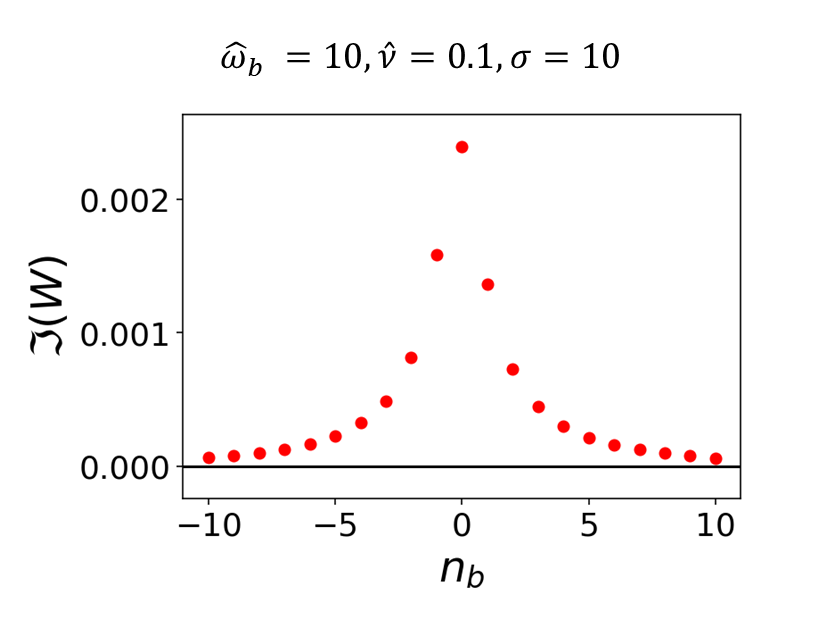}}
	\caption{\small{Imaginary part of the weight function $W_{\nb}(\sigma,\hat{\nu})$ as given by Eq.(\ref{eq:def_Phi_nb_main}), versus the bounce resonance index $\nb$, with $\sigma=10$, $\hat{\nu} =  0.1$, and $\hat{\omega}_b = 10$. The convergence evolves slowly with the number of harmonics. The bounce average component is larger than sidebands, but not by a large factor. }}
	 \label{Fig:Weight_function_Phi_FP_full_low_omegab_highsigma}
\end{figure}

This slow convergence is confirmed in Fig.\ref{Fig:Weight_function_Phi_FP_full_low_omegab_highsigma}, which shows the weight function $W_{\nb}(\sigma,\hat{\nu})$ (imaginary part) vs $\nb$. 
Obviously, in this case, keeping only the harmonic $\nb=0$ is a poor approximation. 
Moreover, quite a high number of harmonics is required to arrive at an accurate value of $W(\sigma,\hat{\nu})$. 
A closer look at values of $W_{\nb}$ at $\sigma=10$ compared to $\sigma=0.1$, with $\hat{\nu}$ held constant, clearly shows that amplitudes of $W_{\nb}$ are much smaller when $\sigma=10$ than when $\sigma=0.1$. 
In fact, the total value of $\Im[W]$ is $ \sim 10^{-2}$ for $\sigma=10$, as compared to $0.25$ for $\sigma=0.1$, which is a mere $4$\% of the former. 
When translated to a tokamak, this result implies that primary helical trapping (low $\sigma$) produces more transport than primary toroidal trapping (bananas, high $\sigma)$, a well-known result. 

\begin{figure}[H]
\centering
	\centerline{\includegraphics[width=10cm]
	{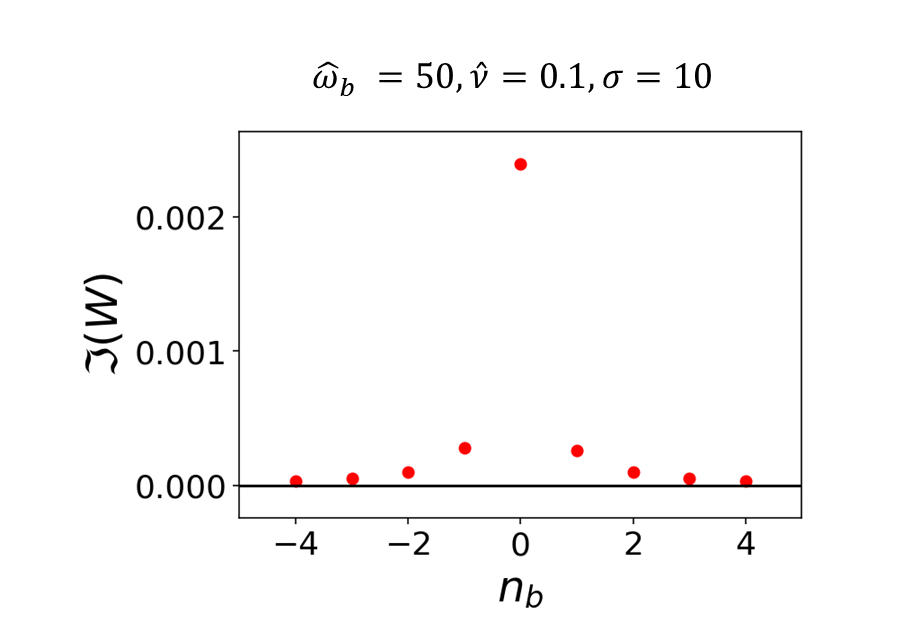}}
	\caption{\small{Imaginary part of the weight function $W_{\nb}(\sigma,\hat{\nu})$ as given by Eq.(\ref{eq:def_Phi_nb_main}), versus the bounce resonance index $\nb$, with $\sigma=10$, $\hat{\nu} =  0.1$, and $\hat{\omega}_b = 50$. The bounce average component dominates over sidebands. }}
	 \label{Fig:Weight_function_Phi_FP_full_high_omegab_highsigma}
\end{figure}

We now turn to more realistic and larger values of the bounce frequency $\hat{\omega}_b = 50$. The weight function is shown on  Fig.(\ref{Fig:Weight_function_Phi_FP_full_high_omegab_highsigma}). As in $\sigma=0.1$, the bounce average component $\nb=0$ clearly prevails over the sidebands.

In summary, the bounce average approximation works well for a high enough $ \hat{\omega}_b$, whatever the value of $\sigma$. Thus, we will proceed to analyse this limit in more detail, particularly to assess the validity of the analytical solutions derived above.

\subsubsection[]{Numerical and analytic solutions in the adiabatic limit $\omegab \gg \omega_d$}
\label{Numerical and analytic solutions in the adiabatic limit}
\begin{figure}[H]
\centering
	\centerline{\includegraphics[width=10cm]
	{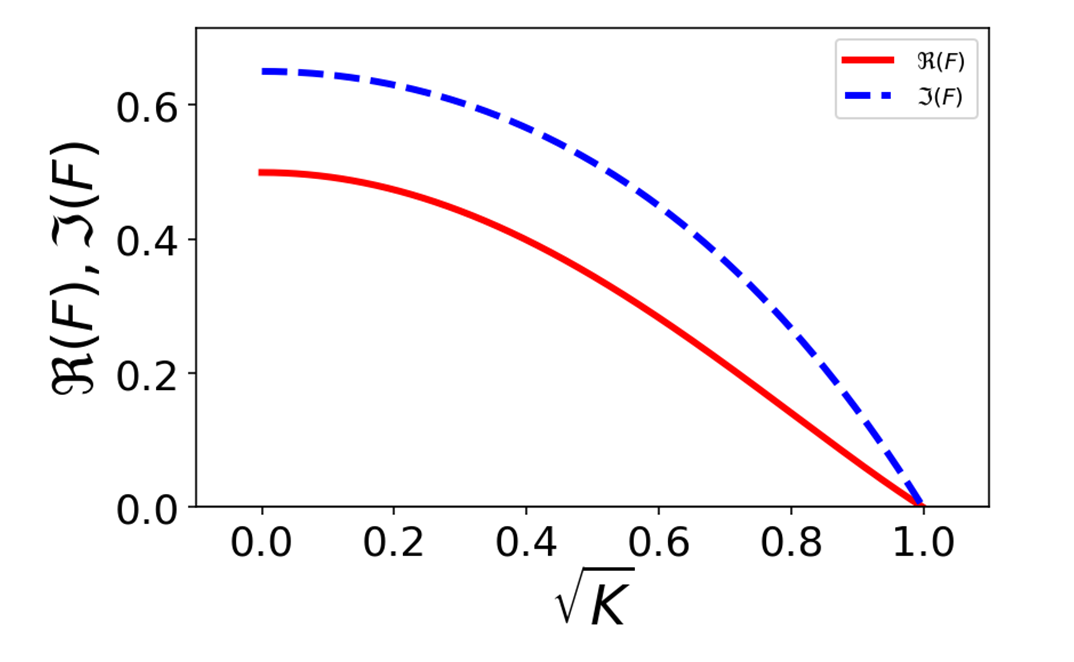}}
	\caption{\small{ Solution of the bounce average Fokker-Planck equation Eq.(\ref{eq:bounce_average_FP_equation_normalised}) in the low $\sigma$ limit, with $\sigma=0.1$ and $\hat{\nu} =  1$. }}
	 \label{Fig:Solution_FP_lowsigma}
\end{figure}

\begin{figure}[H]
\centering
	\centerline{\includegraphics[width=16cm]
	{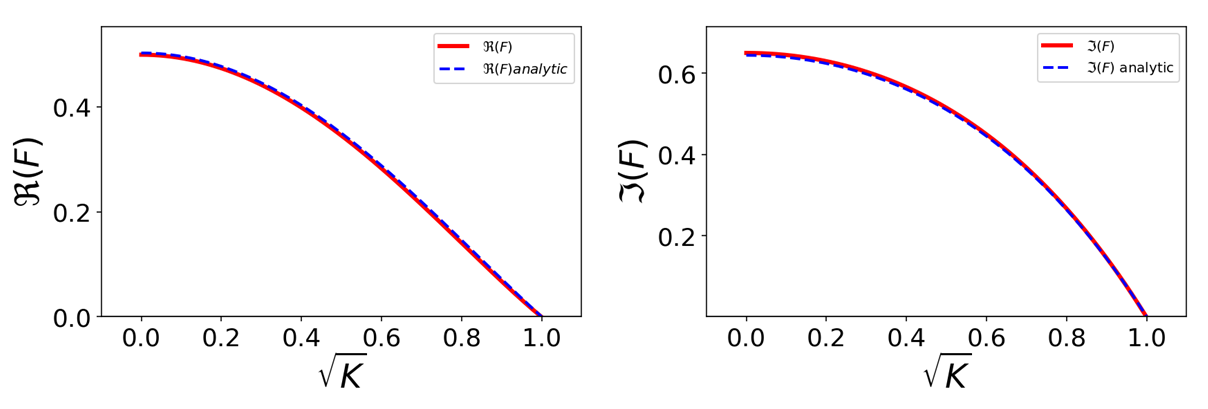}}
	\caption{\small{ Comparison of the numerical solution of the bounce average Fokker-Planck equation Eq.(\ref{eq:bounce_average_FP_equation_normalised}) to the analytic solution Eq.(\ref{eq:solution_bounce_average_FP_low_sigma}) in the low $\sigma$ limit, with $\sigma=0.1$ and $\hat{\nu} =  1$. }}
	 \label{Fig:Solution_FP_lowsigma_vs_analytic}
\end{figure}

\begin{figure}[H]
\centering
	\centerline{\includegraphics[width=8cm]
	{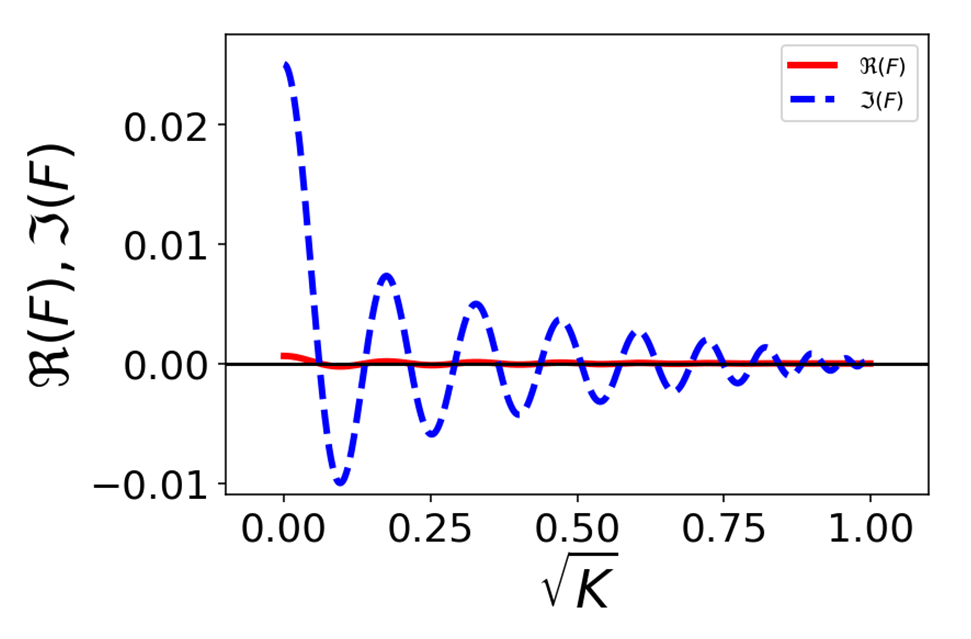}}
	\caption{\small{ Solution of the bounce average Fokker-Planck equation Eq.(\ref{eq:bounce_average_FP_equation_normalised}) in the high $\sigma$ limit, with $\sigma=20$ and $\hat{\nu} =  1$. }}
	 \label{Fig:Solution_FP_highsigma}
\end{figure}

\begin{figure}[H]
\centering
	\centerline{\includegraphics[width=14cm]
	{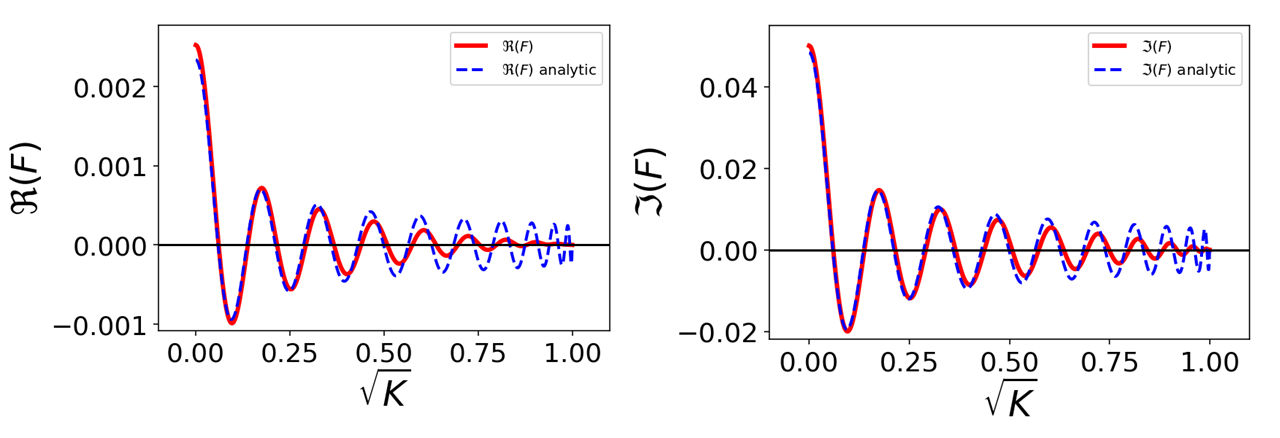}}
	\caption{\small{ Comparison of the numerical solution of the bounce averaged Fokker-Planck equation Eq.(\ref{eq:bounce_average_FP_equation_normalised}) to the analytic solution Eq.(\ref{eq:solution_bounce_average_FP_high_sigma}) in the high $\sigma$ limit, with $\sigma=20$ and $\hat{\nu} =  0.05$. $\sigma$ has been adjusted to ensure boundary condition at $K=1$ --- here $\sigma = 20.329$ instead of $20$.}}
	 \label{Fig:Solution_FP_highsigma_vs_analytic}
\end{figure}

In the adiabatic limit $\omegab \gg \omega_d$, it is sufficient to solve the bounce average equation Eq.(\ref{eq:bounce_average_FP_equation_normalised}). Fig.\ref{Fig:Solution_FP_lowsigma} shows an example for small $\sigma$, here $\sigma=0.1$, and intermediate collisionality $\hat{\nu} =  1$ . 
As mentioned in Section \ref{General case - impact of bounce resonances}, the numerical solution $ \hat{F}(K)$ is smooth in $K$, which is in line with the analytical solution Eq.(\ref{eq:solution_bounce_average_FP_low_sigma}). The normalised solution reads
\begin{equation}
        \hat{F}(K) = 
        \left[ 
           1 - \frac{J_0 \left( 2 \sqrt{-i \frac{K}{\hat{\nu}} } \right)}
                    {J_0 \left( 2 \sqrt{-i \frac{1}{\hat{\nu}} } \right)} 
        \right]
	\label{eq:solution_bounce_average_FP_low_sigma}
\end{equation}
In the case where $\hat{\nu} >1 $, this solution reduces to
\begin{equation}
    \hat{F}(K) = i \frac{1 - K}{1 - i \hat{\nu}}
	\label{eq:solution_large_nuhat_Fhat}
\end{equation}
Comparison of the analytical Eq.(\ref{eq:solution_bounce_average_FP_low_sigma}) and numerical solutions is shown on Fig.\ref{Fig:Solution_FP_lowsigma_vs_analytic}. 
In practice, fair agreement is found whenever $\hat{\nu} >1$, as expected. \\
We now turn to the large $\sigma$ limit. Fig.\ref{Fig:Solution_FP_highsigma} shows the numerical solution when $\sigma=20$ and $\hat{\nu} =  0.05$. 
 The comparison to the corresponding analytical solution at large $\sigma$ Eq.(\ref{eq:solution_bounce_average_FP_high_sigma}) is shown on Fig.\ref{Fig:Solution_FP_highsigma_vs_analytic}. The normalised solution is 
\begin{equation}
		   \hat{F}(K) = 
			   \frac{ 1 }
				      { 1 - i \hat{\nu} \sigma^2 }	J_0(\sigma \thetab) 
	\label{eq:solution_bounce_average_FP_high_sigma}
\end{equation}
where $\thetab = 2 \arcsin K $, up to a slight adjustment of $\sigma$ to satisfy the boundary condition at $K=1$. 
It appears that numerical and analytical solutions are close for deeply trapped particles $K\ll 1$, while some discrepancy is observed near the trapped domain boundary $K=1$, as expected. 
Let us recall that transport coefficients involve integrals of the distribution function weighted by $J_0(\sigma \thetab)$, which is maximum near $\thetab=0$, equivalent to $K=0$. 
Therefore, the largest contribution comes from deeply trapped particles for which agreement is acceptable, which helps to minimise the overall difference between the analytical expression of the entropy production rate and its numerical evaluation. 

\begin{figure}[H]
\centering
	\centerline{\includegraphics[width=10cm]
	{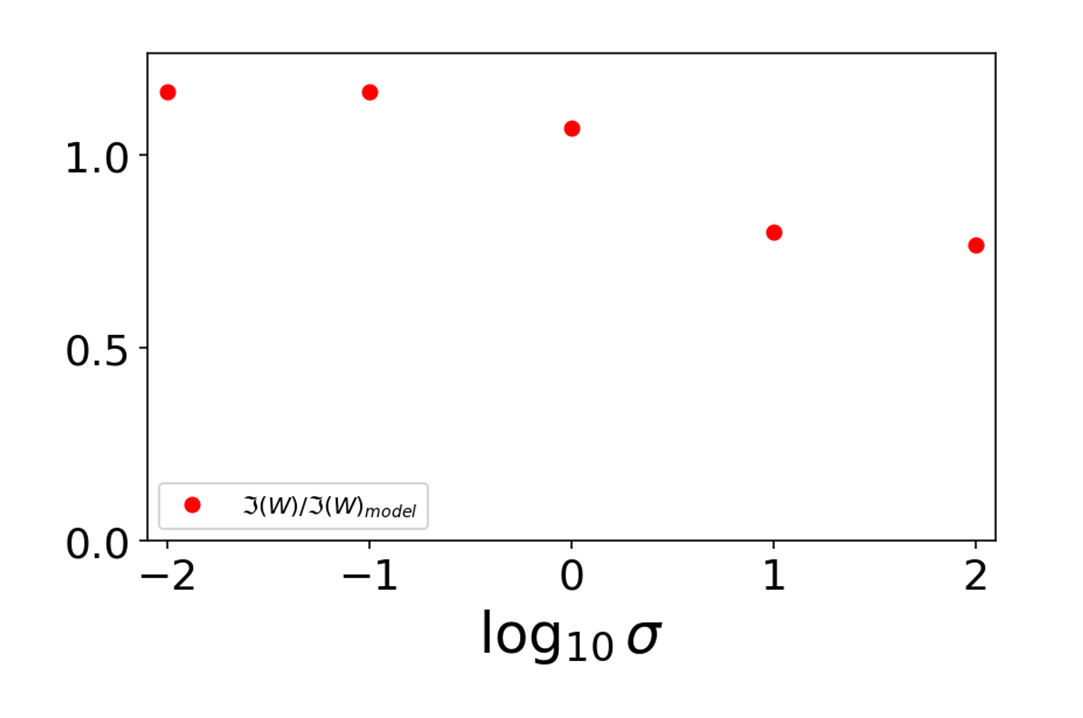}}
	\caption{\small{ Ratio of function $W(\sigma,\hat{\nu})$ numerically computed to model Eq.(\ref{eq:value_Phi_sigma_all_sigma_main}) versus $\sigma$ for large collision frequency  $\hat{\nu} = 50$. }}
	 \label{Fig:test_ratio}
\end{figure}

It is also interesting to explore the variation with collisionality at fixed $\sigma$. Fig.\ref{Fig:test_ratio} shows the ratio of the numerical value of $W(\sigma,\hat{\nu})$ to the model Eq.(\ref{eq:value_Phi_sigma_all_sigma_main}) in the high collisional limit $\hat{\nu} = 50$. We find reasonable agreement for a wide range of value of $\sigma$, for reasons explained above. The ratio ranges between $1.17$ and $0.76$ with increasing $\sigma$. Thus, we can conclude that the analytical model does reproduce the correct order of magnitude and trend, but should not be used for accurate calculations, where the numerical value should be used, or a drift-kinetic code if high accuracy is needed. \\

\begin{figure}[H]
\centering
	\centerline{\includegraphics[width=10cm]
	{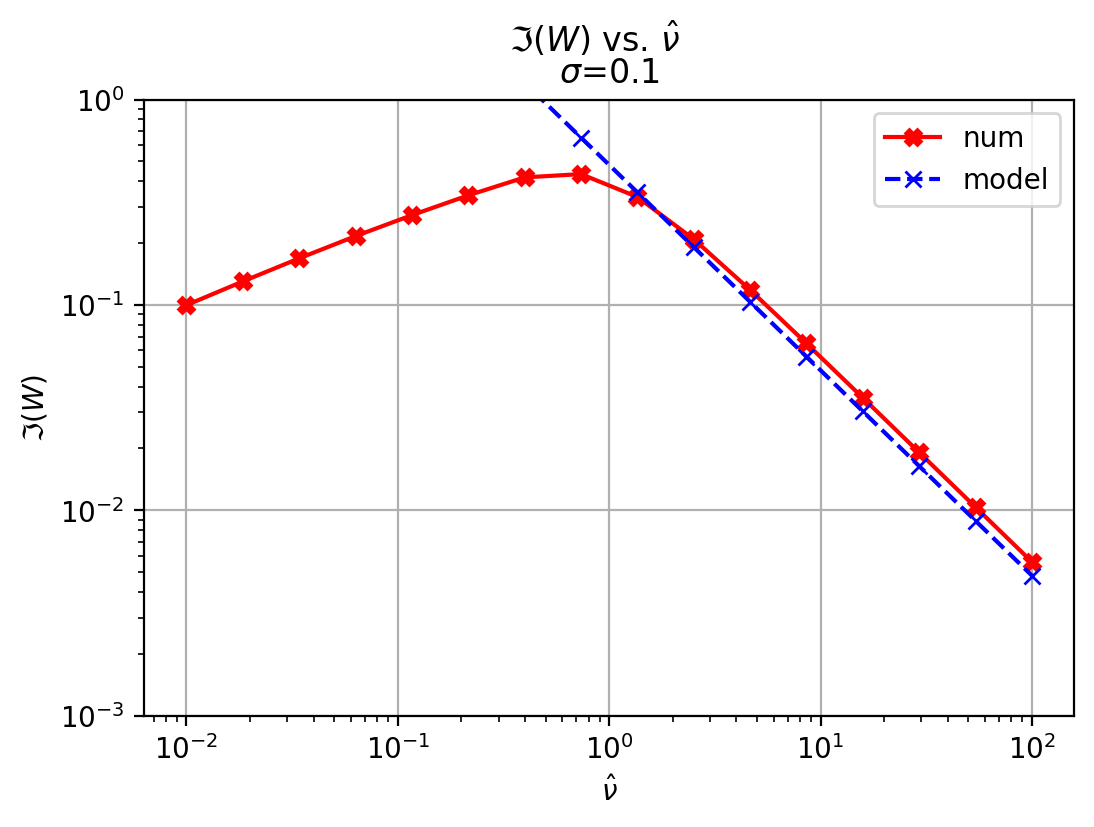}}
	\caption{\small{ Computed function $W(\sigma,\hat{\nu})$ compared to analytic model Eq.(\ref{eq:value_Phi_sigma_all_sigma_main}) versus $\hat{\nu}$ for low value of $\sigma = 0.01$. }}
	 \label{Fig:test_numerical_phi_vs_model_low_sigma}
\end{figure}

Fig.\ref{Fig:test_numerical_phi_vs_model_low_sigma} shows a comparison of the numerical function $W(\sigma,\hat{\nu})$ and the analytical model Eq.(\ref{eq:value_Phi_sigma_all_sigma_main}) versus collisionality at low $\sigma = 0.01$. 
While the agreement is quite satisfactory at large collision frequency, it fails markedly in the weak collisional regime. 
This is due to the boundary layer that appears near $K=1$ that is not well represented in the model, even with a simple Dirichlet boundary condition at $K=1$. 

\begin{figure}[H]
\centering
	\centerline{\includegraphics[width=14cm]
	{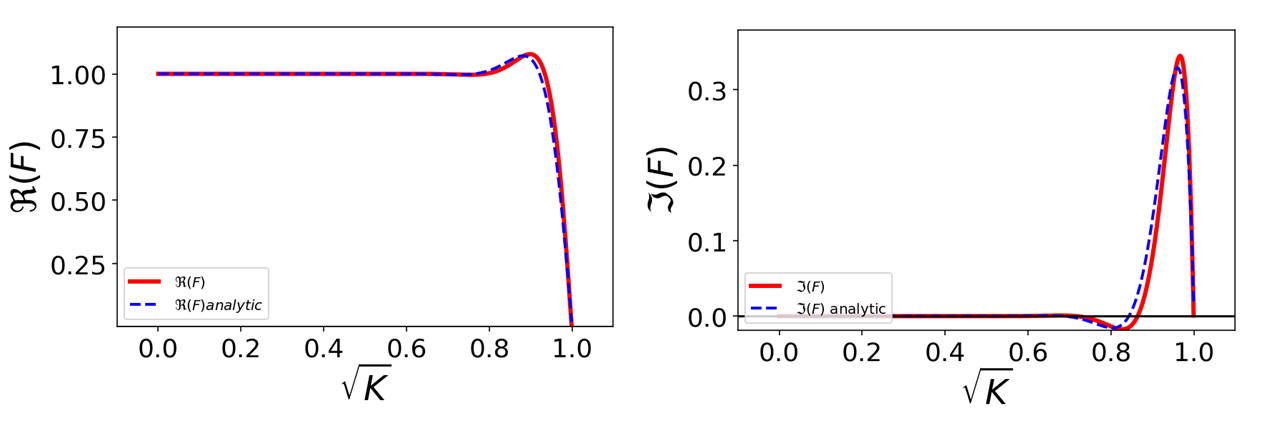}}
	\caption{\small{ Computed function $\hat{F}(K)$ compared to model Eq.(\ref{eq:solution_bounce_average_FP_low_sigma}) for $\hat{\nu}=0.005$ and $\sigma = 0.01$. }}
	 \label{Fig:test_numerical_Fhat_vs_model_low_sigma_lownuhat}
\end{figure}

Fig.\ref{Fig:test_numerical_Fhat_vs_model_low_sigma_lownuhat} compares the computed solution $\hat{F}(K)$ of the Fokker-Planck equation with the analytical expression Eq.(\ref{eq:solution_bounce_average_FP_low_sigma})  
for $\hat{\nu}=0.005$ and $\sigma = 0.01$, which appears to be close to the numerical calculation. 
As before, the explanation lies in the boundary layer that develops near the trapped domain border $K=1$. 
This is the reason why the estimate Eq.(\ref{eq:value_Phi_sigma_all_sigma_main}) does not work well at low values of $\hat{\nu}$. 

\begin{figure}[H]
\centering
	\centerline{\includegraphics[width=10cm]
	{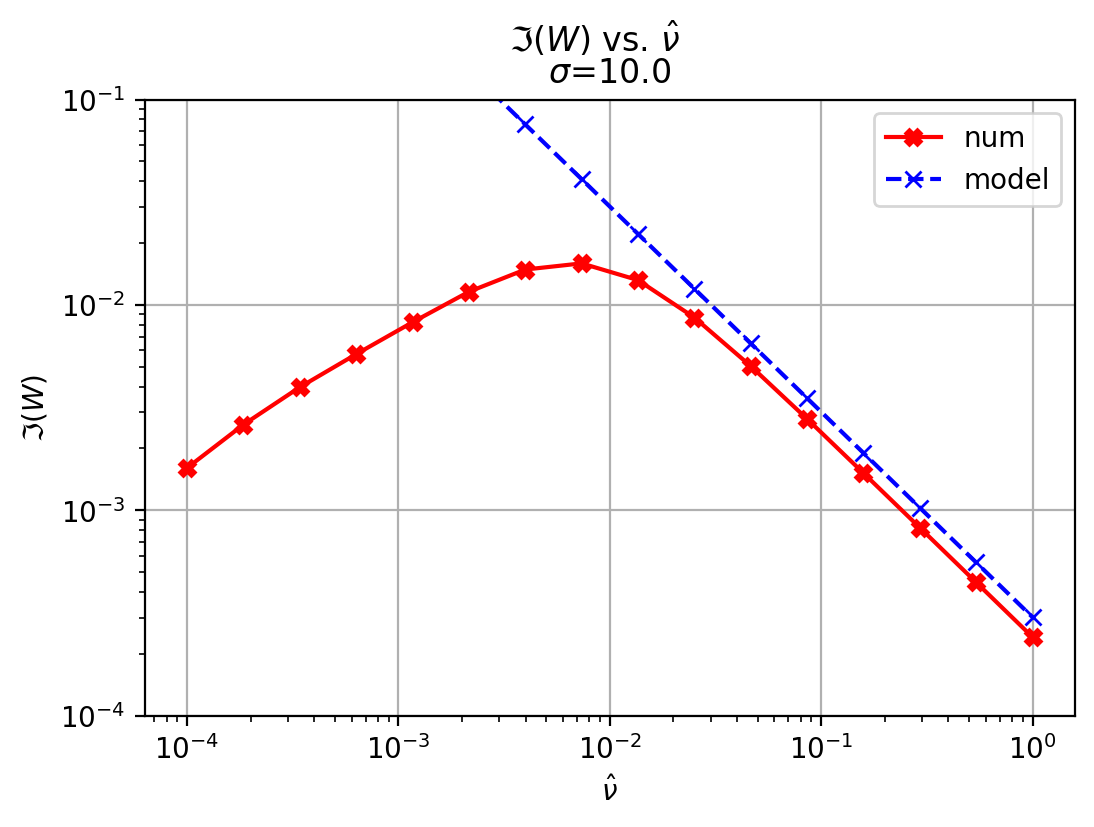}}
	\caption{\small{ Computed function $W(\sigma,\hat{\nu})$ compared to model Eq.(\ref{eq:value_Phi_sigma_all_sigma_main}) versus $\hat{\nu}$ for a high value of $\sigma = 10$. }}
	 \label{Fig:test_numerical_phi_vs_model_high_sigma}
\end{figure}

Fig. \ref{Fig:test_numerical_phi_vs_model_high_sigma} shows a comparison of the numerical function $W(\sigma,\hat{\nu})$ and the model Eq.(\ref{eq:value_Phi_sigma_all_sigma_main}) versus collisionality at high $\sigma = 10$. 
The results are similar to those at lower values of $\sigma$, where agreement is good at high collisionality, but not at low collisionality, for the same reason as before: the development of a boundary layer near the passing/trapped border $K=1$. 
However, it appears that the transition occurs at lower collision frequencies (note the difference in the scale of $\hat\nu$ between Figs. \ref{Fig:test_numerical_phi_vs_model_low_sigma} and \ref{Fig:test_numerical_phi_vs_model_high_sigma}), due to the factor of $\sigma^2$ multiplying the effective detrapping collision frequency, which is $\sigma^2 = 100$ in this case. 

To summarise, analytical solutions work well at high collisionality $\hat{\nu} \ge 1$, but fail markedly at low values of $\hat{\nu} \ll 1$. 
In the small $\hat{\nu}$ limit, the Fokker-Planck equation should be solved for accurate results. 
Nevertheless, this operation is quite fast and accurate as long as the number of harmonics remains reasonable. 
Also, in the large $\sigma$ limit and with small values of the bounce frequency $\omegab / \omega_d $, the bounce average approximation does not work well. 
The Fokker-Planck equation should therefore be solved for each harmonic $\nb$ and the entropy production rate computed accordingly. 
Nevertheless, for large enough values of collisionality $\hat{\nu} > 1$ and bounce frequencies $\omegab / \omega_d > 10$, the bounce average approximation does work well, so analytical results can be relied on with those conditions. 
This model is thus quite fast.

\section{Computation of neoclassical fluxes in tokamaks}
\label{sec:Computation of neoclassical fluxes}

We now apply the derived model to tokamaks in the presence of two magnetic perturbations.

\subsection{Guiding-centre model}
\label{subsec:Guiding-centre model}
The starting point is a tokamak in the large aspect ratio limit $\epsilon(r) = r/R_0 \ll 1$, where $r$ is the minor radius, and $R_0$ the major radius, to which a helical perturbation is added. 
A minimum model for the Hamiltonian that rules the motion of guiding-centers then reads
\begin{equation}
	H = \Heq - \mu B_0 \epsilon(r) \cos(\theta) - 
	       \mu B_0 \delta(r,\theta) \cos( M\theta + N\varphi )
 \label{eq:Hamiltonian_helical}
\end{equation} 
where $\theta$ and $\varphi$ are the poloidal and toroidal angles, $\mu$ is the magnetic moment (a constant of motion), $B_0$ is the magnetic field at major radius $R_0$, and $\delta \ll 1$ is the amplitude of the helical perturbation with poloidal and toroidal wavenumbers $(M,N)$. 
Note that $\delta$ should depend weakly on the poloidal angle $\theta$: if not, the decomposition of $\delta(r,\theta)$ would yield a sum of radially varying functions for each poloidal harmonic.
In other words, the term  $\delta(r,\theta) \cos( M\theta + N\varphi )$ would actually account for multiple perturbations, and the problem would be out of the scope of the model derived throughout the paper.
In practice, this condition is verified in tokamaks.
The unperturbed Hamiltonian is 
\begin{equation}
	\Heq = \frac{1}{2}m_a \vpar^2 + \mu \Beq(r) + e_a \Phieq (r) 
 \label{eq:Hamiltonian_unperturbed}
\end{equation} 
where $\vpar$ is the parallel velocity, $m_a$ and $e_a$ are the particle mass and charge, $\Phieq$ is the unperturbed electric potential and $\Beq$ the equilibrium magnetic field amplitude. 
The unperturbed Hamiltonian thus corresponds to a periodic cylindrical screw pinch where the equations of motion are simply
\begin{equation} 
   \frac{dr}{dt} = 0 \text{~~~~;~~~~}
	\frac{d\theta}{dt} = \frac{\vpar}{q(r) R_0} + \omega_E \text{~~~~;~~~~}
	\frac{d\varphi}{dt} = \frac{\vpar}{R_0}
    \label{eq:motion_GC}
\end{equation} 
where $q(r)$ is the safety factor, and $\omega_E = \frac{1}{r} \frac{d\Phieq}{B_0 dr}$ is the $E\times B$ poloidal drift frequency. 
The ``toroidal'' angle is defined as $\varphi = 2\pi z/L $, where $z$ is the axial coordinate and $L$ is the length of the cylinder.  
In the large aspect ratio limit, the poloidal field $B_\theta(r) = B_0 \epsilon(r)/q(r)$ is smaller than the axial field $B_0$, which implies that $\Beq = B_0 (1 + \Ocal(\epsilon^2))$. 
The amplitude of the $\cos(\theta)$ perturbation in the Hamiltonian Eq.(\ref{eq:Hamiltonian_helical}) is first order in $\epsilon$, such that $\Beq = B_0$ at the same level of approximation. 
This allows us to ignore the magnetic drift due to the unperturbed poloidal field at this level of approximation. 
If desired, a dependence on $r$ can easily be kept in the unperturbed magnetic field $\Beq(r)$, wherein the magnetic drift then adds up to the $E\times B$ drift frequency.

\subsection{Angle/action variables in a tokamak}
\label{subsec:angle_action_tokamak}

The first angle $\alpha_1$ is the gyroangle, which is conjugate to the action
\begin{equation}
    J_1= \frac{m_a \mu}{e_a}
    \label{eq:action_J1}
\end{equation}
and the corresponding Hamiltonian frequency is thus the cyclotron angular frequency 
\begin{equation}
    \Omega_1 = \frac{e_a \Beq}{m_a}
    \label{eq:Omega_1}
\end{equation}
The second and third angles coincide with the poloidal and toroidal angles $\alpha_2 = \theta$ and $\alpha_3 = \varphi$, and the respective Hamiltonian angular frequencies are 
\begin{equation}
    \Omega_2 =  \frac{\vpar}{q(r)R_0} + \omega_E 
    \label{eq:Omega_2}
\end{equation}
and 
\begin{equation}
    \Omega_3 = \frac{\vpar}{R_0}
    \label{eq:Omega_3}
\end{equation}
The parallel velocity $\vpar$ is a constant of motion, as a consequence of the time invariance of the energy in Eq.(\ref{eq:Hamiltonian_unperturbed}) and the particle radial coordinate $r$ that labels magnetic surfaces. 
In the large aspect ratio limit, the second and third actions are
\begin{eqnarray} 
	  J_2 & = & e_a \chi(r)
	\label{eq:action_J2}	\\
		J_3 & = & - e_a \psi(r) + m_a \vpar R_0
 \label{eq:action_J3}		
\end{eqnarray}
where $\chi(r)$ and $\psi(r)$ are fluxes of axial and poloidal magnetic fields respectively, normalised to $2\pi$. 
The toroidal flux is just $\chi(r) = B_0 r^2/2$. The flux of the poloidal magnetic field is such that $d\psi(r) =  B_0  r dr/q(r)$. 
It can thus be expressed as a function of the toroidal flux $\chi$, such that $d\psi/d\chi = 1/q(\chi)$. 
Lastly, the third action $J_3$ coincides with the axial canonical momentum $P_\varphi $. 
The parallel velocity is therefore a function of the second and third actions 
\begin{equation}
   	\vpar(\Jbold) = \frac{1}{m_a R_0} 
   	   \left[ J_3 + \Psi(J_2) \right]
 \label{eq:vpar_vs_J_3andJ2}
\end{equation}
where $d\Psi/dJ_2 =  1/q(J_2)$ or, in other words, $\Psi(J_2) =  e_a \psi(J_2/e_a) $. 
The unperturbed Hamiltonian of guiding-centres in large aspect-ratio tokamaks, expressed in angle/acion coordinates, then reads
\begin{equation}
  \nonumber
   	\Heq(\Jbold) = \frac{1}{2 m_a R_0^2} 
   	   \left[ J_3 + \Psi(J_2) \right]^2 + J_1 \Omega_1 + e_a \Phieq(J_2)  	
 \label{eq:vpar_vs_J}
\end{equation}
where it can be verified that $\Omega_i = \partial_{J_i}\Heq$.

For cases relevant to tokamaks and stellarators, the resonance condition $\Omega(J_2 , J_3) = \Omega(J_2 , G(J_2)) = 0$ reads $\kpar \vpar + n_2 \omega_E = 0$, where $\kpar = (n_2/q(J_2) + n_3)/R_0$.
When the electric potential vanishes i.e. $\Phieq=0$, the resonance condition reduces to $\vpar(J_2 , J_3) = 0$, so that $G(J_2 , J_3) = - \Psi(J_2 , J_3)$.

\subsection{Equilibrium distribution function of the unperturbed problem}
\label{Equilibrium distribution function of the unperturbed problem}

As mentioned in Section \ref{Fokker-Planck equation}, solutions of the collisionless and unperturbed Fokker-Planck equation are arbitrary functions of the actions $\Jbold$.
In the context of tokamaks, local Maxwellians $F_M$ are also solutions of $C[\Feq]=0$, chosen constant on magnetic surfaces, such that 
\begin{equation}
	  F_M(\psi,\vpar,\mu,t) =  
		\frac{n_{\rm eq}(\psi,t)}{\left[2 \pi m_a  \Teq(\psi,t) \right]^{3/2}} 
				\left( 1 + \frac{m_a \vpar \Vpareq(\psi,t)}{\Teq(\psi,t)} \right)
		\exp \left(- \frac{\frac{1}{2} m_a \vpar^2  + \mu B_0}{ \Teq(\psi,t) } 
		        \right)  
	\label{eq:local_Maxwellian}
\end{equation}
where $n_{\rm eq}(\psi,t)$, $\Teq(\psi,t)$, $\Vpareq(\psi,t)$ are the unperturbed density, temperature and parallel velocity, respectively. 
Hence, a reasonable choice that combines both properties is
\begin{equation}
	  \Feq(\Jbold,t) = 
		\frac{ \Ncaleq(\Jbold,t) }{\left[ 2 \pi m_a  \Teq(\Jbold,t) \right]^{3/2}} 
		\exp\left( - \frac{ \Heq( \Jbold) }{\Teq(\Jbold,t)} \right)
		\left( 1 + \frac{ m_a \vpar( \Jbold ) \Wpareq(\Jbold,t) }{ \Teq(\Jbold,t) } \right)
	\label{eq:local_Maxwellian_canonical}
\end{equation}  
where
\begin{equation}
  \nonumber
	  \Ncaleq(\Jbold,t)= n_{\rm eq}(\Jbold,t) 
	    \exp\left( \frac{e_a \Phieq(\Jbold) }{ \Teq(\Jbold,t)} \right)
	\label{eq:def_Ncaleq}
\end{equation}
and $\Wpareq$ is a field that slightly differs from the mean parallel velocity as the third action depends on the minor radius and particle parallel velocity.
This difference is, in fact, the diamagnetic velocity. The velocity $\vpar( \Jbold )$ is a constant of motion that is a proxy of the actual particle parallel velocity $\vpar$ - the latter is usually not a constant of motion except for fast passing particles. The dependences of $\Ncaleq, \Teq, \Wpareq$ on the actions $(J_2,J_3)$ must be chosen such that they depend only on the minor radius. 
While this is not possible in the general case, this constraint is only actually required on resonant surfaces. 

It is straightforward to link the expression Eq.\ref{eq:local_Maxwellian_canonical} with the generic form chosen to derive the theory, i.e. $\Feq(\Jbold,t) = F_0 \exp\left( - \frac{\Heq(\Jbold) - \Ueq(\Jbold,t)}{T_\ast} \right)$, by defining the thermodynamic potential $\Ueq$ as
\begin{equation}
   	\frac{\Ueq(\Jbold,t)}{T_\ast} = \Xi(\Jbold,t) + 
   	      \frac{ m_a \vpar(\Jbold) \Wpareq(\Jbold,t)}{\Teq(\Jbold,t)}
	\label{eq:mean_thermodynamic_potential}
\end{equation}
with
\begin{equation}
   	\Xi(\Jbold,t) = \ln \Ncaleq(\Jbold,t)  
   	      - \frac{3}{2} \ln \Teq(\Jbold,t)  -
   	      \Heq(\Jbold,t)\left( \frac{1}{\Teq(\Jbold,t)} - \frac{1}{T_\ast} \right)
	\label{eq:def_Xi}
\end{equation}
such that
\begin{equation}
 \nonumber
   	 \frac{\partial \Xi }{\partial \psi} = 
      \frac{\partial \ln \Ncaleq}{\partial \psi}  +
   	      \left( \frac{\mu B_0}{\Teq} - \frac{3}{2} \right) 
   	        \frac{\partial \ln \Teq}{\partial \psi} 
	\label{eq:def_Xi_partial_derivative_psi_explicit_helical}
\end{equation}
and
\begin{equation}
 \nonumber
   \Wpareq  =
   	    \Vpareq + \frac{M}{M + Nq} 
   	    \frac{R_0\Teq}{e_a} 
   	       \frac{\partial \Xi}{\partial \psi}
	\label{eq:generalised_force_balance_explicit_helical}
\end{equation}

\subsection{Total entropy production rate}
\label{Total entropy production rate}
The simple guiding-centre Hamiltonian system presented in Section \ref{subsec:Guiding-centre model} describes the particle dynamics in a tokamak with an externally introduced helical perturbation or a very schematic stellarator with a single helicity.
As a reminder, in addition to this ``helical'' perturbation\footnote{which also includes the ripple perturbation corresponding to the case where the poloidal wavenumber of the perturbation is zero.} $\mu B_0 \delta \cos(M\theta + N\varphi)$, the curvature of the tokamak is responsible for another magnetic perturbation $\mu B_0 \epsilon \cos(\theta)$ and referred to as ``toroidal'', leading to the Hamiltonian Eq.(\ref{eq:Hamiltonian_helical}).
Each of them can be the primary perturbation in turn, while the other becomes the secondary perturbation. 

Conventional neoclassical theory predicts 4 collisional transport contributions \cite{Mynick_2006, Galeev_1979, Kovrizhnykh_1999}.

\begin{itemize}
    \item Two contributions are due to only one perturbation, called ``primary'', and correspond to situations which exhibit a symmetry.
    \begin{enumerate}
        \item A helically symmetric contribution with $\epsilon=0$ and finite $\delta$, which corresponds to a helical perturbation in a screw pinch and where transport is mainly produced by helically trapped particles. 
        It carries two asymptotic regimes in collisionality  $\nustar$ for transport coefficients, generically dubbed $C_\Gamma$:
        \begin{itemize}
            \item At low $\nustar$: the ``helically trapped'' regime where $C_\Gamma \propto \nu$ (with $\nu$ the collision frequency).
            \item At high $\nustar$: the ``helically symetric-plateau'' regime where $C_\Gamma$ is independent of $\nu$.
        \end{itemize}
        \label{enum:hel_I}
        \item A toroidally symmetric (``axisymmetric'') contribution with $\delta=0$ and finite $\epsilon$ instead, which actually corresponds to the standard neoclassical transport in a tokamak \cite{Hinton_Hazeltine_1976,Hirshman_Sigmar_1981}. 
        It also carries two similar asymptotic regimes:
        \begin{itemize}
            \item At low $\nustar$: the ``banana'' regime where $C_\Gamma \propto \nu$.
            \item At high $\nustar$: the ``plateau'' regime where $C_\Gamma$ is independent of $\nu$.
        \end{itemize}
        \label{enum:tor_I}
    \end{enumerate}
In both cases, collisions are instrumental to producing transport, since particles remain well confined in primary perturbations in the absence of collisions. 

\item The two remaining contributions are due to the effect of a ``secondary" perturbation on top of the equilibrium obtained with the counterpart primary perturbation. 
    \begin{enumerate}
    \setcounter{enumi}{2}
        \item A contribution arises from the effect of the toroidal perturbation on helically trapped particles.
        This contribution comes from the action of the vertical magnetic drift due to the toroidal perturbation on the {\em helically} trapped particles.
        In the adiabatic regime $\hat{\omega}_b \gg 1$, three regimes appear:
        \begin{itemize}
            \item At very low $\nustar$: the ``$\nu$ superbanana'' regime comes from the effect of the $E \times B$ drift on the helically trapped particles which compensate the effect of the vertical magnetic drift.
            We recall that this regime is not captured by our model for reasons detailed in Section \ref{sec:Numerical_computation_of_the_weight_function}.
            \item At low $\nustar$: the ``$\sqrt{\nu}$ superbanana'' regime, similar to the previous one, that scales as $C_\Gamma \propto \sqrt{\nu}$.
            \item At high $\nustar$: the ``$1/\nu$ superbanana'' regime which corresponds to helically trapped particles with uncompensated magnetic drift where $C_\Gamma$ is inversely proportional to $\nu$.
        \end{itemize}
        These scaling hold as long as no "superbanana resonance" $\omega_d = 0$ occurs. This is typically the case when the $E\times B$ drift contribution to the superbanana precession frequency prevails over its magnetic drift counterpart so that $\omega_d $ is nearly constant. This paper focuses on this limit. When a superbanana resonance occurs, the scalings change to linear $\nu$ and plateau \cite{Kovrizhnykh_1999}. Details can be found in \cite{Shaing_2010,Sun_2019}, as well numerical evidence from as gyrokinetic and drift kinetic simulations \cite{Matsuoka_2017}. 
        \label{enum:tor_II}
        \item The other contribution comes from the effect of the helical perturbation on {\em toroidally} trapped particles (``bananas''). 
        It is due to the magnetic drift being modified due to the helical perturbation, which modifies the particle motion, most notably at the banana tip.
        In the adiabatic regime $\hat{\omega}_b \gg 1$, three regimes appear:
        \begin{itemize}
            \item At very low $\nustar$: the ``$\sqrt{\nu}$ banana-drift'' regime, analogous to the ``$\nu$ superbanana'' regime (thus, not present in our model).
            \item At low $\nustar$: the ``$1/\nu$ banana-drift'' regime that occurs when $\nu$ is lower than the bounce frequency. 
            In that case, the transport is maximised as, on average, the radial displacement of the banana tip is not compensated back because a collision occurs before.
            In that case, the scaling goes as $C_\Gamma \propto \nu^{-1}$.
            \item At high $\nustar$, the ``ripple-plateau\footnote{The name comes from the pioneer literature on rippled-tokamak, that was then extended to helical perturbation. Yet, the terminology persists.}'' where multiple collisions occur during a bounce period, limiting the effect at the banana tip. 
            In that case, the transport is independent of $\nu$.
        \end{itemize}
        \label{enum:hel_II}
    \end{enumerate}
    In both cases, transport is closely related to uncompensated magnetic drift along trapped trajectories. 
    Therefore, in the absence of collisions or compensating drifts, transport would be disastrous, resulting in very large diffusion, or rather, ballistic motion. 
    Hence, collisions actually help by limiting transport, since they reduce losses by re-confining particles.
    Note that in the non-adiabatic regime, i.e. $\hat{\omega}_b < 1$, the aforementioned scalings do not necessarily hold, and one must use the exact derived model instead of the analytical estimation.
\end{itemize}

Each of these contributions, and their associated transport regimes, are encapsulated in the total resonant entropy production rate which, in tokamaks, is the sum 
\begin{equation}
    \dot{S}_\text{res} = 
    \dot{S}_\text{hel,I} + \dot{S}_\text{tor,II}
    + 
    \dot{S}_\text{tor,I} + \dot{S}_\text{hel,II} 
  \label{eq_resonant_rate_total}
\end{equation}
where 
\begin{itemize}
    \item $\dot{S}_\text{hel,I}$ and $\dot{S}_\text{tor,I}$ are the entropy production rates that would result from the presence of only one perturbation, i.e. helical or toroidal\footnote{$\dot{S}_\text{tor,I}$ is the total neoclassical contribution in an axisymmetric tokamak.}, respectively.
    They correspond to bullets \ref{enum:hel_I} and \ref{enum:tor_I} above.
    \item $\dot{S}_\text{tor,II}$ and $\dot{S}_\text{hel,II}$ are the entropy production rates caused by the secondary toroidal and helical perturbations, respectively, when the corresponding primary perturbations are helical and toroidal, respectively.
    They correspond to bullets \ref{enum:tor_II} and \ref{enum:hel_II} above.
\end{itemize}
Each of these contributions is explicitly given in the following Section \ref{subsec:Explicit_expressions_of_geometrical_parameters_and_frequencies} in Eqs.(\ref{eq_resonant_rate_helical},\ref{eq_resonant_rate_toroidal_secondary},\ref{eq_resonant_rate_toroidal_primary},\ref{eq_resonant_rate_ripple_secondary}). 

\subsection{Velocity-dependent collision frequency}
\label{subsec:Velocity_dependant_collision_frequency}
Naturally, collisions play a major role in the neoclassical physics in tokamaks.
In the previous Section, we defined multiple frequencies, namely the detrapping frequency $\nu_d$ Eq.(\ref{eq:nud}), the effective detrapping frequency $\nueff$ Eq.(\ref{eq:effective_collision_frequency}) and the normalized collision frequency $\hat{\nu}$ Eq.(\ref{eq:nuhat}).
Each of these play an important role in the calculation of the total entropy production rate.
They are all related to the key dimensionless parameter $\nu_\Omega = D_\Omega / \omegab^3$ defined in Eq.(\ref{eq:nu_Omega}) that we call ``collisionality'' such that

\begin{equation}
    \nu_d = \frac{\omegab}{2}\nu_\Omega
    \text{~~;~~}
    \nueff = \frac{\omegab}{2}\nu_\Omega \max{\left( 1,\sigma^2 \right)}
    \text{~~;~~}
    \hat{\nu} = \frac{\hat{\omega}_b}{2}\nu_\Omega
\end{equation}

In the context of tokamaks, the scattering in frequency is given by
\begin{equation}
	\nonumber
    D_\Omega =  
   	   \frac{ n_h^2 }{q^2 R_0^2}
   	     \frac{1}{2} 
          \frac{\left \langle \Delta \vpar^2 \right \rangle}
               {\Delta t}  
  \label{eq_scattering_Omega}
\end{equation}
where 
\begin{equation}
    n_h = n_2 + n_3 q
  \label{eq:n_h}
\end{equation}
and
\begin{equation}
	\nonumber
    \frac{1}{2}
        \frac{\left \langle \Delta \vpar^2 \right \rangle}
             {\Delta t}  = 
        \nu_c(v) v^2 v_T^2
  \label{eq_parallel_velocity_scattering}
\end{equation}
where $v_T$ is the thermal velocity, the normalised velocity is 
\begin{equation}
    v = \sqrt{\frac{\mu B_0}{T_{\rm eq}}}
\end{equation}
and the bounce frequency reads
\begin{equation}
    \omegab = \frac{n_h}{q R_0} \sqrt{\dfrac{h}{m_a}}
\end{equation}
such that
\begin{equation}
   \nu_\Omega = \frac{qR_0}{n_h v_T} 
      \frac{\nu_c(v) v^2 }{\left( \frac{h}{\Teq} \right)^{3/2}}
 \label{eq:nu_Omega_tokamak}
\end{equation}

where the collision frequency $\nu_c(v) = \nu \nubar_v(v)$ is defined with the thermal collision frequency $\nu$ expressed as
\begin{equation}
  \nu = \frac{4\sqrt{\pi}}{3} \frac{ e_a^4 }{\left( 4 \pi \epsilon_0 \right)^2} 
	\frac{n_{\rm eq}}{m_a^2v_{T}^3}\ln\Lambda 
  \label{eq_def_nu_thermal}
\end{equation}
and the velocity dependent normalized collision frequency $\nubar_v(v)$ given by \cite{Hinton_Hazeltine_1976,Hirshman_Sigmar_1981}
\begin{equation}
	\nonumber
  \nubar_v(v) = \frac{3}{4}\sqrt{2\pi} \frac{1}{v^3}\left(\text{erf}(v)-G(v)\right)
  \label{eq_def_nubar}
\end{equation}
with
\begin{equation}
	\nonumber
  \text{erf}(v) = \frac{2}{\sqrt{\pi}} \int_{0}^{v} dx \exp\left( -x^2 \right)
  \label{eq_def_Phi}
\end{equation}
and
\begin{equation}
	\nonumber
  G(v)=\frac{1}{2v^2} \left( \text{erf}(v) - v \frac{d\ \text{erf}(v)}{dv} \right)
  \label{eq_def_G_v}
\end{equation}

With this definition, we see that the commonly used collisionality $\nustar$ --- associated with the toroidal perturbation -- is specifically the value of $\nu_\Omega$ for conventional neoclassical theory \cite{Hinton_Hazeltine_1976,Hirshman_Sigmar_1981}, with $n_2=1$, $n_3=0$, and $v=1$, which gives
\begin{equation}
  \nustar = \frac{\nu q R_0}{v_T \epsilon^{3/2}}
  \label{eq_def_nustar_tor}
\end{equation}

\subsection{Explicit expressions of geometrical parameters and frequencies}
\label{subsec:Explicit_expressions_of_geometrical_parameters_and_frequencies}

In Sections \ref{sec:Primary_entropy_production_rate} and \ref{sec:Secondary_entropy_production_rate}, where the model is developed, multiple geometrical parameters and frequencies have been defined in the general case of Hamiltonian systems with two perturbations.
Their explicit expressions are summarized in Table \ref{tab:overview_main_parameters_tokamak}.

Here are some details to understand how the expression for the drift frequency $\omega_d$ has been obtained.
It is given by $\omega_d = n_d \Omegad$, with $n_d = n'_2 n_3 - n'_3 n_2$ and $\Omegad$ is defined in Eq.(\ref{eq:Omega_d}). 
The dominant contribution to $\omega_d$ comes from $n_d \partial_{\Jd}\Heq$, while another comes from the dependence of the parallel adiabatic invariant on $\Jd$.
Both are readily computed using the expression of the drift action $\Jd$ versus $(J_2,J_3)$ (see Eq.(\ref{eq:Jd_3rd_drift_action})). 
Because $\vpar=0$ at the resonant point, transit frequencies do not contribute to $\omega_d$, so the electric drift can no longer be ignored for this calculation. 
The first piece of $\omega_d$ comes from the electric drift, and the second part comes from the magnetic drift, which results from the dependence of the perturbed Hamiltonian on the toroidal flux $\chi$. 
Calculating the drift frequency is quite intricate; however, an estimate can be obtained in the case of deeply trapped particles, where we take $\Heq(\Jbold_R) = e_a \Phieq(J_2/e_a) - h(J_2/e_a)$ to be the unperturbed Hamiltonian, such that
\begin{equation}
    \omega_d = 
       \frac{ n_d q }{ n_h } \frac{\partial}{\partial \chi} \left( \phieq(\chi) -  \frac{1}{e_a} h(\chi) \right)
  \label{eq:omega_d_tokamak}
\end{equation}

In practice, the drift frequency is calculated in the limit of deeply trapped particles (the precession frequency of barely trapped particles can make the calculation unwieldy).

When the primary perturbation is helical, it reads
\begin{equation}
    \omega_d = 
      \frac{ N q } { M + Nq } \frac{1}{ r}
      \left( \frac{1}{B_0 } \frac{\partial \phieq}{\partial r}   +  
            \frac{\mu }{e_a}  \frac{\partial \delta }{\partial r} \right)
  \label{eq:omega_d_case_helical}
\end{equation}

When the primary perturbation is toroidal, it reads

\begin{equation}
    \omega_d = -
      \frac{ N q } { r } 
      \left( \frac{1}{B_0 } \frac{\partial \phieq }{\partial r}   -  
            \frac{\mu }{e_a R_0}  \right)
  \label{eq:omega_d_case_toroidal}
\end{equation}

In the following, we will consider the case where the $E\times B$ drift is dominant compared to the precession due to the magnetic drift.
This approximation excludes super-banana resonances \cite{Shaing_2010}. 

\setlength{\tabcolsep}{0.3em}
{\renewcommand{\arraystretch}{3.}
\begin{table}[!ht]
\resizebox{\textwidth}{!}{%
\centering
\begin{tabular}{|c|c|c|c|}
\hline
                                                                                                                                        & General                                                                                   & \shortstack{\\ Primary helical \\ $h=\mu B_0 \delta$ \\ $(n_2, n_3) = (M, N)$}                                                                       & \shortstack{\\ Primary toroidal \\ $h=\mu B_0 \epsilon$ \\ $(n_2, n_3) = (1, 0)$}                               \\ \hline
Primary wavenumber $n_h$                                                                                                                & $n_2 + n_3 q$                                                                             & $M_h = M + Nq$                                                                                                                                       & $1$                                                                                                             \\ \hline
Secondary wavenumber $n_h'$                                                                                                             & $n_2' + n_3' q$                                                                           & $1$                                                                                                                                                  & $M_h$                                                                                                           \\ \hline
\shortstack{Drift wavenumber $n_d$ \\ Eq.(\ref{eq:n_d})}                                                                                & $n'_2 n_3 - n_2 n'_3$                                                                     & $N$                                                                                                                                                  & $-N$                                                                                                            \\ \hline
\shortstack{Primary resonant freq. $\Omega$ \\ Eq.(\ref{eq:hamilton_eq_angle}) + see Section \ref{subsec:angle_action_tokamak}}         & $n_h \dfrac{\vpar}{q R_0}$                                                                & $M_h \dfrac{v_\parallel}{q R_0}$                                                                                                                     & $\dfrac{v_\parallel}{q R_0}$                                                                                    \\ \hline
\shortstack{Secondary resonant freq. $\Omega'$ \\ Eq.(\ref{eq:hamilton_eq_angle}) + see Section \ref{subsec:angle_action_tokamak}}      & $n'_h \dfrac{\vpar}{q R_0}$                                                               & $\dfrac{v_\parallel}{q R_0}$                                                                                                                         & $M_h \dfrac{v_\parallel}{q R_0}$                                                                                \\ \hline
\shortstack{Primary curvature $C$ \\ Eq.(\ref{eq:curvature})}                                                                           & $\dfrac{n_h^2 }{m_a q^2 R_0^2}$                                                           & $\dfrac{M_h^2}{m_a q^2 R_0^2}$                                                                                                                       & $\dfrac{1}{m_a q^2 R_0^2}$                                                                                      \\ \hline
\shortstack{Secondary curvature $C^\star$ \\ Eq.(\ref{eq:curvature_hybrid})}                                                            & $\dfrac{n_h n_h'}{m_a q^2 R_0^2}$                                                         & $\dfrac{M_h}{m_a q^2 R_0^2}$                                                                                                                         & $\dfrac{M_h}{m_a q^2 R_0^2}$                                                                                    \\ \hline
\shortstack{Curvature ratio $\sigma$ \\ Eqs.(\ref{eq:sigma_vs_C_Cstar}, \ref{eq:definition_sigma_ratio})}                               & $\dfrac{n_h'}{n_h} = \dfrac{C^\star}{C}$                                                  & $\dfrac{1}{M_h}$                                                                                                                                     & $M_h$                                                                                                           \\ \hline
\shortstack{Bounce frequency $\omegab$ \\ Eq.(\ref{eq:bounce_frequency})}                                                              & $\sqrt{Ch}$                                                                               & $\dfrac{M_h}{q R_0} \sqrt{\dfrac{\mu B_0 \delta}{m_a}}$                                                                                              & $\dfrac{1}{q R_0} \sqrt{\dfrac{\mu B_0 \epsilon}{m_a}}$                                                         \\ \hline
\shortstack{Drift frequency $\omega_d$ \\ Eqs.(\ref{eq:omega_d_tokamak}, \ref{eq:omega_d_case_helical}, \ref{eq:omega_d_case_toroidal})} & $\dfrac{ n_d q }{ n_h e_a} \dfrac{\partial (e_a \phieq(\chi) - h(\chi)) }{\partial \chi}$ & $\dfrac{Nq}{r M_h} \left( \dfrac{1}{B_0} \dfrac{\partial \phi_{\rm eq}}{\partial r} + \dfrac{\mu}{e_a} \dfrac{\partial \delta}{ \partial r} \right)$ & $\dfrac{Nq}{r} \left( \dfrac{1}{B_0} \dfrac{\partial \phi_{\rm eq}}{\partial r} - \dfrac{\mu}{e_a R_0} \right)$ \\ \hline
\shortstack{Kinetic collisionality $\nu_\Omega$ \\ Eqs.(\ref{eq:nu_Omega}, \ref{eq:nu_Omega_tokamak})}                                  & $ \nu_\Omega = \frac{qR_0}{n_h v_T} 
      \frac{\nu_c(v) v^2 }{\left( \frac{h}{\Teq} \right)^{3/2}} $                          & $\dfrac{\nu_c q R_0}{v M_h \delta^{3/2}}$                                                                                                            & $\dfrac{\nu_c q R_0}{v \epsilon^{3/2}}$                                                                         \\ \hline
\shortstack{Detrapping freq. $\nu_d$ \\ Eq.(\ref{eq:nud})}                                                                        & $\dfrac{\omegab \nu_\Omega}{2}$                                                          & $\dfrac{\nu_c}{2 \delta}$                                                                                                                            & $\dfrac{\nu_c}{2 \epsilon}$                                                                                     \\ \hline
\shortstack{Effective detrapping freq. $\nu_{\rm eff}$ \\ Eq.(\ref{eq:effective_collision_frequency})}                                   & $\nu_d \max{(1, \sigma^2)}$                                                               & $\dfrac{\nu_c}{2 \delta}$                                                                                                                            & $M_h^2 \dfrac{\nu_c}{2 \epsilon}$                                                                               \\ \hline
\end{tabular}
}
\caption{Explicit expression of the main frequencies and geometrical parameters necessary to compute the entropy production rate in tokamak. 
The first column gives the general expression, the second is for the case where the primary perturbation is helical and the secondary is toroidal, and the third is the other way around.}
\label{tab:overview_main_parameters_tokamak}
\end{table}
}

\subsection{General framework}
\label{General framework}
\subsubsection{Trapping condition, effective wells and form factors}
\label{subsubsection:Trapping_condition_and_effective_well}
%
As discussed in Section \ref{Discriminating trapping in primary and secondary perturbations}, the trapping condition in the primary perturbation must be refined in the presence of a secondary perturbation when the curvature ratio $\sigma$ is large. 
In practice, this applies when the primary perturbation is helical, and reads
\begin{equation}
	\nonumber
    Y(r,\theta)= \alpha(r) \left|\sin\theta\right| < 1
      \label{eq_local_trapping}
\end{equation}
where
\begin{equation}
	\nonumber
    \alpha(r) = \frac{\epsilon(r)}{M_h \delta(r) }
      \label{eq_alpha_parameter}
\end{equation}
This trapping condition is space-dependent, and defines a domain in $(r,\theta)$ where the primary perturbation is helical, i.e. the primary perturbation is toroidal outside of this domain.
Inside this domain, the effective perturbation amplitude $\delta_\text{eff}$ becomes the depth $\delta_{\rm depth}$ of the magnetic field lines formed by the successive minima and maxima due to the helical perturbation \cite{Yushmanov_1990}.
It is expressed as:
\begin{equation}
     2 \delta_{\rm depth}(r, \theta) = \frac{B_{\rm max}}{B_{\rm min}} - 1 = 2 \delta(r, \theta)\left(\sqrt{1-Y^2}-Y\arccos Y\right)
  \label{eq_Deltaeff}
\end{equation}
Thus, in the following, the helical perturbation amplitude reads as follows:
\begin{equation}
    \delta_{\rm eff}(r,\theta) =
    \left\{
    \begin{array}{cc}
         &  \delta(r,\theta) \text{ if } Y \ge 1 \\
         &  \delta_{\rm depth}(r,\theta) \text{ if } Y < 1
    \end{array}
    \right.
\end{equation}

Finally, to account for the fact that the trapping condition is poloidally dependent, i.e. it is not constant on a flux-surface, we introduce the following ``form factors''
\begin{eqnarray} 
    \nonumber
    G_0(r) &=& \int_{Y<1}\frac{d\theta}{2\pi}  
    \label{eq:form_factors_G0}              
    \\
    \nonumber
    G_1(r) &=& \int_{Y<1}\frac{d\theta}{\pi} \sin^2\theta
    \label{eq:form_factors_G1}
\end{eqnarray}

They appear as pre-factors in the entropy production rate when the primary perturbation is helical. 

\subsection{Explicit expressions of contributions to the total resonant production rate}

In Section \ref{sec:Primary_entropy_production_rate}, we derived an expression Eq.(\ref{eq:value_entropy_production_collisional_summary}) for the resonant entropy production rate due to only one perturbation and that we can use to compute $\dot{S}_\text{hel,I}$ and $\dot{S}_\text{tor,I}$.
Similarly, in Section \ref{sec:Secondary_entropy_production_rate}, we derived the resonant entropy production rate due to a secondary perturbation on top of the equilibrium associated with a primary perturbation.
This led to an expression Eq.(\ref{eq:value_entropy_production_collisional_secondary_collisional}) that holds in high collisionality regime, and Eq.(\ref{eq:entropy_production_resonant_secondary_final_covariant}) for low collisionality regime.
As the latter need the numerical solving of 1-dimensional second-order differential equations, an analytical form unifying both regimes and valid in the adiabatic limit $\omegab \gg \omega_d$ is proposed in Eq.(\ref{eq:entropy_production_resonant_secondary_covariant_main}).
We use this expression to get an analytical form of resonant entropy production rates in tokamaks.

Entropy production rates take the form of integrals over the phase space of kinetic functions.
For tokamaks, the phase space integration element reads $d\gamma = dV_x dV_p$ where $dV_x = 4\pi^2 R_0 r dr$ is the volume element in real space and $V_p = 2 \pi m^3 v_\perp dv_\perp dv_\parallel = 2 \pi m^3 v_T^2 du dv_\parallel$ is the volume element in velocity space where the change of variable $u=v_\perp^2/(2 v_T^2)= \mu B_0 / \Teq$ has been used.

As discussed in Section \ref{subsec:Entropy_production_and_variational_principle}, entropy production rates are integrals of transverse --- here, radial --- fluxes that are proportional to the gradient of a thermodynamic potential. 
It is then attractive to make a reference diffusion coefficient appear in their expressions.
We choose the ``plateau'' value in standard neoclassical theory in a tokamak \cite{Hinton_Hazeltine_1976,Hirshman_Sigmar_1981} defined as
\begin{equation}
	\nonumber
     D_\text{p} = \frac{q R_0 v_D^2 }{v_T}
  \label{eq_def_plateau_diffusion_coefficient}
\end{equation}
where $v_D = \frac{\Teq}{e_a B_0 R_0}$.

Finally, before giving the explicit form of contributions to the entropy production rate, the relevant kinetic functions appearing in integrals are computed in Table \ref{tab:kinetic_functions}.

\setlength{\tabcolsep}{0.3em}
{\renewcommand{\arraystretch}{2.}
\begin{table} [H]
  \centering
  \begin{tabular}{|c|c|c|}
  \hline
                                                                            & Primary helical                                                                & Primary toroidal \\ \hline
    $d\gamma \Feq $                                                      & $\dfrac{d^{3}\mathbf{x}}{\sqrt{2\pi}} n_{\rm eq}  \dfrac{d \vpar}{v_T} du e^{-u}$  & $\dfrac{d^{3}\mathbf{x}}{\sqrt{2\pi}} n_{\rm eq}  \dfrac{d \vpar}{v_T} du e^{-u}$ \\ \hline
    $d\gamma \delta(\Omega) \Feq $                                       & $\dfrac{d^{3}\mathbf{x}}{\sqrt{2\pi}} n_{\rm eq} \dfrac{qR_0}{M_h v_T}  du e^{-u}$ & $\dfrac{d^{3}\mathbf{x}}{\sqrt{2\pi}} n_{\rm eq} \dfrac{qR_0}{v_T}  du e^{-u}$    \\ \hline
    $d\gamma \delta(\Omega) \omegab \Feq $                              & $\dfrac{d^{3}\mathbf{x}}{\sqrt{2\pi}} n_{\rm eq}  du u^{1/2} e^{-u}$              & $\dfrac{d^{3}\mathbf{x}}{\sqrt{2\pi}} n_{\rm eq}  du u^{1/2} e^{-u}$   \\ \hline
    $\dfrac{1}{T_\ast} \nbold \cdot \partial_\Jbold \Ueq $                   & $\dfrac{1}{\Teq} M_h \dfrac{\Wpareq}{q R_0}$                                   & $\dfrac{1}{\Teq} \dfrac{\Wpareq}{q R_0}$  \\ \hline
    $\dfrac{1}{T_\ast} (\nbold' - \nbold \sigma) \cdot \partial_\Jbold \Ueq$ & $\dfrac{Nq}{M_h} \dfrac{1}{e_a} \partial_\psi \Xi$                             & $ - Nq \dfrac{1}{e_a} \partial_\psi \Xi$ \\ \hline
\end{tabular}
  \caption{Kinetic functions used to compute the contribution to the entropy production rate.}
  \label{tab:kinetic_functions}
\end{table}
}

\subsubsection{Helical primary perturbation}

Introducing the notation $M_h = n_h = M + N q$ and $\nubar(u)=\nubar_v(\sqrt{u})$, the primary entropy production rate when the perturbation is helical is given by
 \begin{eqnarray}
  \nonumber 
			\dot{S}_\text{hel,I} &=& \frac{1}{2} \sqrt{\frac{\pi}{2}} \int dV_x \: n_\text{eq} \: D_\text{p} \: 
       \frac{M^2}{M_h} G_0 \left( \frac{\delta_{\rm eff}}{\epsilon} \right)^2 
       \int_{0}^{+\infty} du e^{-u} u^{2}  K_\text{hel,I}(r,u)  \\
      & &  \left[
	       \frac{1}{\Ncaleq}\frac{d \Ncaleq}{d r} + 
	          \left( u - \frac{3}{2} \right)  \frac{1}{\Teq}\frac{d\Teq}{d r} +
						\frac{M_h}{M} \frac{e_a B_p}{\Teq} \Vpareq 
	      \right]^2			
   \label{eq_resonant_rate_helical} 
\end{eqnarray} 
where
\begin{equation}
	\nonumber
      K_\text{hel,I}(r,u) 
      =
      \frac{2}{\pi} \Lambda_{\rm I}(r,u)
      =
      \min\left( 1 , 
	     \frac{4}{\pi} \mathcal{I} 
			\frac{\nustar}{M_h} \left(\frac{\epsilon}{\delta_{\rm eff}}\right)^{3/2}
			\frac{\nubar(u)}{u^{1/2}} \right)
  \label{eq_def_KhelI}
\end{equation}
with $\mathcal{I}=1.38$. 
The secondary entropy production rate is given by
\begin{eqnarray}
  \nonumber 
	\dot{S}_\text{tor,II} &=&   \left( \frac{2}{\pi}\right)^{3/2} 
	\int dV_x \: n_\text{eq} \: D_p \:
	\left (\frac{\delta_{\rm eff}}{\epsilon} \right)^{3/2}
	G_1 \left( \frac{N q}{M_h} \right)^2 \\
  & & \int_{0}^{+\infty} du e^{-u} u^{5/2} \frac{K_\text{tor,II}}{\nustar \nubar(u)}
    \left[
	   \frac{1}{\Ncal}\frac{d \Ncaleq}{d r} + 
	      \left( u - \frac{3}{2} \right)  \frac{1}{\Teq}\frac{d\Teq}{d r} 
	  \right]^2
  \label{eq_resonant_rate_toroidal_secondary} 
\end{eqnarray}  
where  
\begin{equation}
	\nonumber
    K_\text{tor,II} 
    =
    \frac{\pi}{2} \Lambda_{\rm II}
    =\min\left( \frac{8}{9},\frac{M_h}{2}  \right)
  \label{eq_def_KtorII}
\end{equation}

\subsubsection{Toroidal primary perturbation}
Similarly, the primary entropy production rate when the perturbation is toroidal is 
\begin{eqnarray}
	 \nonumber 
     \dot{S}_\text{tor,I} &=&  \frac{1}{2} \sqrt{\frac{\pi}{2}} \int dV_x \: n_\text{eq} \: D_\text{p} 
	       \int_{0}^{+\infty} du e^{-u} u^{2} K_\text{tor,I}(r,u) \\
      & &  
	   \left[  
		   \frac{1}{\Ncaleq}\frac{d \Ncaleq}{d r}  + 
		   \left( u -\frac{3}{2} \right) \frac{1}{\Teq} \frac{d\Teq}{d r}
       + \frac{e_a B_p}{\Teq} \Vpareq
		\right]^2
   \label{eq_resonant_rate_toroidal_primary}
\end{eqnarray}  
where 
\begin{equation}
	\nonumber
  K_\text{tor,I}(r,u)
  =
  \frac{2}{\pi} \Lambda_{\rm I}(r,u)
  =
  \min\left( 1,\frac{4}{\pi}\mathcal{I}\frac{\nustar \nubar(u)}{u^{1/2}}\right)
  \label{eq_def_KtorI}
\end{equation}

The secondary entropy production rate is 
\begin{eqnarray}
   \nonumber  
	   \dot{S}_\text{hel,II} &=&   \left(\frac{2}{\pi}\right)^{3/2} 
	   \int dV_x n_\text{eq} D_\text{p}  \frac{N^2 q^2 }{M_h^3} 
    	\left( \frac{\delta_{\rm eff}}{\epsilon} \right)^2 \\
  & & \int_{0}^{  +\infty }  du  
    	e^{-u} u^{5/2}\frac{K_\text{hel,II}(r,u)}{\nustar \nubar(u)}
       \left[
        	\frac{1}{\Ncaleq}\frac{d \Ncaleq}{d r} +
	       \left( u - \frac{3}{2} \right) \frac{1}{\Teq}\frac{d\Teq}{d r} 
	     \right]^2
  \label{eq_resonant_rate_ripple_secondary}
\end{eqnarray}
where
\begin{equation}
	\nonumber
  K_\text{hel,II}(r,u) = \max\left( 1 , 
	      \frac{\pi^2}{8} 
				M_h^2 \frac{\nustar \nubar(u)}{u^{1/2}} \right)
  \label{eq_def_KhelII}
\end{equation}
is a smooth function connecting the two collisionality regimes derived in Section \ref{subsec:Weakly_collisional_case_secondary} and \ref{subsec:Highly_collisional_case_secondary}.

\subsection{Computing neoclassical fluxes}
\label{Computing neoclassical fluxes}

Fluxes and thermodynamic forces can then be properly defined from the resonant entropy production rate using the expression
\begin{equation}
  \dot{S}_\text{res} = - \frac{1}{2} \int dV_x n_\text{eq}
    \left(
     \frac{\Gamma}{n_\text{eq}}
      \frac{1}{\Ncaleq}\frac{d \Ncaleq}{d r}
      + \frac{\Mcal}{ m_a v_{T}}
      \frac{\Vpareq}{v_{T}}
       + \frac{Q}{n_\text{eq} \Teq}
     \frac{1}{\Teq}\frac{d\Teq}{d r}
    \right)
  \label{eq_force_flux_entropy}
\end{equation}
Identification provides the particle flux $\Gamma$, heat flux $Q$ and a magnetic braking force $\Mcal$, usually referred to as Neoclassical Toroidal Viscosity (NTV). 
When several species are present, these expressions must be summed over all species. 
Extra terms can also be added that account for inter-species friction and equipartition terms. 
Ultimately, minimisation with respect to toroidal velocity gives a friction rate \cite{Shaing_1993}.

The expressions of $\dot{S}_\text{hel,I}$, $\dot{S}_\text{tor,II}$, $\dot{S}_\text{tor,I}$, and $\dot{S}_\text{hel,II}$, respectively given by Eqs.(\ref{eq_resonant_rate_helical},\ref{eq_resonant_rate_toroidal_secondary},\ref{eq_resonant_rate_toroidal_primary},\ref{eq_resonant_rate_ripple_secondary}), coincide with those found and tested against drift/gyrokinetic codes in the past \cite{Varennes_2023}, in the special case of $M=0$ which corresponds to a magnetic ripple perturbation in the tokamak. 
These expressions will be confronted with first-principles simulations in a forthcoming paper. 

\section{Conclusion}
\label{Conclusion}

In summary, a unified framework is proposed for the computation of neoclassical equilibria combining a variational approach based on the principle of minimum entropy production with a Hamiltonian description of gyrocentre motion. 
This formulation provides a transparent identification of the respective contributions of different classes of trapped particles in magnetic well minima. 
In particular, the dominant mechanisms responsible for the largest transport coefficients are clearly identified and quantitatively assessed. 
The principle of minimum entropy production is derived from first-principles arguments, and the entropy production rate is naturally decomposed into contributions from ``primary'' and ``secondary'' perturbations. 
Trapping in the primary perturbations recovers the standard resonant banana-plateau contribution to neoclassical transport, whereas secondary perturbations give rise to the well-known $1/\nu$ regime. 
The resulting expressions are consistent with those previously reported in the literature. 
The analysis also recovers the role of magnetic and electric drifts in restoring an acceptable scaling with collisionality through de-trapping induced by precession.

A central element of the approach is the resolution of the Fokker-Planck equation for each bounce-resonance harmonic. 
When the ratio of bounce to precession frequencies is smaller than a threshold of order ten, several harmonics must be retained. 
By contrast, in the adiabatic limit where this ratio becomes large, the dominant contribution originates from the bounce-averaged distribution function. 
In all regimes, the numerical method developed here provides a fast and accurate evaluation of the entropy production rate, and therefore of the associated neoclassical transport coefficients. 
Furthermore, analytical solutions can be obtained in the sufficiently collisional regime, characterized by an effective detrapping frequency larger than the precession frequency. 
In this limit, the agreement between analytical and numerical results is satisfactory, providing an even faster route for the evaluation of neoclassical transport. 
These analytical expressions have already been benchmarked in the case of a rippled tokamak.

More generally, the present work demonstrates that neoclassical fluxes can be cast into a compact entropy-production formalism whose minimization directly yields the transport matrix. 
This opens the way to the development of fast and flexible reduced models for the evaluation of transport in tokamaks subject to magnetic ripple or non-resonant magnetic perturbations.

\section{Acknowledgements}
\label{Acknowledgements}
This research is supported by the National Research Foundation, Singapore.
We acknowledge the financial support of the Cross-Disciplinary Program on Numerical Simulation of CEA, the French Alternative Energies and Atomic Energy Commission.

\pagebreak
\appendix
\textbf{\Large APPENDICES}\\
\section{Building a functional based on entropy production}
\label{appendix:functional_entropy_production}

A functional $\Scal \left( \Ueq , \Ueqdag \right)$ is built following \cite{Samain_1997, Garbet_2009, Garbet_2010b}
\begin{equation}
  \Scal_t + \Scalres + \Scalcoll = \Scal
 \label{eq:entropy_production_generalised_appendix}
\end{equation}
where
\begin{eqnarray} 
  \Scal_t \left( \Ueq , \Ueqdag \right) &=&  
	  \frac{2}{T_\ast^2}  
	  \int d\gamma \: \Feq 		\:	
		\Ueqdag \: \frac{\partial \Ueq}{\partial t} =
		\left\langle  \Ueqdag \: \big| \: \partial_t \Ueq \right\rangle   
 \label{eq:entropy_production_generalised_t}	\\
  \Scalres \left( \Ueq , \Ueqdag \right) &=& 
	 -  \frac{2}{T_\ast^2}  
	  \int d\gamma \: \Feq 
		\frac{\partial \Ueqdag}{\partial \Jbold} 
		  \cdot \Gammabold_U[\Ueq] = -
		\left\langle  \partial_{\Jbold} \Ueqdag \: \big| \:  \Gammabold_U \right\rangle
 \label{eq:entropy_production_generalised_resonant} \\
  \Scalcoll \left( \Ueq , \Ueqdag \right) &=&  
	  - \frac{2}{T_\ast^2}  
	  \int d\gamma \: \Feq \:  
	   \Ueqdag \: \Ccal\left[\Ueq\right]  = -
   \left\langle \Ueqdag \: \big| \:   \Ccal\left[\Ueq\right] \right\rangle 
 \label{eq:entropy_production_generalised_collisional}
\end{eqnarray}
where the bracket $ \left\langle \: \big| \: \right\rangle 	 $ is a scalar product defined as
\begin{equation}
 \nonumber
   \left\langle \Ueqdag \: \big| \: \Ueq \right\rangle = 
       \frac{2}{T_\ast^2}  
	  \int d\gamma \: \Feq \: \Ueqdag \cdot \Ueq 
  \label{eq:scalar_product}
\end{equation} 
The extremum of the functional $\Scal$ for all variations of $\Ueqdag$ near $\Ueq$ yields the transport equation Eq.(\ref{eq:kinetic_equation_mean_U}). 
This is readily demonstrated by writing $\Ueqdag = \Ueq + \delta \Ueqdag$ and recasting the functional variation $\delta \Scal$ as
\begin{equation}
 \nonumber
  \delta \Scal = 
	\frac{2}{T_\ast^2}  
	  \int d\gamma \: \Feq \: \delta \Ueqdag
	    \left( \frac{\partial \Ueq }{\partial t}
			 + \nabla_\Jbold
		      \cdot \Gammabold_U 
			 -  \Ccal\left[\Ueq\right] 
		  \right)
 \label{eq:entropy_production_variation}
\end{equation} 
where an integration by parts has been performed using the aforementioned boundary conditions, while still assuming that derivatives of $\Feq$ are weak compared with those of $\Ueq$ or $\Ueqdag$. 
In other words, $\Feq$ is simply a weight in the phase space. 

Eq.(\ref{eq:kinetic_equation_mean_U}) is equivalent to stating $\delta \Scal = 0$ for all $\delta \Ueqdag$.
It formally reads as a condition of the null functional derivative 
\begin{equation}
 \nonumber
  \frac{\delta \Scal}{\delta \Ueqdag} = 	   
      \frac{\Feq}{T_\ast^2} 
      \left(\frac{\partial \Ueq }{\partial t}
			      + \nabla_\Jbold
		        \cdot \Gammabold_U 
		     	   -  \Ccal\left[\Ueq\right] \right) = 0 
 \label{eq:null_functional_derivative}
\end{equation} 
This variational principle is thus a ``weak formulation'' of Eq.(\ref{eq:kinetic_equation_mean_U}) based on the entropy production rate Eq.(\ref{eq:entropy_production_rate_split_eq}) with contributions of Eqs.(\ref{eq:entropy_production_generalised_t},\ref{eq:entropy_production_generalised_resonant},\ref{eq:entropy_production_generalised_collisional}).
It can be simplified by considering a functional of the form
\begin{equation}
 \nonumber
   \Scal\left(\Ueq,\Ueqdag\right) = 
	  \frac{2}{T_\ast^2}  
	  \int d\gamma \: \Feq \: \Ueqdag \Dcal[\Ueq] =
	  \left\langle \Ueqdag \: \big| \: \Dcal[\Ueq] \right\rangle
 \label{eq:functional_self_adjoint}
\end{equation}
where $\Dcal$ is an operator. Stating $\delta \Scal / \delta \Ueqdag=0$ is equivalent to the condition $\Dcal[\Ueq]=0$. 
Let us now assume that the operator $\Dcal$ is self-adjoint, and consider a slightly different functional 
\begin{equation}
 \nonumber
   \Tcal\left(\Ueqdag\right) =
     \frac{1}{2} \Scal\left(\Ueqdag,\Ueqdag\right) =
	  \frac{1}{T_\ast^2}  
	  \int d\gamma \: \Feq \: \Ueqdag \Dcal[\Ueqdag] = 
	    \left\langle \Ueqdag \: \big| \:  \Dcal[\Ueqdag] \right\rangle
 \label{eq:functional_self_adjoint_variant}
\end{equation}
Using the self-adjoint property of $\Dcal$, its variation reads
\begin{equation}
 \nonumber
   \delta \Tcal = \left\langle \delta \Ueqdag \: \big| \:  \Dcal[\Ueqdag] \right\rangle 
     +  \left\langle \Ueqdag \: \big| \:  \Dcal[\delta \Ueqdag] \right\rangle =
     2 \left\langle \delta \Ueqdag \: \big| \:  \Dcal[\Ueqdag] \right\rangle 
 \label{eq:functional_self_adjoint_variation}
\end{equation}
so that
\begin{equation}
 \nonumber
   \left. \frac{\delta \Tcal}{\delta \Ueqdag} \right|_{\Ueqdag = \Ueq}  =
    2\ \frac{\Feq}{T_\ast^2} \ \Dcal[\Ueqdag] = 
    \frac{\delta \Scal}{\delta \Ueqdag}
 \label{eq:functional_self_adjoint_derivative}
\end{equation}
Therefore, a variational principle based on the functional $\Tcal(\Ueqdag)$ yields the same result as the one derived for the functional $\Scal(\Ueq,\Ueqdag)$. 
The advantage of the former is that only one field, $\Ueqdag$, is used. 
This trick does not apply to $\Scal_t(\Ueq,\Ueqdag)$.

On the other hand, more compact formulations based on this procedure can be used for the resonant and collisional contributions $\Scalres$ and $\Scalcoll$. 
The self-adjointness of the Boltzmann collision operator in the angle/action space is demonstrated in \cite{Garbet_2009}, and the corresponding functional is a quadratic function of partial derivatives of $\Ueqdag$ in the action space. 
In Sections \ref{sec:Primary_entropy_production_rate} and \ref{sec:Secondary_entropy_production_rate}, the self-adjointness for resonant processes is demonstrated
and it is shown that the flux $\Gammabold_U$ is a linear combination of $\partial_\Jbold \Ueq$ components, i.e. is diffusive in the action space
\begin{equation}
    \Gamma_{Ui} = - D_{ik} \frac{\partial \Ueq}{\partial J_k}
 \label{eq:diffusion_flux}
\end{equation} 
This operator is obviously self-adjoint. 
Hence, both the collisional and resonant entropy production rates involve gradients of $\Ueqdag $ in the action space. 
However, a subtle point is that there is a hidden dependence on $\Ueq$ via the equilibrium distribution $\Feq$. 
This dependence should never be used to compute functional derivatives for reasons given above. 
Hence the arguments of the functionals $\Scalres$ and $\Scalcoll$ are formally $\Ueq$ and $ \partial_\Jbold\Ueqdag$, bearing in mind that only the latter dependence should be considered for variation calculus. To simplify the notations, only the dependence on $ \partial_\Jbold\Ueqdag$ is explicitly mentioned. The expressions of these functionals read
\begin{eqnarray} 
  \Scalres\left( \partial_\Jbold\Ueqdag \right) &=& 
	 -  \frac{1}{T_\ast^2}  
	  \int d\gamma \: \Feq 
		\frac{\partial \Ueqdag}{\partial \Jbold} 
		  \cdot \Gammabold_U[\Ueqdag]  
  \label{eq:entropy_production_generalised_resonant_bis_appendix} \\
  \Scalcoll\left( \partial_\Jbold\Ueqdag \right) &=&  
	  - \frac{1}{T_\ast^2}  
	  \int d\gamma  \: \Feq \:
	   \Ueqdag \: \Ccal\left[\Ueqdag\right]  
	\label{eq:entropy_production_generalised_collisional_bis}
\end{eqnarray}
The evolution equation on $\Feq$ Eq.(\ref{eq:kinetic_equation_mean}) is equivalent to computing an extremum of the functional $\Scal[\partial_\Jbold\Ueqdag]$ with respect to $\Ueqdag $, calculated at $\Ueqdag = \Ueq$. 
The extremum of the entropy functional is zero at the extremum, $\Scal=0$, and therefore, $\Stdot = - \Scalres - \Scalcoll$. 
Hence, the actual entropy production rate can be found from resonant and collisional contributions computed at the extremum, which is a convenient property. 
Further, for time-independent potentials $\Ueq$, finding an extremum of $\Scal$ is equivalent to finding a minimum of entropy production rate. 
Notations can be even more simplified by replacing $\Ueqdag$ by $\Ueq$ in the expressions Eq.(\ref{eq:entropy_production_generalised_resonant_bis}) and Eq.(\ref{eq:entropy_production_generalised_collisional_bis}), with the aforementioned caveat regarding extremum computation, which should be performed with respect to $\partial_\Jbold\Ueq$ rather than the $\Ueq$ term hidden in the weight $\Feq$. 

\section{Primary perturbation and island in the phase space}
\label{Primary perturbation and island in the phase space}
In the presence of a primary perturbation, a new set of variables $\thetabold$ and $\Ibold$ can be defined in the vicinity of a specific point $\Jboldstar$ that lies on the resonant surface $\Omega=0$, or equivalently $J_3 = G(J_2)$. 

Defining the rotational transform $\iota(J_2) = - \frac{dG(J_2)}{dJ_2}$, the new variables are
\begin{eqnarray}
  \nonumber
	  \theta_2 & = & n_2 \alpha_2 +  n_3 \alpha_3	            \\
  \nonumber
    \theta_3 & = & 		\frac {\alpha_2 - \iota_{\ast} \alpha_3} 
		              {\iota_{\ast} n_2 + n_3}    
	\label{eq:link_theta_vs_alpha}
\end{eqnarray} 
\begin{eqnarray} 
   	I_2 & = &  \frac{\iota_{\ast} \Jtilde_2 + \Jtilde_3}{\iota_{\ast} n_2 + n_3} 	 \\
		I_3 & = & n_3 \Jtilde_2 - n_2 \Jtilde_3
 \label{eq:link_I_vs_J_linearised}
\end{eqnarray}
where $\Jtildebold = \Jbold - \Jboldstar $, and $\iota_{\ast} = \iota(J_{2\ast })$, hence it is now a parameter. 
It can be verified that $(\thetabold, \Ibold)$ is a set of conjugate variables since $\{ I_j, I_k\} = 0$, $\{ \theta_j, \theta_k\} = 0$ and $\{ \theta_j, I_k\} = \delta_{jk}$ whatever the indices $(j,k) \in [2,3]$. 
This property can also be demonstrated by constructing the generating function that relates $(\thetabold, \Ibold)$ to $(\alphabold, \Jbold)$. 
Note however that $\theta_3$ is not an angle since $\iota_{\ast}$ is not necessarily a rational number. 
The angle $\theta_2$ appears to be the argument of the primary perturbation, while the variable $\theta_3$ characterises the second degree of freedom in the angle space, such that the new system of coordinates is conjugate. 
The set of variables $(\thetabold, \Ibold)$ is invertible, thus allowing moving back and forth from old to new set of conjugate coordinates.
The relationship between angles can be inverted easily 
\begin{eqnarray} 
 \nonumber
  \alpha_2 & = & n_3 \theta_3 + 
		\frac {\iota_{\ast} \theta_2} 
		      {\iota_{\ast} n_2 + n_3}               \\
 \nonumber
	 \alpha_3 & = & - n_2 \theta_3 +  
	   \frac { \theta_2} 
		       {\iota_{\ast} n_2 + n_3}
 \label{eq:link_alpha_vs_theta}
\end{eqnarray}
In a similar way, the relationship between action variables is inverted in a convenient form
\begin{equation}
    \Jbold = \Jboldstar + I_2 \nbold + I_3 \tbold 
	\label{eq:link_J_vs_I}
\end{equation} 
where the vector 
\begin{equation}
  \nonumber
		\tbold = \frac{1}{n_2 \iota_{\ast} + n_3} 
		\left( 
		 \begin{array}{c} 
		   1  \\ - \iota_{\ast} 
		\end{array}
		\right)
  \label{eq:vecteur_tangent_covariant}
\end{equation}
is tangent to the resonant surface at $\Jbold = \Jboldstar$ in the $(J_2,J_3)$ space\footnote{This is really verified by noting that $\partial_\Jbold \Omega$ is a vector normal to the resonant surface. 
Since $\Omega_R(J_2)=0$ all along the resonant surface, it readily appears that $\left. \left(\partial_{J_2} \Omega - \iota  \partial_{J_3} \Omega  \right) \right|_{\Jbold = \Jboldstar} = 0 $, hence $\tbold \cdot \left. \partial_\Jbold \Omega \right|_{\Jbold = \Jboldstar}  = 0$, which demonstrates that $\tbold$ is tangent to the resonant surface at $\Jbold = \Jboldstar$.}. 
Defining $\Jbold_R$ as $\Jbold_R = \Jboldstar + I_3 \tbold$, it appears that Eq.(\ref{eq:link_J_vs_I}) can be recast as $\Jbold = \Jbold_R + I_2 \nbold$, where $\Jbold_R$ lies on the resonant surface. 
The nature of the actions $(I_2,I_3)$ becomes clear: $I_3$ characterises the dynamics along the resonant surface in the vicinity of $\Jboldstar$, whereas $I_2$ measures the dynamics across the resonant surface, and varies in time as
\begin{equation}
	\nonumber
    \frac{ d I_2 }{ dt }  =  - h \sin \theta_2
	\label{eq:dIdt}
\end{equation}
The frequency $\Omega$ can be Taylor expanded near $\Jbold_R$ as $\Omega(\Jbold)  =  C(\Jbold_R) I_2 $, where
\begin{equation}
    C(\Jbold_R)  =  \left. n_i n_j 
					  \frac{\partial^2 \Heq}
		             {\partial J_i J_j} 
				  \right|_{\Jbold = \Jbold_R} 
	\label{eq:curvature_appendix}
\end{equation} 
where $C$ can be called the ``curvature'' of the Hamiltonian. In principle, it depends on the actions on the resonant surface $\Jbold_R$; however, in all cases of interest for neoclassical applications, it turns out that $C$ is a constant, and we will consider it so in the following. 
As long as $h(\Jbold)$ does not vary much in the area of interest, the evolution equation of the phase is then simply
\begin{equation}
	\nonumber
    \frac{ d \theta_2 }{ dt }  =  C I_2
	\label{eq:phase_island_reduced}
\end{equation}
A new invariant of motion $H_I$ is built, namely
\begin{equation}
	\nonumber
    H_I  =  \frac{1}{2} C I_2^2 - h \cos \theta_2
	\label{eq:island_Hamiltonian}
\end{equation} 
which is in fact a Hamiltonian.
The action $I_2$ can be replaced by the pulsation $\Omega = C I_2$, so that $d_t\theta_2 = \Omega$ and the perturbed Hamiltonian can be recast as
\begin{equation}
  \nonumber
    H_\Omega = \frac{1}{2} \Omega^2 - \omegab^2 \cos\theta_2 
	\label{eq:Hamiltonian_near_resonance_Omega}
\end{equation}

\section{Hamiltonian secondary perturbation and entropy production}
\label{Hamiltonian secondary perturbation and entropy production}
\subsection{Expressing the secondary frequency at the primary resonant surface}
\label{appendix:Expressing_the_secondary_frequency_at_the_primary_resonant_surface}
The phase $\xi' = \nbold' \cdot \alphabold$ of the secondary perturbation can be expressed as a function of the variables $\thetabold$ using the relationship between $\alphabold$ and $\thetabold$, i.e.
\begin{equation}
    \xi' = \sigma \theta_2 + n_d \theta_3 
	\label{eq:phase_secondary_perturbation}
\end{equation}
where
\begin{equation}
	\nonumber
    n_d = n'_2n_3 - n_2 n'_3
	\label{eq:definition_nbar2}
\end{equation}
and
\begin{equation}
    \sigma = \frac{n'_2 \iota_{\ast} + n'_3}{n_2 \iota_{\ast} + n_3}
	\label{eq:definition_sigma_ratio}
\end{equation}
where $\iota_{\ast}=-\frac{dG}{d_{J_2}}(\Jboldstar)$ with $G$ the function relating $J_2$ and $J_3$ defined in Eq.(\ref{eq:def_G}) and $\Jboldstar$ a point on the primary resonance surface.
The parameter $\sigma$ can be given a physical interpretation as follows.
Let us use the expansion Eq.(\ref{eq:vecteur_tangent_covariant}) of $\Jbold$ near the specific point $\Jboldstar$ on the primary resonant surface, hence $\Omega (\Jboldstar)$. 
The resonant frequency $\Omega$ reads 
\begin{equation}
	\nonumber
   \Omega(\Jbold) = \Omega (\Jboldstar) +
	    I_2
			\left( \nbold \cdot
		   \left.  
			    \frac{\partial \Omega}
		           {\partial \Jbold} 
			  \right|_{\Jbold = \Jboldstar} 
			\right)
			 + I_3
		 \left( \tbold \cdot
		   \left.  
			    \frac{\partial \Omega}
		           {\partial \Jbold} 
			  \right|_{\Jbold = \Jboldstar}  
			\right)
	\label{eq:Omega_variation_resonant_surface}
\end{equation}
where $\tbold$ is the tangent vector given by Eq.(\ref{eq:vecteur_tangent_covariant}). 
Since $\Omega(\Jbold_R) = 0$ on the primary resonant surface for any $\Jbold_R$, and therefore any $I_3$, the derivative along the tangent direction  $\left. \tbold \cdot \partial_{\Jbold}\Omega \right|_{\Jbold = \Jboldstar} $ must vanish. 
Another way to derive this relation is to note that $\tbold$ is tangent to the resonant surface at $\Jbold = \Jboldstar$, while $\left. \partial_{\Jbold}\Omega \right|_{\Jbold = \Jboldstar}$ is normal to the resonant surface. 
Hence their scalar product must be zero. 
This imposes the following constraint
\begin{equation}
	\nonumber
		\left.  
			\frac{\partial \Omega}
		           {\partial J_2} 
		\right|_{\Jbold = \Jboldstar}  
    = \iota_{\ast}
		\left.  
			\frac{\partial \Omega}
		           {\partial J_3} 
		\right|_{\Jbold = \Jboldstar}  
	\label{eq:constraint_resonance}
\end{equation}
Let us now expand the resonant frequency $\Omega'$ in a similar way
\begin{equation}
	\nonumber
   \Omega'(\Jbold) = \Omega' (\Jboldstar) +
	    I_2
			 \left( \nbold \cdot
		    \left.  
			    \frac{\partial \Omega'}
		           {\partial \Jbold} 
			   \right|_{\Jbold = \Jboldstar} 
			 \right)
			 + I_3 
		 \left( \tbold \cdot
		   \left.  
			    \frac{\partial \Omega'}
		           {\partial \Jbold} 
			  \right|_{\Jbold = \Jboldstar}  
			\right)
	\label{eq:Omega_prime_variation_resonant_surface}
\end{equation}
Combining these relations, one finds the constraint $\Omega'(\Jbold) = \Omega'(\Jbold_R) + \sigma \Omega(\Jbold)$
where $\Jbold_R = \Jboldstar + I_3 \tbold$. 
The frequencies associated with the primary and secondary perturbation are thus expressed as $\Omega = C I_2$ and $\Omega' = \Omega'(\Jbold_R) + C^{\ast} I_2$
where
\begin{equation}
	\nonumber
    C  =  \left. n_j 
					  \frac{\partial \Omega}
		             {\partial J_j} 
				  \right|_{\Jbold = \Jboldstar} 
			= \left( 
			    n_2 \iota_{\ast}  + n_3 
				\right) 
				\left.  
			   \frac{\partial \Omega}
		          {\partial J_3} 
			  \right|_{\Jbold = \Jboldstar}
	\label{eq:curvature_vs_iota}
\end{equation} 
and
\begin{equation}
	\nonumber
    C^{\ast}  =
			   \left. n'_j 
					  \frac{\partial \Omega}
		             {\partial J_j} 
				  \right|_{\Jbold = \Jboldstar} 
			= \left( 
			    n'_2 \iota_{\ast} + n'_3 
				\right) 
				\left.  
			   \frac{\partial \Omega}
		          {\partial J_3} 
			  \right|_{\Jbold = \Jboldstar}
	\label{eq:curvature_secondary}
\end{equation}

\subsection{Expressing a local Maxwellian in action/angle variable}
\label{appendix:Expressing_a_local_Maxwellian_in_action_angle_variable}

From Appendix \ref{Primary perturbation and island in the phase space} and Eqs.(\ref{eq:link_I_vs_J_linearised}), fields that live on the primary resonant surface depend on $\Jbold = \Jboldstar + I_2 \nbold + I_3 \tbold $, where $I_3  =  n_3 \Jtilde_2 - n_2 \Jtilde_3$. 
$\Jboldstar$ lies on the resonant surface $\Omega = \nbold \cdot \Omegabold=0$. 
Assuming a small electric drift frequency compared to the transit frequency, the resonant condition reduces to $\vpar=0$, such that $\Jboldstar$ is just a label of a magnetic surface. An equivalent invariant can thus be defined up to a constant as in
\begin{eqnarray} 
 \nonumber
	 I_3 & = & \psi - \psi_\ast  - \frac{n_2}{q_\ast n_3 + n_2} \frac{1}{e_a} R_0 m_a \vpar    \\
	   & = & \frac{1}{e_a}  
	       \frac{ 1 }
	            { \left. \left( n_3 \frac{d\chi}{dr} + 
	              n_2 \frac{d\psi}{dr} \right) \right|_{r=r_\ast} }
	     \left[ n_3 \left( J_2 - J_{2\ast} \right) -
	            n_2 \left(J_3 - J_{3\ast}\right) \right] 
  \label{eq:K_invariant}
\end{eqnarray} 
Thus, the function $\Xi$ Eq.(\ref{eq:def_Xi}) that appears in the thermodynamic potential $\Ueq(\Jbold,t)$ Eq.(\ref{eq:mean_thermodynamic_potential}) is a function of $\Heq$ and $I_3$. Since the resonant surface is located at $\vpar = 0$, $\psi_R = \psi_\ast + I_3$ is a magnetic flux that spans the resonant surface in the action space, in fact the third coordinate of $\Jbold_R$. The first line of Eq.(\ref{eq:K_invariant}) can be inverted to express $\psi_R$ as a function of $(\psi,\vpar)$
\begin{equation}
 \nonumber
   	 \psi_R  = \psi - 
       \frac{n_2}{q_\ast n_3 + n_2} \frac{1}{e_a} R_0 m_a \vpar
\label{eq:inversion_psi_vs_actions}
\end{equation}
The last term is proportional to $R_0 m_a \vpar/e_a $ and is small compared to $\psi_R = \psi_\ast +  I_3$. The potential $\Xi$ is written as a function of $\Psi_R$ and $\Heq$ which can expanded near $\psi_R$. All calculations are run in the vicinity of $\psi_R$, i.e. for small values of $R_0 m_a \vpar/e_a$. The reference temperature $T_\ast$ is chosen as the temperature at the surface $\psi=\psi_\ast$, hence $T_\ast =  T(\psi_\ast)$.
The definition of the thermodynamic potential Eqs.(\ref{eq:mean_thermodynamic_potential}, \ref{eq:def_Xi}) can then be rewritten as $\Ueq = U_{\rm eq,\psi}  +  U_{\rm eq,\vpar}$, where
\begin{eqnarray}
 \nonumber 
   	\frac{1}{ T_\ast } U_{\rm eq,\psi}(\Heq,\psi,t) &=& \Xi(\Heq,\psi,t) \\
  \nonumber   	
   	&=& 
   	\ln \Ncaleq(\psi,t)  
   	      - \frac{3}{2} \ln \Teq(\psi,t)  +
   	      \Heq(\Jbold)\left( \frac{1}{\Teq(\psi,t)} - \frac{1}{T_\ast} \right)
	\label{eq:mean_thermodynamic_potential_psi}
\end{eqnarray}
and
\begin{equation}
 \nonumber
   	\frac{1}{ T_\ast }  U_{\rm eq,\vpar}(\Heq,\psi,t) =
   	   - \frac{n_2}{n_h} 
   	    \frac{R_0}{e_a} 
   	       \frac{\partial \Xi(\Heq,\psi,t)}{\partial \psi}
           m_a \vpar
   	    + \frac{\Wpareq(\Heq,\psi,t)}{\Teq(\psi,t)} 
   	  m_a \vpar(\Jbold) 
	\label{eq:mean_thermodynamic_potential_vpar}
\end{equation}
with
\begin{equation}
 \nonumber
   	 \frac{\partial \Xi(\Heq,\psi,t)}{\partial \psi} = 
      \frac{\partial \ln \Ncaleq}{\partial \psi}  +
   	      \left( \frac{E}{\Teq}- \frac{3}{2} \right) 
   	        \frac{\partial \ln \Teq}{\partial \psi} 
	\label{eq:def_Xi_partial_derivative_psi}
\end{equation}
where $E(\Heq,\psi,t) = \Heq - e_a \Phieq(\psi,t)$ is the kinetic energy. The effective density $\Ncaleq$ is defined as $\Ncaleq = n_{\rm eq} \exp\left( e_a \Phieq /\Teq \right)$ such that
\begin{equation}
   \nonumber
  \frac{1}{\Ncaleq}\frac{d \Ncaleq}{d \psi} = \frac{1}{n_{\rm eq} }\frac{d n_{\rm eq}}{d \psi}
	+\frac{e_a}{\Teq}\frac{d\Phieq}{d \psi}
  \label{eq_definition_Ncal}
\end{equation}
All these expressions must be calculated at $\psi=\psi_R$, a constant of motion. 
In principle the function $\vpar(\Jbold) $, a constant of motion, differs from the actual parallel velocity $\vpar$. However, since the mean parallel velocity is carried by passing particles, we will assume that  $\vpar(\Jbold) \simeq \vpar$.
In fact, the part of the radial derivative of $\Xi$ that depends on gradients is responsible for the diamagnetic current. 
The canonical distribution function can now be identified with the local Maxwellian Eq.(\ref{eq:local_Maxwellian}) if
\begin{equation}
   \Wpareq(\Heq,\psi,t)  =
   	    \Vpareq(\psi,t) + \frac{n_2}{n_h} 
   	    \frac{R_0\Teq(\psi,t)}{e_a} 
   	       \frac{\partial \Xi(\Heq,\psi)}{\partial \psi}
	\label{eq:generalised_force_balance}
\end{equation}
Then, the last step is to compute the partial derivatives of the thermodynamic potential on the resonant surface $\vpar=0$. 
Partial derivatives of the thermodynamic potential with the actions are also needed. 
As only the resonant values at $\vpar=0$ are needed, we use the expression Eq.(\ref{eq:mean_thermodynamic_potential}) of the thermodynamic potential, the definition Eq.(\ref{eq:def_Xi}) of the function $\Xi(\Jbold,t)$, and the expression of the parallel velocity versus the actions $(J_2,J_3)$ Eq.(\ref{eq:vpar_vs_J_3andJ2}) to find
\begin{eqnarray}
  \nonumber
	  \frac{1}{T_\ast} \frac{\partial \Ueq}{\partial J_2} & = & 
	       \frac{q n_3}{n_h} \frac{1}{e_a} \frac{\partial \Xi}{\partial \psi} 
	       + \frac{1}{\Teq } \frac{\Wpareq}{qR_0 } 	\\
  \nonumber
	  \frac{1}{T_\ast} \frac{\partial \Ueq}{\partial J_3} & = & 
	      - \frac{q n_2}{n_h} \frac{1}{e_a} \frac{\partial \Xi}{\partial \psi} 
	       + \frac{1}{\Teq } \frac{\Wpareq}{R_0 }     
	\label{eq:partial_derivatives_U_J2J3}
\end{eqnarray}
Hence,
\begin{equation}
 \nonumber
   	\frac{1}{T_\ast} \nbold \cdot \frac{\partial \Ueq}{\partial \Jbold} = 
   	     m_a \nbold \cdot \frac{\partial \vpar}{\partial \Jbold} \Wpareq =
   	      \frac{1}{\Teq} n_h \frac{\Wpareq}{q R_0}
	\label{eq:partial_derivative_Ueq_primary}
\end{equation}
From Eq.(\ref{eq:generalised_force_balance}), $\Wpareq$ can be replaced in terms of the toroidal velocity and gradients to give
\begin{equation}
 \nonumber
   \frac{1}{T_{\ast}} \nbold \cdot \frac{\partial \Ueq}{\partial \Jbold} = 
   	  \frac{1}{\Teq} n_h \frac{\Vpareq }{q R_0 } 
   	  + \frac{n_2}{q} \frac{1}{e_a}
   	      \frac{\partial \Xi}{\partial \psi}
	\label{eq:partial_derivative_new}
\end{equation}
Moving on to the secondary perturbation, we can use the relation
\begin{equation}
 \nonumber
   	\nbold' - \nbold \sigma = \frac{n_d}{n_h} \left(q \ebold_{2} - \ebold_{3}\right)
	\label{eq:secondary_nd}
\end{equation}
and the definition $n_d = n'_2 n_3 - n'_3 n_2$ to obtain
\begin{equation}
 \nonumber
   	\frac{1}{T_\ast} (\nbold' - \nbold \sigma) \cdot \frac{\partial \Ueq}{\partial \Jbold} = 
   	     \frac{n_d}{n_h} \frac{q}{e_a} \frac{\partial \Xi}{\partial \psi}
	\label{eq:partial_derivative_Ueq_secondary}
\end{equation}

\subsection[]{Secondary entropy production rate in the weakly collisional regime $\nueff \ll \omegab $}
\label{appendix:weakly_collisional_regime_exact}

\subsubsection{Solving the Fokker-Planck equation}
\label{appendix:Solving_Fokker_Planck_equation_weakly_coll_exact}
In that regime, solving the Fokker-Planck equation is faster in the new set of conjugate variables $(\alphab, \alphad, \Jb, \Jd)$ Eq.(\ref{eq:Fokker_Planck_linearised}). 
The perturbed distribution function is expanded in Fourier series, as was done for the Hamiltonian Eq.(\ref{eq:fourier_series_vs_alphabar})
\begin{equation}
	\nonumber
    \Ftilde' = e^{i n_d \alphad}
      \sum_{\nb = - \infty}^{+\infty} F'_{\nb} e^{i \nb \alphab} + C.C. 		
	\label{eq:fourier_series_vs_alphabar_distribution}
\end{equation}
The Fokker-Planck equation then reads
\begin{equation}
    - i n_d 
		\frac{\partial \Ffrakeq}{\partial \Jd} 
		 H'_{\nb}
    + i \left( \omega_d + \nb \Omegab \right)F'_{\nb} 
		= \Ccal[\Ftilde']_{\nb}
	\label{eq:bounce_average_FP_equation}
\end{equation}
As done previously, it is anticipated that most of the entropy production comes from the fast variation of the distribution near the primary resonant surface. The collision operator is essentially a diffusion operator
\begin{equation}
  \nonumber
		\Ccal[F] = \frac{1}{2}
		\frac{\partial }{\partial I_2}
		  \left( 
			  \frac{ \left\langle  \Delta I_2^2 \right\rangle}{\Delta t} 
		    \frac{\partial F}{\partial I_2}
			\right) = \frac{1}{2}
		\frac{\partial }{\partial \Omega}
		  \left( 
			  \frac{\left\langle  \Delta \Omega^2 \right\rangle}{\Delta t} 
		    \frac{\partial F}{\partial \Omega}
			\right)		
  \label{eq:collision_operator_diffusion}
\end{equation}
It is recalled that calculating a harmonic $\nb$ is equivalent to compute a Fourier transform over the angle $\alphab$. 
Hence
\begin{eqnarray} 
 \nonumber
 		\Ccal[ \Ftilde' ]_{\nb}  &=&  
			\int_{0}^{2\pi} \frac{d \alphab}{ 2 \pi }
			 e^{ - i \nb \alphab}
		  \frac{\partial }{\partial \Omega}
		   \left( D_\Omega
		     \frac{ \partial F' }
				      {\partial \Omega}
			 \right)   \\
 \nonumber
	   & = & \sum_{n'_b=-\infty}^{+\infty}	\int_{0}^{2\pi} \frac{d \alphab}{ 2 \pi }
			 e^{ + i \left( n'_b - \nb \right) \alphab}
		  \frac{\partial }{\partial \Omega}
		   \left( D_\Omega
		     \frac{ \partial F'_{n'_b} }
				      {\partial \Omega}
			 \right)
   \label{eq:collision_operator_n3_eq_0_full}
\end{eqnarray}
where $D_\Omega = \text{lim}_{\Delta t \rightarrow 0} \frac{\left\langle \Delta \Omega^2 \right\rangle}{2 \Delta t}$. The r.h.s of this relation is an approximation. 
This expression can be improved in several ways. 
First it is assumed that the collisional operator essentially regularises each Fourier component independently from the others, in other words $F'_{n'_b}$ is replaced by $F'_{\nb}$ in $\Ccal[ \Ftilde' ]_{\nb}$. 
Mathematically this would result from a faster variation of the oscillating exponential in $\alphab$ compared to the other terms.
In the special case where the bounce component $\nb=0$ prevails over the others, this approximation is well justified. 
Second the diffusion coefficient $D_\Omega$ is supposed not to depend on the angle $\alphab$, a reasonable approximation in general --- for Coulombian collisions, $D_\Omega$ depends on energy only. 
The integration element $d\alphab$ is restated using the relationship $dt = d\alphab/ \Omegab = d\theta_2/\Omega$, hence $d\alphab= \Omegab d\theta_2/\Omega $. 
Also the chain derivative rule $\partial_\Omega = \Omega \partial_{H_\Omega}$ is used when moving to the variable $H_\Omega$ instead of $\Omega$, hence assuming that ``fast'' variations are along the $H_\Omega$ variable, or equivalently the trapping parameter $K$. 
Moreover $F'_{\nb}$  depends only on $H_\Omega$ (thus not on $\alphab$) and $2 \pi C \Jb= \oint d \theta_2 \: \Omega$. The $\nb$ component of the collision operator then reads
\begin{equation}
  \nonumber
		\Ccal[ \Ftilde' ]_{\nb}  =  
 		    D_\Omega \Omegab
		   \frac{\partial }{\partial H_\Omega}
		    \left[
				 \left( 
				    \oint d \theta_2 \: \Omega  			     
				 \right) \;
				    \frac{ \partial F'_{\nb} }
				          { \partial H_\Omega }	
				\right] =
		     D_\Omega \Omegab
		   \frac{\partial }{\partial H_\Omega}
		    \left[C \Jb \;
				    \frac{ \partial F'_{\nb} }
				          { \partial H_\Omega }	
				\right] 				  
  \label{eq:collision_operator_n3_eq_0_reworked}
\end{equation}
The Fokker-Planck equation Eq.(\ref{eq:bounce_average_FP_equation}) then reads
\begin{equation}
 \nonumber
    \left( \omega_d + \nb \Omegab \right) F'_{\nb} + 
      i  C D_\Omega \Omegab
		   \frac{\partial }{\partial H_\Omega}
		    \left[\Jb \;
				    \frac{ \partial F'_{\nb} }
				          { \partial H_\Omega }	
				\right]
		= n_d \frac{\partial \Ffrakeq}{\partial \Jd} H'_{\nb}
	\label{eq:bounce_average_FP_equation_final}
\end{equation}
Using Eqs.(\ref{eq:expression_Jbar2}, \ref{eq:expression_Omegabar2}, \ref{eq:expression_hamiltonian_secondary_bounceaverage}), this equation can be rewritten with respect to the dimensionless energy variable $K = (H_\Omega + \omegab^2 )/( 2 \omegab^2 )$ 
\begin{equation}
    \left( \omega_d + i  \frac{2 \nb \omegab}{R(K)}  \right)
      \: F'_{\nb}(K) +
      i \frac{D_\Omega}{ \omegab^2} 
       \frac{1}{R(K)} 
		   \frac{\partial }{\partial K}
		    \left[ Q(K) \;
				    \frac{ \partial F'_{\nb} }
				          { \partial K }	
				\right]	
		= - n_d \frac{\partial \Ffrakeq}{\partial \Jd} \: 
		   \frac{h'}{2} \: L_{\sigma,\nb}(K) 
	\label{eq:bounce_average_FP_equation_final_vsK}
\end{equation}
with $R = 2 \partial_K Q$. For reasons given in Section \ref{Entropy production rate for a secondary perturbation}, the boundary condition at the border of the trapped domain $K=1$ is chosen as $F'_{\nb}(1)=0$, which holds for rippled tokamaks and small helical perturbations, at least for the bounce average component $\nb=0$. 
It is convenient to introduce 
a normalised distribution function 
\begin{equation}
		\hat{F}_{\nb}(K) = - \frac{ 2 \omega_d }
		                    { n_d \frac{\partial \Ffrakeq}{\partial \Jd} \: h' }
		             F'_{\nb}(K)
  \label{eq:normalised_distribution_function_appendix}
\end{equation} 
The normalised Fokker-Planck equation becomes
\begin{equation}
    i \hat{\nu} \left( G(K)   
		    \frac{ \partial^2 \hat{F}_{\nb}(K) }{\partial K^2}
		      + \frac{ \partial \hat{F}_{\nb}(K) }{ \partial K} \right) + 
		      \left( 1 + \frac{2 \nb  \hat{\omega}_b}{R(K)}  \right)  \hat{F}_{\nb}(K)
		= L_{\sigma,\nb}(K) 
	\label{eq:full_FP_equation_normalised_appendix}
\end{equation}
where $G(K) = \frac{2 Q(K)}{R(K)}$. 
Eq.(\ref{eq:full_FP_equation_normalised_appendix}) can be solved numerically using a standard finite difference scheme for arbitrary $\hat{\nu}$, and $\sigma$ parameters, and any $\nb$. 
It is nevertheless useful to devise approximate solutions to be compared with.
\subsubsection{Secondary entropy production rate}
\label{appendix:secondary_entropy_rate_exact}

Expanding all perturbed fields in Fourier series while keeping only the $\pm n_d $ and all $\nb$ components, the entropy production rate functional reduces to 
\begin{equation}
  \nonumber
	  \Scalres \left(\partial_\Jbold\Ufrakeqdag \right)	= 
	  		\frac{1}{T_\ast^2}  \sum_{\nb=-\infty}^{+\infty}
	       \int d\gamma \; \Ffrakeq \;
		     \left\{ \Htilde' ,  \Ufrakeqdag\right\}_{\nb}^{\ast} 
		       \Utilde'_{\nb} + C.C.
  \label{eq:entropy_production_resonant_secondary_bis}
\end{equation}
Nothing that 
\begin{equation}
  \nonumber
  \left\{ \Htilde' ,  \Ufrakeqdag\right\}_{\nb} =  i n_d   
	  \frac{\partial \Ufrakeqdag }{\partial  \Jd}  \Htilde'_{\nb} =
	  i n_d  \frac{\partial \Ufrakeqdag }{\partial  \Jd} 
	    \frac{h'}{2} L_{\sigma,\nb}(K)
 \label{eq:Poisson_bracket_bounce_average}
\end{equation}
and using Eq.(\ref{eq:normalised_distribution_function_appendix})
\begin{equation}
		U'_{\nb}(K) = - \frac {1}{ \omega_d } 
		     n_d \frac{\partial \Ufrakeq}{\partial \Jd} 
		\frac{h'}{2}  \hat{F}_{\nb}(K)  
  \label{eq:Uprime_vs_Fhat}
\end{equation}
one finds
\begin{equation}
  \nonumber
	  \Scalres  \left(\partial_\Jbold\Ufrakeqdag \right)  = 
	   \frac{1}{2 T_\ast^2}  \sum_{\nb=-\infty}^{+\infty}
	  \int d\gamma \: \Ffrakeq \:  h'^2 \: \frac{1}{\omega_d}
		  \left[ 
		     n_d  \frac{\partial \Ufrakeqdag}{ \partial \Jd}  
			\right]^2 L_{\sigma,\nb}(K) \Im \left[ \hat{F}_{\nb}(K)\right]   
  \label{eq:entropy_production_resonant_secondary_general}
\end{equation}
Using the relationship $\Jb = \lambda_m Q(K) $, and $\partial_K Q(K) =  R/2$, the element of integration reads 
\begin{equation}
	  d\gamma =  d\Scal d\Jb = \frac{1}{2}  d\Scal  \lambda_m R(K) dK 
  \label{eq:volume_integration_vs_K}
\end{equation}
where the integration element along the resonant surface is $d\Scal = (2\pi)^2 \: d\theta_3  \: dI_1 \: dI_3 $. The integral in $K$ can be isolated, and tabulated against $\hat{\nu}$ and $\sigma$, i.e.
\begin{equation}
	  \Scalres  \left( \partial_\Jbold\Ufrakeqdag \right)  = 
	   \frac{1}{2 T_\ast^2}  \Im
	  \int d\Scal \lambda_m \: \Ffrakeq \:  h'^2 \: \frac{1}{\omega_d}
		  \left[ 
		     n_d  \frac{\partial \Ufrakeqdag}{ \partial \Jd}  
			\right]^2  W(\sigma,\hat{\nu})
  \label{eq:entropy_production_resonant_secondary_general_final}
\end{equation}
where $W(\sigma,\hat{\nu})$ is a dimensionless ``weight'' function defined as
\begin{equation}
  \nonumber
	  W(\sigma,\hat{\nu}) = \frac{1}{2} \sum_{\nb=-\infty}^{+\infty}
	    \int_0^{1} dK R(K)
	         L_{\sigma,\nb}(K) \hat{F}_{\nb}(K)
  \label{eq:def_Phi}
\end{equation}
In general, the function $W(\sigma,\hat{\nu})$ must be computed numerically. 
However, exact results are available in the bounce average limit $\nb=0$ as presented in Appendix \ref{appendix:Analytical_form_of_the_weight_function_in_the_bounce_average_limit_at_high_collisionality}.

\subsection[]{Analytical expression of the secondary entropy production rate in the weakly collisional and adiabatic limit $\omegab \gg \omega_d$}
\label{appendix:secondary_entropy_rate_analytical}

\subsubsection{Solution to the Fokker-Planck equation}
\label{Solution to the Fokker-Planck equation}
If the frequency $\Omegab \sim \omegab$ is much larger than $\omega_d = n_d \Omegad$, the response of the harmonics $\nb \ne 0$ are smaller than those with $\nb = 0$ in the ratio $\omega_d / \omegab$. 
In this limit, only the single Fourier component $(\nb = 0 , n_d)$ of the distribution function, i.e. $F'_{0}$, needs be calculated. 
This is a common approximation for neoclassical transport in stellarators and rippled tokamaks. 
The Fokker-Planck equation becomes
\begin{equation}
    i \hat{\nu} \left( G(K)   
		    \frac{\partial^2 \hat{F}(K)}{\partial K^2}
		      + \frac{\partial \hat{F}(K)}{\partial K} \right) + \hat{F}(K)
		= L_{\sigma}(K) 
	\label{eq:bounce_average_FP_equation_normalised_appendix}
\end{equation}
where $\hat{F}$ stands for $\hat{F}_{\nb=0}$, and similarly $L_{\sigma}$ denotes $L_{\sigma,\nb=0}$.

The bounce average collision operator becomes much simpler when computed in the deeply trapped regime.
In this limit $R(K)\simeq 2$, $Q(K)\simeq K$, $G(K)=K$, so that Eq.(\ref{eq:bounce_average_FP_equation_final_vsK}) becomes
\begin{equation}
 \nonumber
   \hat{F} + i \hat{\nu}
		   \frac{\partial }{\partial K}
		    \left[K \;
				    \frac{ \partial \hat{F} }
				          { \partial K }	
				\right]	
		=  L_{\sigma}(K) 
	\label{eq:bounce_average_FP_equation_homogeneous_trapped}
\end{equation}
In the case where $\sigma \ll 1$, then $L_{\sigma}(K) \simeq 1$. A general solution can be built as a series in $K/\hat{\nu}$, where $\hat{\nu} = \frac{\nu_d} {\omega_d} $ - details can be found in Appendix \ref{Collisional boundary layer}. When $\hat{\nu}>1$, an approximate solution is
\begin{equation}
	\nonumber
    \hat{F} =  \frac{ 1 - K } { 1 - i \hat{\nu} }
	\label{eq:solution_large_nueff_F0prime}
\end{equation}
For $\sigma \gg 1$, the variable $K$ is not that convenient for practical calculations. It is replaced by the bounce angle $\thetab$. Using $L_{\sigma}(K) \simeq J_0( \sigma \thetab )$, and $K = \frac{\thetab^2}{4}$, the following expression is obtained
\begin{equation}
  \nonumber
		\Ccal[ \Ftilde' ]_{0} = \nu_d
		 	\frac{1}{ \thetab }		
       \frac{\partial }{\partial \thetab}
		    \left(
				  \thetab \;
				   \frac{ \partial F'_{0} }
				        { \partial \thetab }
			  \right)
  \label{eq:collision_operator_n2_eq_0_final}
\end{equation}
The Fokker-Planck equation is thus rephrased as
\begin{equation}
	\nonumber
    \hat{F} 
		+ i \hat{\nu} 
			\frac{1}{ \thetab }		
      \frac{\partial }{\partial \thetab}
		    \left(
				  \thetab \;
				   \frac{ \partial \hat{F}  }
				        { \partial \thetab }
			 \right) = J_0( \sigma \thetab )
	\label{eq:component_n2_eq_0_detailed}
\end{equation}
The solution $\hat{F} $ is requested to vanish at the trapped/passing boundary $\thetab = \pi$. 
The function $J_0(\sigma \thetab) $ is an exact solution of this differential equation, up to a multiplicative constant. 
However, it only approximately satisfies the boundary condition $J_0(\sigma \theta_m) \simeq 0$. 
Indeed an approximate expression of the Bessel function $J_0$ for large arguments is obtained from Eq.(\ref{eq:Bessel_function_asymptotics})
\begin{equation}
		J_0(\sigma \thetab) \simeq 
		\sqrt{ \frac{2}{ \pi \sigma \thetab } } 
		\cos \left( \sigma \thetab - \frac{\pi}{4} \right)
  \label{eq:Bessel_asymptotics}
\end{equation}
So it oscillates fast and is equal to zero via a change of the upper bound $\theta_m$ by a small shift of order\footnote{More precisely, if $k$ is the closest integer to $\frac{2}{\pi} \sigma \theta_m - \frac{1}{2}$, then $\theta_m$ should be chosen equal to $\frac{\pi}{2 \sigma} \left(k + \frac{1}{2}\right)$. 
Hence $\theta_m$ is shifted by $\nu \frac{\pi}{2 \sigma}$, where $\nu$ is a number between $0$ and $1$.} $1/\sigma$. 

In fine the two limits $\sigma \ll 1$ and $\sigma \gg 1$ can be summarised in a compact form
\begin{itemize}
	\item $\sigma \ll 1$
	  \begin{equation}
		   \hat{F}(\thetab) =  
			   \frac{1}
				      { 1 - i \hat{\nu} }	
		      \left(1 - \frac{\thetab^2}{4}\right) 
	   \label{eq:solution_sigma_le_1}
    \end{equation}
	\item $\sigma \gg 1$
	  \begin{equation}
		   \hat{F}(\thetab) = 
			   \frac{ 1 }
				      { 1 - i \hat{\nu} \sigma^2 }	
		      J_0(\sigma \thetab) 
	   \label{eq:solution_sigma_ge_1}
    \end{equation}	
\end{itemize}
We note that for the case $\sigma\ll 1$, the boundary condition $F'_{0}=0$ at $K=1$ translates to an approximate condition $\thetab = \theta_m = 2$, hence the variable $K$ should be preferred to the bounce angle $\thetab$ whenever possible.

\subsubsection{Analytical expression of the secondary entropy production rate}
For deeply trapped particles, the limit solutions Eqs.(\ref{eq:solution_sigma_le_1},\ref{eq:solution_sigma_ge_1}) of the Fokker-Planck equation can be used, to yield 
\begin{equation}
  \nonumber
	  \Scalres \left( \partial_\Jbold\Ufrakeqdag \right)  = 
	   \frac{1}{2 T_\ast^2} \Im  
	  \int d\gamma \Ffrakeq h'^2
			   \frac{ 
				       \left( 
							  n_d 
								\frac{\partial \Ufrakeqdag}
								     {\partial \Jbar_3} 
							 \right)^2
							}
				      { 
							  \omega_d - i \nu_d 
							}	
				\left( 1 - \frac{\thetab^2}{4} \right)
  \label{eq:entropy_production_resonant_secondary_low_sigma}
\end{equation}
when $\sigma \ll 1$, and 
\begin{equation}
  \nonumber
	  \Scalres \left( \partial_\Jbold\Ufrakeqdag \right)  = 
	   \frac{1}{2 T_\ast^2} \Im  
	  \int d\gamma \Ffrakeq h'^2
			   \frac{ 
				       \left( 
							  n_d 
								\frac{\partial \Ufrakeqdag}
								     {\partial \Jbar_3} 
							 \right)^2
							}
				      { 
							  \omega_d - i \nu_d \sigma^2 
							}	
		      J_0^2(\sigma \thetab) 
  \label{eq:entropy_production_resonant_secondary_high_sigma}
\end{equation}
when $\sigma \gg 1$. 
Let us spend some time on the element of phase space integration. 
It reads
\begin{equation}
	  d\gamma =  d\Scal d\Jb = 
		   d\Scal \frac{\thetab d\thetab}{2} \lambda_m
  \label{eq:volume_integration}
\end{equation}
The second equality is obtained by using $\Jb = \lambda_m \thetab^2/4 $, where $\lambda_m = 2 \sqrt{ h/C }$. 
A tractable expression of the entropy production rate functional is therefore
\begin{equation}
	  \Scalres  \left(\partial_\Jbold\Ufrakeqdag \right)  = 
	   \frac{1}{2 T_\ast^2} \Re  
	   \int d\Scal \lambda_m \Ffrakeq h'^2
			   \frac{ 
				       \left( 
							  n_d 
								\frac{\partial \Ufrakeqdag}
								     {\partial \Jd} 
							 \right)^2
							}
				      { 
							  \omega_d - i \nu_d \text{Max} ( 1 , \sigma^2 )
							}	
		      \Lambda_{\rm II}(\sigma) 
  \label{eq:entropy_production_resonant_secondary_3}
\end{equation}
where
\begin{eqnarray}
   \Lambda_{\rm II}(\sigma) &=& \frac{1}{2} 
	   \int_{0}^{\theta_m} d\thetab \thetab
	   \left\{
	     \begin{array}{c}
          \left( 1 - \dfrac{\thetab^2}{4} \right) 
					\text{  if } \sigma \ll 1 
                    \\[8pt]
					J_0^2(\sigma \thetab) 
					\text{  if } \sigma \gg 1
     \end{array}
    \right.
  \label{eq:definition_Lambda}
\end{eqnarray}
Asymptotic expressions of $\Lambda(\sigma)$ can be found in Appendix \ref{appendix:Analytical_form_of_the_weight_function_in_the_bounce_average_limit_at_high_collisionality}.
These entropy production rate functionals can easily be expressed in their covariant form using the method details in Appendix \ref{appendix:Covariant_expression_of_the_entropy_production_rate}.

\subsection{Covariant expression of the entropy production rate}
\label{appendix:Covariant_expression_of_the_entropy_production_rate}
It is of interest to derive expressions of the entropy production rate that are independent of the choice of variables, called here ``covariant''.

For the "unperturbed" set of variables, the objective is to compute a covariant form of $\nbold' \cdot  \partial_\Jbold \Ufrakeq$. 
The function $\Ufrakeq$ must be a function $\Ueq$ of the actions on the primary resonant surface only $\Jbold = \Jbold_R$. 
Using the relation $ \Jbold = \Jbold_R + \nbold I_2$, it appears that 
\begin{equation}
  \Ufrakeq(\Jbold) = \Ueq(\Jbold  - \nbold I_2(\Jbold)) 
	  \simeq \Ueq(\Jbold) - I_2(\Jbold) 
	  \left(\nbold \cdot \nabla_{\Jbold} \Ueq(\Jbold) \right)
  \label{eq:covariant_tangent_derivative_appendix}
\end{equation}
One then has
\begin{equation}
  \nonumber
		\nbold' \cdot  \frac{\partial \Ufrakeq}{\partial \Jbold} =
		\nbold' \cdot  \frac{\partial \Ueq}{\partial \Jbold} -
		\left( \nbold  \cdot  \frac{\partial \Ueq}{\partial \Jbold} \right) 
		\left( \nbold' \cdot  \frac{\partial I_2}{\partial \Jbold} \right) =
		(\nbold' - \sigma \nbold)  \cdot  \frac{\partial \Ueq}{\partial \Jbold}
  \label{eq:derivative_Ubareq_appendix}
\end{equation}
where the expression Eq.(\ref{eq:link_I_vs_J_linearised}) of $I_2 (\Jbold)$ has been used.
This allows to write the entropy production rate in the high collisionality limit as
\begin{equation}
  \nonumber
    \Scalres \left[ \partial_\Jbold\Ueqdag \right] =  
	    \frac{1}{T_\ast^2} \frac{\pi}{2}
	    \int d\gamma \; \delta( \nbold' \cdot \Omegabold ) \; 
	    \Feq(\Jbold)  \; h'^2 \;      
			   \left( (\nbold' - \sigma \nbold)  \cdot 
	          \frac{\partial \Ueqdag}{\partial \Jbold} 
			   \right)^2 
    \label{eq:appendix_value_entropy_production_collisional_secondary_collisional}
\end{equation}
For the set of variables associated with the first perturbation, let us first note that the element of surface along a resonant surface can be written
\begin{eqnarray} 
  \nonumber
    \lambda_m d\Scal & = & (2\pi)^2 d\theta_3  dI_1 dI_3 2 \sqrt{ \frac{h}{C} }\\
  \nonumber
     & = &  (2\pi)^2 d\theta_3 dI_1	dI_2 dI_3  
     \delta(\Omega) C 2 \sqrt{ \frac{h}{C} }\\
  \nonumber
     & = &  (2\pi)^2 d\theta_3 dI_1 dI_2 dI_3  
     \delta(\Omega) 2 \omegab \\ 
  \nonumber
     & = &  2 d\gamma \delta(\Omega)  \omegab 	
	 \label{eq:surface_element_covariant}
\end{eqnarray}
Of course, the $\delta$ function enforces all functions to be calculated at $I_2=0$, which leaves the action $I_3$ as an integration variable. 

The old actions are related to the new ones via the equation Eq.(\ref{eq:link_J_vs_I}), i.e. reads $\Jbold = \Jboldstar + I_2 \nbold + I_3 \tbold $, where $\tbold = \frac{1}{n_h} ( \frac{1}{\iota_{\ast}} \ebold_2 -  \ebold_3)$ is a tangent vector along the primary resonant surface, and $n_h = n_2  + n_3 /\iota_{\ast}$ is a primary helical wavenumber. Moreover, using $n_d = n'_2 n_3 - n_2 n'_3$ and $\sigma = \frac{n'_h}{n_h}$ with $n'_h = n'_2  + n'_3 /\iota_{\ast}$. The following relationship is found  
\begin{equation}
    \nbold' - \sigma \nbold = 
      \frac{n_d}{n_h} \left( \frac{1}{\iota_{\ast}} \ebold_2 -  \ebold_3 \right) =  n_d \tbold
	\label{eq:nprime_minus_sigman}
\end{equation}
The gradient along a the primary resonant function of the thermodynamic potential can then be computed 
\begin{equation}
		\left. 
		      \frac{\partial \Ufrakeq}{\partial \Jd} 
				\right|_{I_2=0} =
		\left.  
		      \frac{\partial J_2}{\partial I_3} 
				\right|_{I_2=0} \frac{\partial \Ueq}{\partial J_2} +
		\left.  
		      \frac{\partial J_3}{\partial I_3} 
				\right|_{I_2=0} \frac{\partial \Ueq}{\partial J_3} = 	
   \frac{1}{n_h} \left[ \frac{1}{\iota_{\ast}} \frac{\partial \Ueq}{\partial J_2} -  
       \frac{\partial \Ueq}{\partial J_3} \right] = \tbold \cdot \nabla \Ueq       
  \label{eq:derivative_U_resonant_surface}
\end{equation}
and therefore
\begin{equation}
		n_d \left. 
		      \frac{\partial \Ufrakeq}{\partial \Jd} 
				\right|_{I_2=0} =
		 (\nbold' - \nbold \sigma) \cdot 
		 \left.
		    \frac{ \partial \Ueq }
				     { \partial \Jbold }
		 \right|_{\Jbold = \Jbold_R}
  \label{eq:derivative_U_covariant}
\end{equation}
The identity Eq.(\ref{eq:derivative_U_covariant}) can be applied to the Hamiltonian $ \Hfrakeq$, thus providing a practical expression of the drift  frequency $\omega_d$
\begin{equation}
		\omega_d = 
		n_d \left. 
		      \frac{\partial \Hfrakeq}{\partial \Jd} 
				\right|_{I_2=0} =
		 (\nbold' - \nbold \sigma) \cdot 
		 \left.
		    \frac{ \partial \Heq }
				     { \partial \Jbold }
		 \right|_{\Jbold = \Jbold_R}
  \label{eq:drift_frequency_covariant_expression}
\end{equation}
Combining all these expressions, a covariant expression of the entropy production rate functional is found Eq.(\ref{eq:entropy_production_resonant_secondary_general_final}) reads
\begin{eqnarray} 
  \nonumber
	  \Scalres  \left( \partial_\Jbold\Ufrakeqdag \right)  &=& 
	   \frac{1}{T_\ast^2}  
	  \int \: d\gamma \:  \delta(\Omega) \omegab \: \Feq \:  h'^2 \:  \\
	& & \frac{1}{\omega_d} \Im \left[ W
	       \left( 
		           \sigma , 
		           \frac{\nu_d}
		                {\left(\nbold' - \nbold \sigma \right) 
		                  \cdot \Omegabold } 
		     \right) \right]
		\left( 
	  	\left(\nbold' - \nbold \sigma \right)  \cdot 
			  \frac{\partial \Ueqdag}{\partial \Jbold} 
		\right)^2
  \label{eq:entropy_production_resonant_secondary_final_covariant_appendix}
\end{eqnarray}
In principle, all these integrals should only cover a small volume in the vicinity of the reference point $\Jbold = \Jboldstar$, located on the resonant surface $\Omega=0$. 
Indeed, the calculation was run via an expansion in the neighbourhood of this point. 
This is not a great difficulty given that neoclassical theory is in fact local. 
Hence, the formula above can be first computed in the vicinity of $\Jbold = \Jboldstar$, and then extended to the whole action space by continuity. 
In practice, this means that these expressions can be used at any arbitrary point.

\section{Collisional boundary layer}  
\label{Collisional boundary layer}
In this Section, we consider the special limit of a bounce average Fokker-Planck equation, and deeply trapped particles in the primary perturbation.
The equation to be solved is
\begin{equation}
    \omega_d  \: F'_{0}(K) + i \frac{D_\Omega}{2 \omegab^2} 
		   \frac{\partial }{\partial K}
		    \left[K \;
				    \frac{ \partial F'_{0} }
				          { \partial K }	
				\right]	
		= -  n_d \frac{\partial \Ffrakeq}{\partial \Jd} \: \frac{h'}{2}   
	\label{eq:bounce_average_FP}
\end{equation}
It is convenient to introduce a normalised collision frequency $\hat{\nu} = \frac{D_\Omega} {2 \omegab^2 \omega_d} $, a normalised energy variable $k = \frac{K}{\hat{\nu}}$, and the normalised distribution function 
\begin{equation}
		\hat{F}(K) = - \frac{ 2 \omega_d F'_{0}}
		                    { n_d \frac{\partial \Ffrakeq}{\partial \Jd} \: h' }
  \label{eq:normalised_distribution_function}
\end{equation}
The equation Eq.(\ref{eq:bounce_average_FP}) becomes after normalisation
\begin{equation}
	\nonumber
    \hat{F} + i  
		   \frac{\partial }{\partial k}
		    \left[k \;
				    \frac{ \partial \hat{F} }
				          { \partial k }	
				\right]	= 1
	\label{eq:bounce_average_FP_normalised}
\end{equation}
The solution of the homogeneous equation that is regular at $k=0$ is $J_0 \left( 2 \sqrt{-ik} \right)$, where $J_0$ is the Bessel function of the first kind and zeroth order. 
Moreover $\hat{F}=1$ is obviously a special solution of Eq.(\ref{eq:bounce_average_FP_normalised}). Hence $\hat{F} = 1 - C J_0 \left( 2 \sqrt{-ik} \right)$, where $C$ is a constant complex number. 
The latter is readily determined using the boundary condition $\hat{F}=0$ at $K=1$, hence at $k_c=1/\hat{\nu}$. 
Hence, the solution is
\begin{equation}
	\nonumber
    \hat{F}(k) 	= 1 - \frac{J_0 \left( 2 \sqrt{-ik} \right)}{J_0 \left( 2 \sqrt{-ik_c} \right)} 
	\label{eq:solution_bounce_average_FP_normalised}
\end{equation}
A link can be established with the solution derived by \cite{Beidler_1987} (in a more general case). 
Using Frobenius method, a series can be devised for the Bessel function near the origin, that reads
\begin{equation}
    J_0(x) = \sum_{m=0}^{+\infty}   \frac{(-1)^m}{\left[ m!\right]^2}
                \left( \frac{x}{2} \right)^{2m}
	\label{eq:Bessel_function_series}
\end{equation}
Replacing $x$ by $2 \sqrt{-ik}$, it appears readily that the $J_0 \left( 2 \sqrt{-ik} \right) = h(k) + i g(k)$ where\footnote{The function $h$ should not be confused with the Hamiltonian}
\begin{equation} 
	\nonumber
     g = \sum_{n=0}^{+\infty}  (-1)^n \frac{k^{2n+1}}{\left[(2n+1)!\right]^2} \text{~~~~;~~~~} 
	   h = \sum_{n=0}^{+\infty}  (-1)^n \frac{k^{2n}}{\left[(2n)!\right]^2}
	\label{eq:def_g_h}
\end{equation} 
These functions verify the following identities
\begin{eqnarray} 
	\nonumber
    & &  \frac{\partial }{\partial k}
		       \left(k \; \frac{ \partial g } { \partial k } \right) - h(k) =0
	 \label{eq:PDE_g} \\
	\nonumber	 
    & &  \frac{\partial }{\partial k}
		       \left( k \; \frac{ \partial h } { \partial k } \right)  + g(k)  =  0 		 
	 \label{eq:PDE_h}
\end{eqnarray}
The general solution of the bounce Fokker-Planck equation is thus
\begin{equation}
	\nonumber
        F'_{0}(K) = -  n_d \frac{\partial \Ffrakeq}{\partial \Jd} \: \frac{h'}{2} 
        \left[ 
           1 - \frac{J_0 \left( 2 \sqrt{-i \frac{K}{\hat{\nu}} } \right)}
                    {J_0 \left( 2 \sqrt{-i \frac{1}{\hat{\nu}} } \right)} 
        \right]
	\label{eq:solution_bounce_average_FP_normalised_general}
\end{equation}
The series Eq.(\ref{eq:Bessel_function_series}) rapidly converges whenever $\hat{\nu}\gg 1 $. 
Conversely, the convergence is slow in the low collisionality limit $\hat{\nu} \ll 1 $. 
The reason is the formation of a collision boundary layer near $K=1$. 
In the medium to strong collision regime, a proxy for the function $f$ is
\begin{equation}
	\nonumber
    \hat{F} = \frac{1 - \hat{\nu} k}{1 - i \hat{\nu}}
	\label{eq:solution_large_nuhat_f}
\end{equation}
or equivalently
\begin{equation}
	\nonumber
    F'_{0}(K)= -  n_d \frac{\partial \Ffrakeq}{\partial \Jd} \: \frac{h'}{2}
        \frac{1 - K}{\omega_d - i \nu_d}
	\label{eq:solution_large_nuhat_F0prime}
\end{equation}
\section{Analytical form of the weight function in the bounce average limit at high collisionality.}
\label{appendix:Analytical_form_of_the_weight_function_in_the_bounce_average_limit_at_high_collisionality}

In the bounce average limit $\nb=0$, an analytic expression of the weight function $W$ can be drawn when $\sigma=0$, since $L_{0}(K) = 1$ and $\hat{F}(K)$ is given by Eq.(\ref{eq:solution_large_nuhat_Fhat}). 
When the collisionality $\hat{\nu}$ is large enough, the expression of the weight $W$ carries two asymptotic limits.

\subsection[]{Limit $\sigma \rightarrow 0$}
In that case
\begin{equation}
    W(\sigma,\hat{\nu}) =  \frac{1}{1 - i \hat{\nu}} \Lambda_0
	\label{eq:value_Phi_sigma0}
\end{equation}
where
\begin{equation}
  \nonumber
	  \Lambda_0 =  \frac{1}{2} \int_{0}^{1} dK R(K) ( 1 - K ) =
	     \int_{0}^{1} dK ( 1 - K )
			  \int_{-\thetab}^{+\thetab} \frac{d\theta_2}{2\pi}
		      \frac{1}{ \sqrt{ K - \sin^2 \left( \frac{\theta_2}{2} \right) } }
	 \label{eq:calculation_Lambda__volume_modified}
\end{equation}
The integrals can be permuted to yield the following sequence of steps
\begin{eqnarray} 
  \nonumber
    \Lambda_0 & = &	
			\int_{-\pi}^{\pi} \frac{d\xi}{2\pi}
	    \int_{\sin^2 \left( \frac{\xi}{2} \right)}^{1} dK 
		  \frac{1 - K}
			     { \sqrt{ K - \sin^2 \left( \frac{\xi}{2} \right) } } \\
  \nonumber
     & = & \frac{2}{\pi} 
			\int_{0}^{\frac{\pi}{2}} du
	    \int_{\sin^2 u}^{1} dK 
		  \left( 
			   \frac{1 - \sin^2 u}
			         { \sqrt{ K - \sin^2 u  } }
					- \sqrt{ K - \sin^2 u  }
		  \right) \\
	  \nonumber
     & = & \frac{2}{\pi} 
			\int_{0}^{\frac{\pi}{2}} du
		  \left[ 
			   2 \left( 1 - \sin^2 u \right)
			   \sqrt{ K - \sin^2 u }
					- \frac{2}{3} \left( K - \sin^2 u \right)^{3/2}
		  \right]_{\sin^2 u}^1	\\
	  \nonumber
     & = & \frac{8}{3\pi} 
			\int_{0}^{\frac{\pi}{2}} du \cos^3u \\
	  \nonumber
     & = & \frac{16}{9\pi} 	
	 \label{eq:calculation_Lambda_K}
\end{eqnarray}

\subsection[]{Limit $\sigma \rightarrow +\infty$}
In that case
\begin{equation}
  \nonumber
    W(\sigma,\hat{\nu}) =  \frac{1}{1 - i \hat{\nu} \sigma^2 } \Lambda_{\infty}(\sigma) 
	\label{eq:value_Phi_sigma_large}
\end{equation}
where
\begin{equation}
  \nonumber
	  \Lambda_\infty =  \frac{1}{2} \int_{0}^{\theta_m} d\thetab
	      \thetab J_0^2(\sigma \thetab) = \frac{1}{2\sigma^2} 
	        \int_{0}^{2\sigma} dz z J_0^2(z) 
	 \label{eq:calculation_Lambda_high_sigma}
\end{equation}
in the limit of deeply trapped particles, since $\theta_m=2$ in this approximation. 
A primitive of $z J_0^2(z)$ is $\frac{1}{2} z^2 [J_0^2(z) + J_1^2(z)]$, hence
\begin{equation}
  \nonumber
	  \Lambda_\infty =  \left[ J_0^2(2\sigma) + J_1^2(2\sigma) \right] 
	   \simeq \frac{1}{\pi \sigma}
	 \label{eq:calculation_Lambda_high_sigma_exact}
\end{equation}
using asymptotic expressions of $J_0,J_1$ at large arguments. 

\subsection[]{A proxy weight function for all $\sigma$ when $\hat{\nu} \gg 1$}

A numerical computation of $W$ indicates that expressions derived above are accurate only for $\hat{\nu} \gg 1$.
In this limit, an expression covering all values of $\sigma$ is
\begin{equation}
    W(\sigma,\hat{\nu}) = \frac{i}{\hat{\nu}(\sigma)}  \Lambda_{\rm II}(\sigma) 
	\label{eq:value_Phi_sigma_all_sigma}
\end{equation}
where 
\begin{equation}
    \frac{1}{\Lambda_{\rm II}(\sigma)} = \frac{1}{\Lambda_0} + \frac{1}{\Lambda_{\infty}(\sigma)}
\end{equation}
and
\begin{equation}
    \hat{\nu}(\sigma) = \text{Max} ( 1 , \sigma^2 ) \: \hat{\nu}
\end{equation}

\newpage
\bibliographystyle{unsrt}
\bibliography{bibliography_neoclassical}

\end{document}